%% file: mgr.tex
\documentclass[11pt,a4paper,twoside,openright]{report}
\usepackage[twoside,left=3.7cm]{geometry}
\usepackage[utf8]{inputenc}
\usepackage{polski}
\usepackage[english,polish]{babel}
\usepackage{amsmath}
\usepackage{lmodern}
\usepackage{graphicx}
\usepackage{changepage}
\usepackage{placeins}
\usepackage{lscape}
\usepackage{afterpage}
\usepackage{emptypage}
\usepackage{geometry}
\usepackage{wasysym}
\usepackage{array}
\usepackage{relsize}
\usepackage{caption}
\usepackage{subcaption}
\usepackage{sectsty}
\usepackage{float}
\usepackage{wrapfig}
\usepackage{gensymb}
\usepackage{hyperref}
\usepackage[onehalfspacing]{setspace}
\usepackage[numbers,sort&compress]{natbib}
\usepackage{appendix}
\RequirePackage{xcolor}
\hypersetup{
    colorlinks,
    linkcolor={blue!50!black},
    citecolor={red!50!black},
    urlcolor={blue!80!black}
}
\title{Drift Chamber Track Reconstruction for the P349 Antiproton Experiment}

\sectionfont{\fontsize{13}{15}\selectfont}
\subsectionfont{\fontsize{11}{15}\selectfont}

\begin{document}
\thispagestyle{empty}
\begin{center}
\begin{large}
\textbf{Jagiellonian University in Cracow}\\
Faculty of of Physics, Astronomy and Applied Computer Science\\
\vspace{120pt}
\textbf{Dominika Alfs}\\
\end{large}
\vspace{36pt}
{\Large \textbf{Drift Chamber Track Reconstruction for the P349 Antiproton Experiment}\\ \vspace{12pt}}
Master Thesis
\end{center}
\vspace{138pt}
\begin{flushright}
Supervised by\\
dr inż. Marcin Zieliński\\
Institute of Physics\\
Division of Nuclear Physics
\end{flushright}
\vspace{24pt}
\begin{center}
Kraków, 2017
\end{center}
\newpage
\thispagestyle{empty}
\null
\newpage
\thispagestyle{empty}
\null
\vspace{60pt}
\begin{center}
{\Large \textbf{Abstract}}
\end{center}
The aim of this thesis was to perform the drift chamber calibration and charged particles 3d track reconstruction for the P349 antiproton polarization experiment. A~dedicated procedures were designed, implemented and tested on the experimental data for the the D1 drift chamber. The calibration consisted of the drift time offsets determination, estimation of the initial drift time - space relations by means of the homogeneous irradiation method and the iterative procedure for the time - space relations optimization. Calibration curves for all wire planes of the detector were determined. The obtained uncertainties of the hit position reconstruction are in the range of 150 - 220 $\rm \mu m$. Furthermore, based on the prepared 3d track reconstruction angular distribution of tracks passing through the drift chamber were determined.
\vspace{50pt}

\begin{center}
{\Large \textbf{Streszczenie}}
\end{center}
Celem pracy było przeprowadzenie kalibracji komory dryfowej oraz przygotowanie procedury do trójwymiarowej rekonstrukcji torów cząstek na potrzeby eksperymentu P349 dotyczącego określenia stopnia polaryzacji antyprotonów w procesie produkcji. W ramach kalibracji określone zostały offsety widm czasu dryfu, wyznaczono krzywe kalibracyjne dla każdej płaszczyzny detekcyjnej metodą jednorodnego naświetlenia, a~następnie przeprowadzono ich iteracyjną optymalizację. Otrzymane niepewności odległości przejścia cząstki naładowanej od drutu czułego mieszczą się w zakresie 150 - 220 $\mu m$. Ponadto, na podstawie rekonstrukcji zdarzeń w trzech wymiarach otrzymano rozkłady kątowe cząstek przechodzących przez komorę dryfową D1.
\vspace{24pt}
\selectlanguage{english}
\null
\newpage
\thispagestyle{empty}
\tableofcontents
\thispagestyle{empty}
\newpage
\cleardoublepage
\setcounter{page}{1}
\include{Motivation}
\include{MeasurementOfPolarization}
\include{ExperimentalSetup}
\include{TrackIdentification}

\include{CalibrationProcedure}
\include{TrackReconstruction}
\include{Conclusions}

\begin{appendices}
\chapter{Structure of the data analysis program}
\begin{figure}[!hb]
\begin{center}
\includegraphics[width=\textwidth]{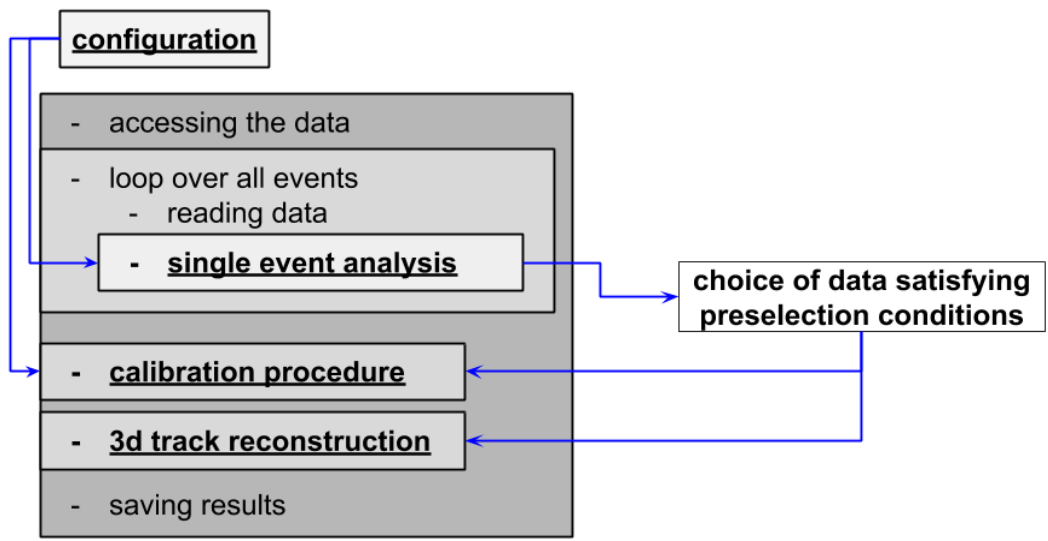}
\end{center}
\caption{Scheme of the program for the calibration and track reconstruction. Details of the part \textbf{\underline{Signle event analysis}} are shown in the Fig. \ref{scheme-single-event}}\label{scheme-code}
\end{figure}

In order to perform the calibration procedure and track reconstruction a~C++~\citep{isocpp} program was prepared. Its structure was designed in a~way which allows for simple introducing modifications and extensions. The program structure is shown in Fig. \ref{scheme-code}.

\begin{figure}[!hb]
\begin{center}
\includegraphics[width=0.95\textwidth]{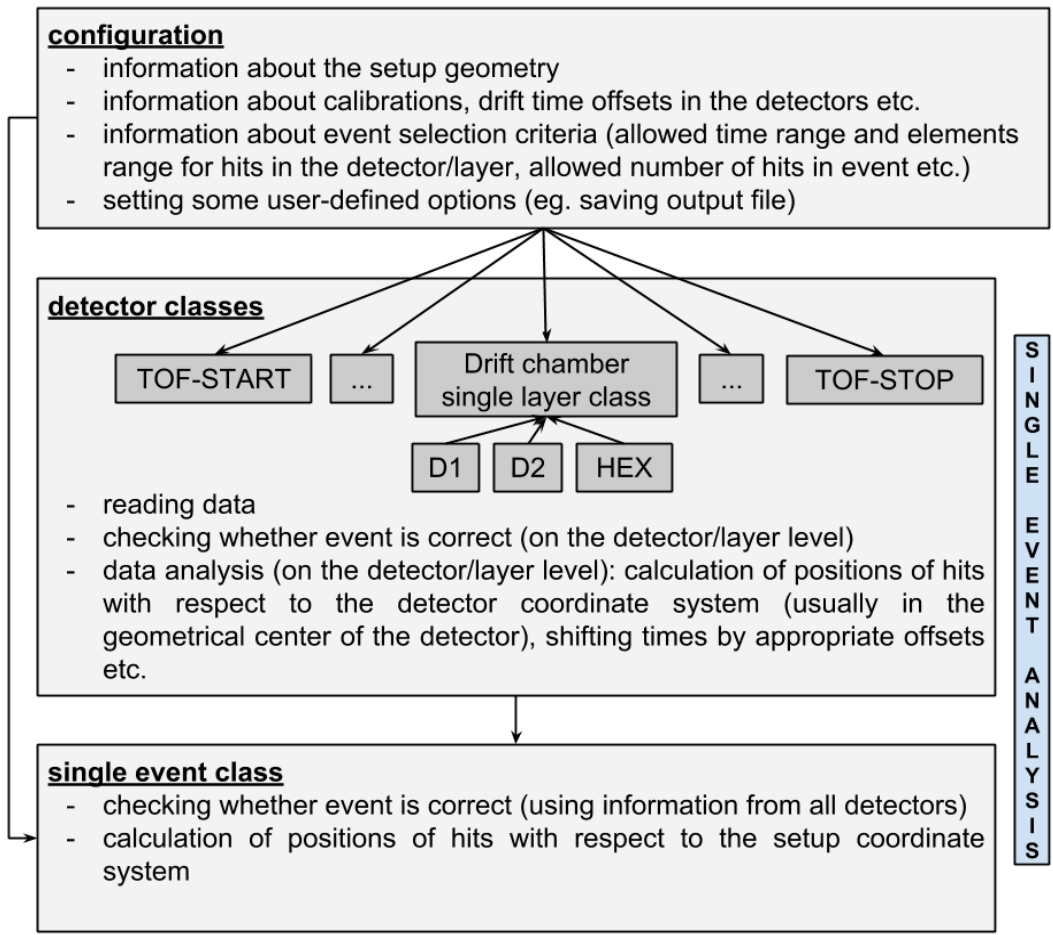}
\end{center}
\caption{Scheme of the part of the program for the single event analysis.}\label{scheme-single-event}
\end{figure}

Program inputs are:
\begin{itemize}
\item paths to any number of root files of the same structure with raw experimental data,
\item path to tree with data in the root file,
\item number of events to process or requirement to process \textit{all} events available in the provided root files,
\item name of the output file.
\end{itemize}

For the correct program operation g++ version 4.8.4,  Root Data Analysis Framework~\citep{root} version 5.34/26 and Boost~\citep{boost} version 1.54 is needed.

\chapter{Results of calibration for all wire planes in the D1 drift chambers}
Results of the calibration procedure for all wire planes in D1 drift chamber after 7 iterations are presented.

\begin{adjustwidth*}{-2cm}{-2cm}
\begin{figure}[!h]
\begin{center}
\begin{subfigure}[t]{0.45\textwidth}
\includegraphics[width=\textwidth]{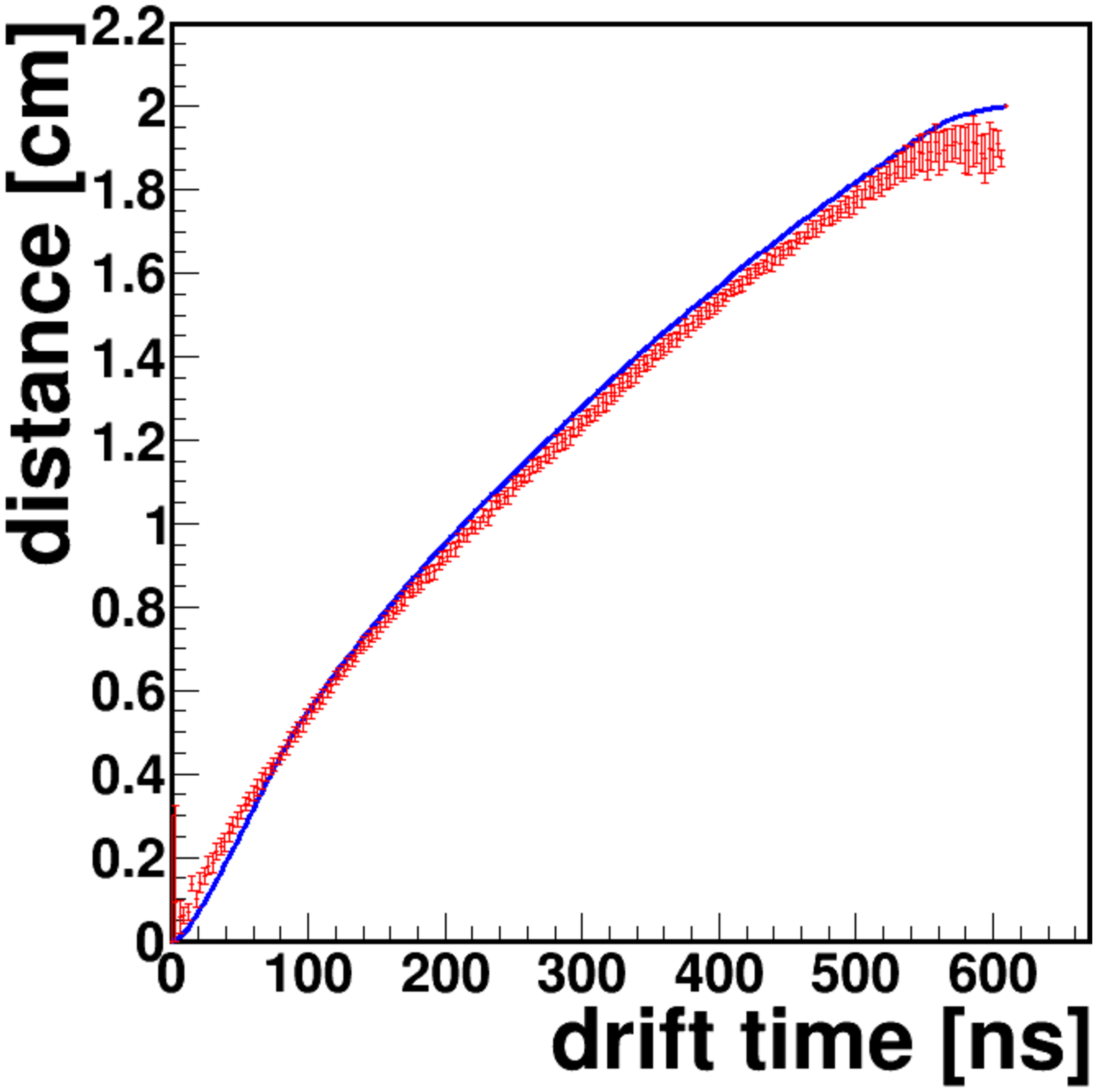}
\caption{Drift time - space relation.}
\end{subfigure}\hspace{25pt}
\begin{subfigure}[t]{0.45\textwidth}
\includegraphics[width=\textwidth]{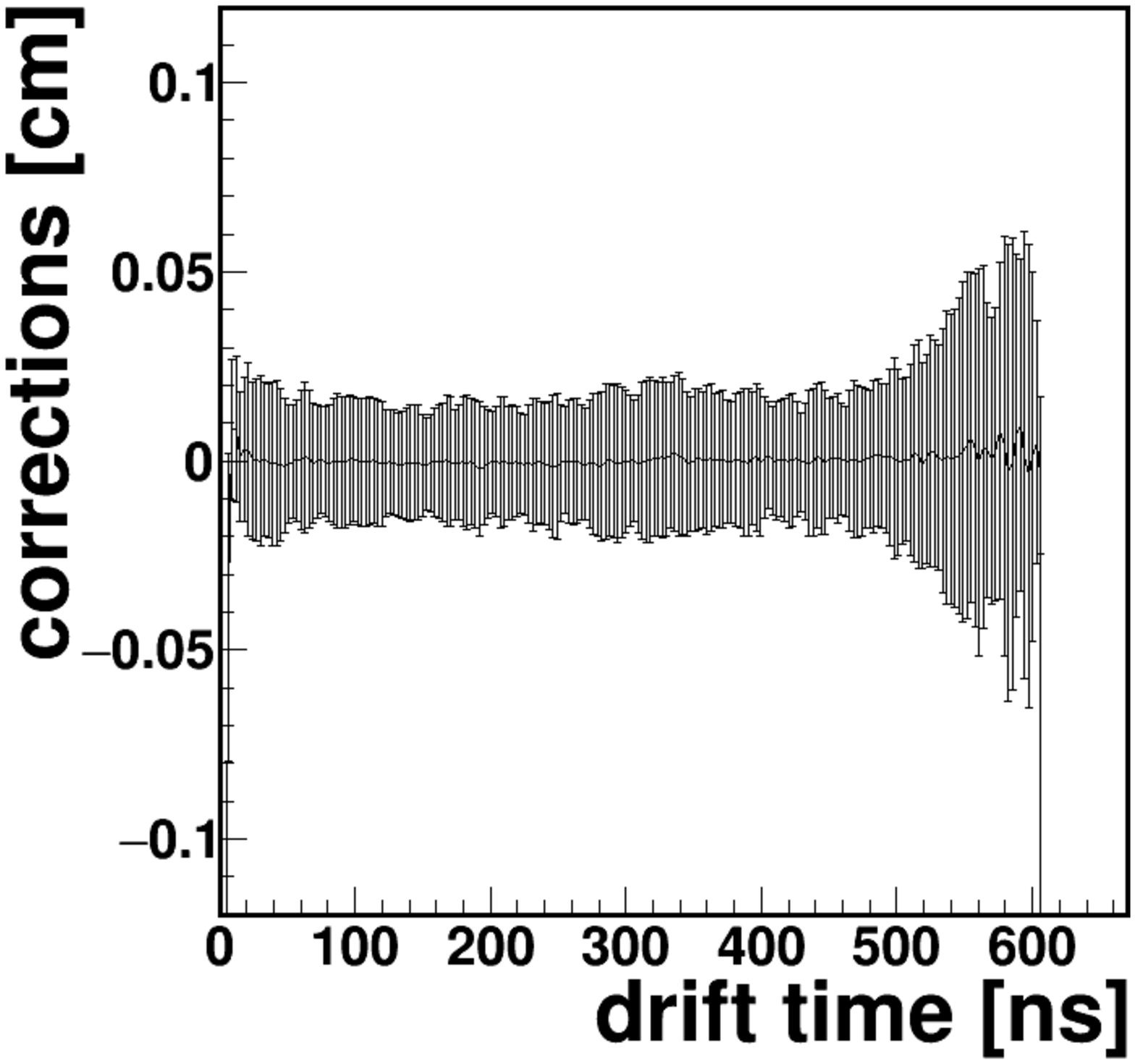}
\caption{Corrections to the drift time - space relation.}
\end{subfigure}
\caption{1st wire plane.}
\end{center}
\end{figure}
\vspace{-15pt}
\begin{figure}[!h]
\begin{center}
\begin{subfigure}[t]{0.45\textwidth}
\includegraphics[width=\textwidth]{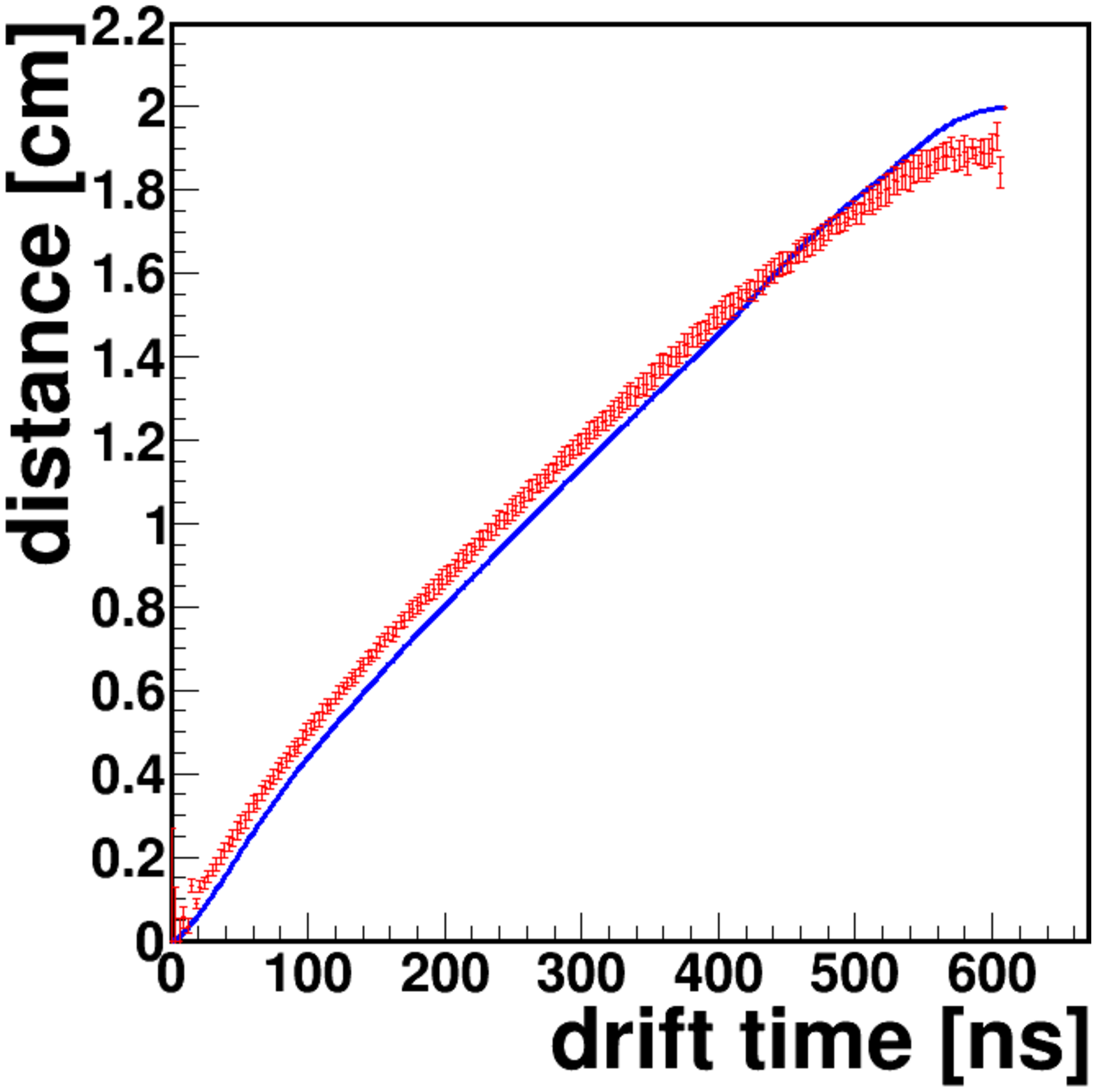}
\caption{Drift time - space relation.}
\end{subfigure}\hspace{25pt}
\begin{subfigure}[t]{0.45\textwidth}
\includegraphics[width=\textwidth]{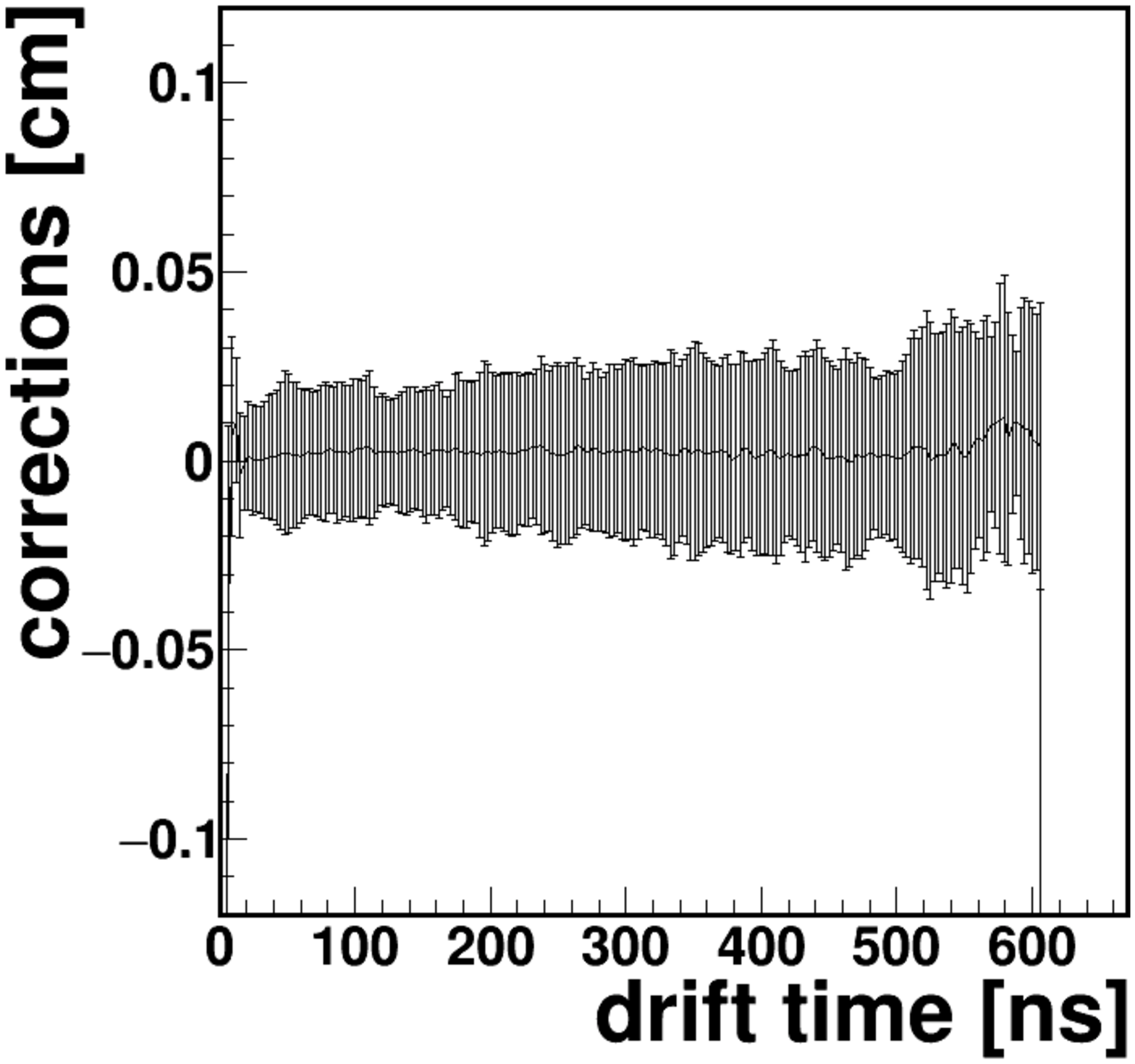}
\caption{Corrections to the drift time - space relation.}
\end{subfigure}
\caption{2nd wire plane.}
\end{center}
\end{figure}
\begin{figure}[!h]
\begin{center}
\begin{subfigure}[t]{0.45\textwidth}
\includegraphics[width=\textwidth]{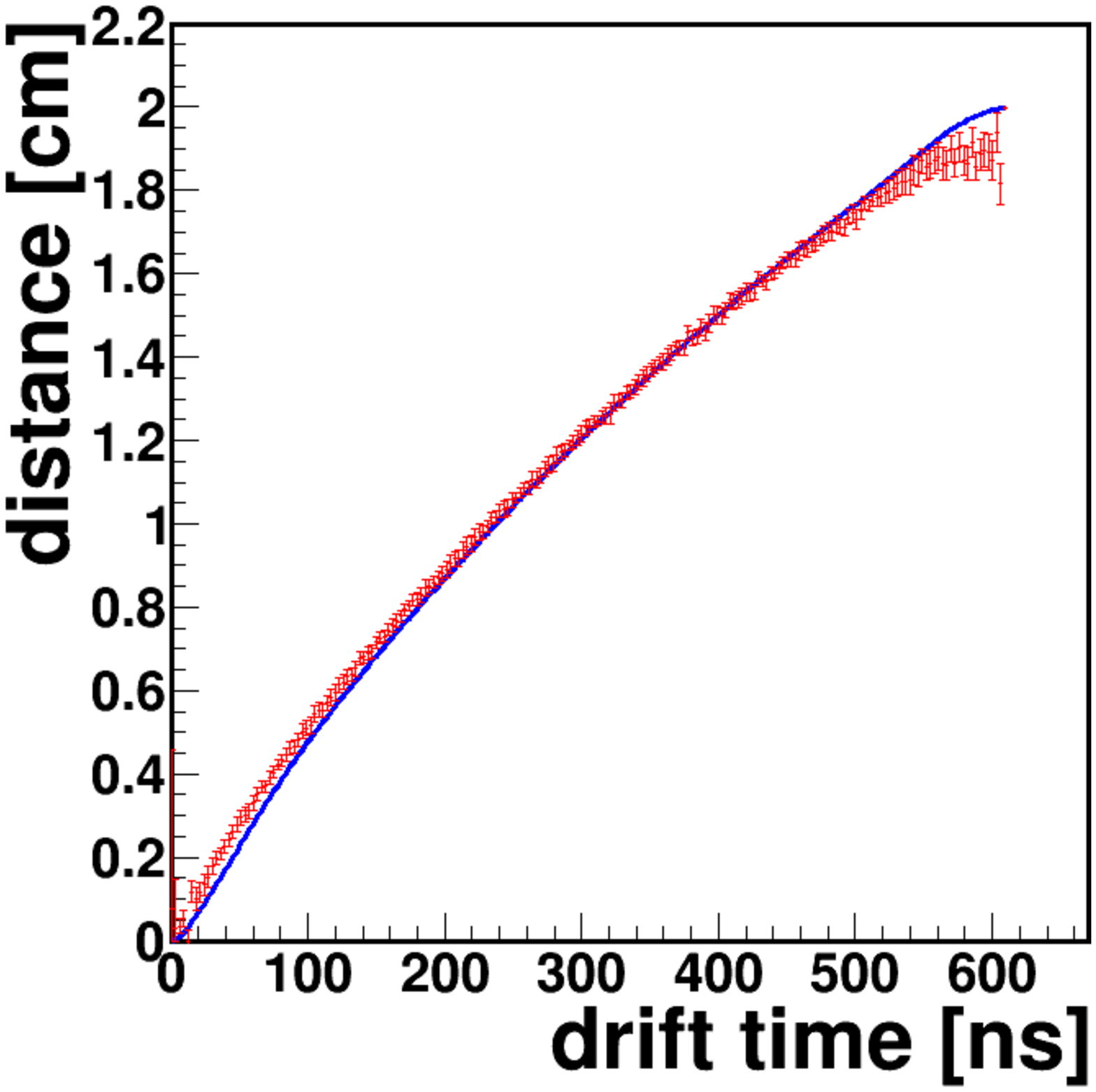}
\caption{Drift time - space relation.}
\end{subfigure}\hspace{25pt}
\begin{subfigure}[t]{0.45\textwidth}
\includegraphics[width=\textwidth]{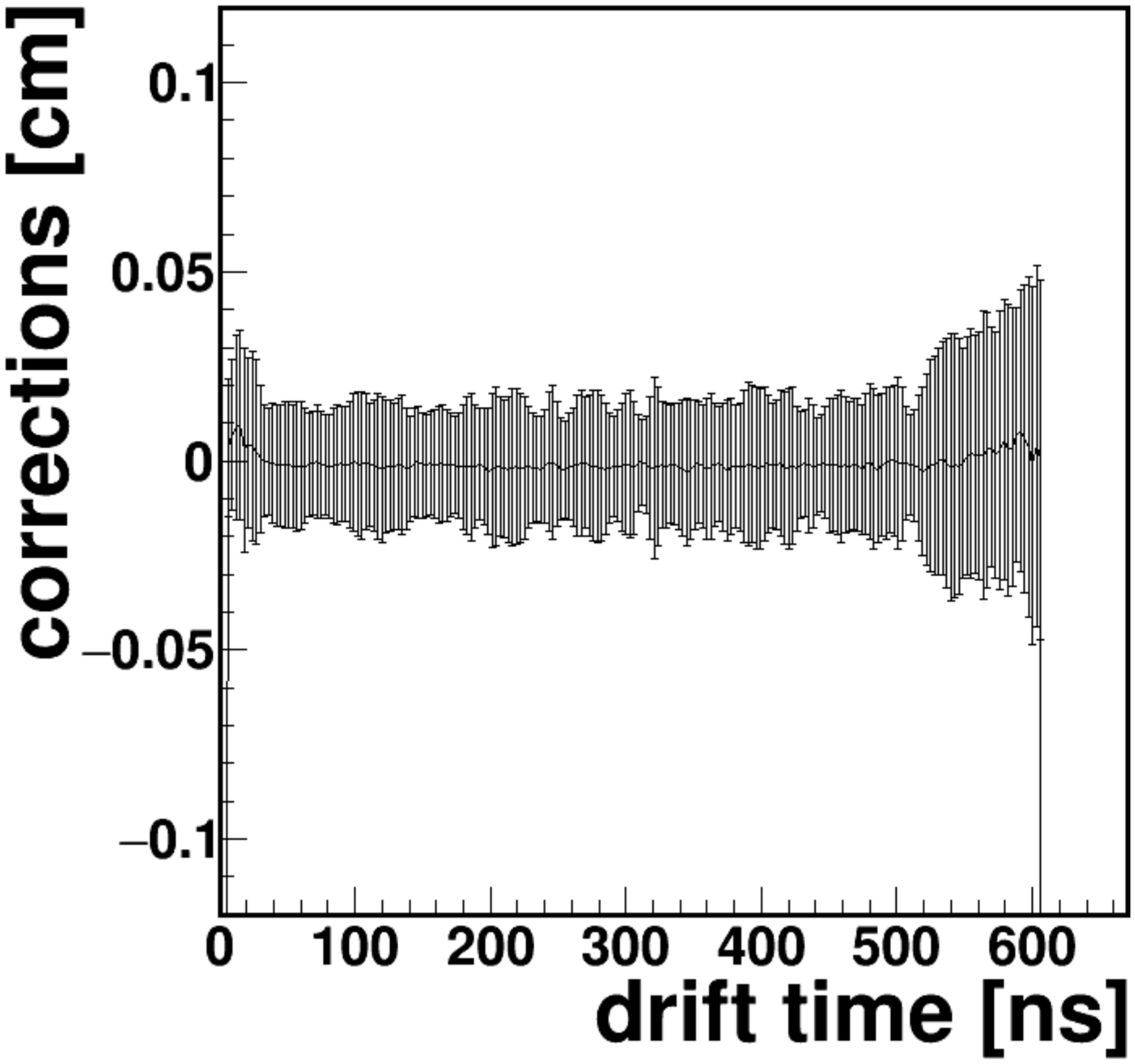}
\caption{Corrections to the drift time - space relation.}
\end{subfigure}
\caption{3rd wire plane.}
\end{center}
\end{figure}
\begin{figure}[!h]
\begin{center}
\begin{subfigure}[t]{0.45\textwidth}
\includegraphics[width=\textwidth]{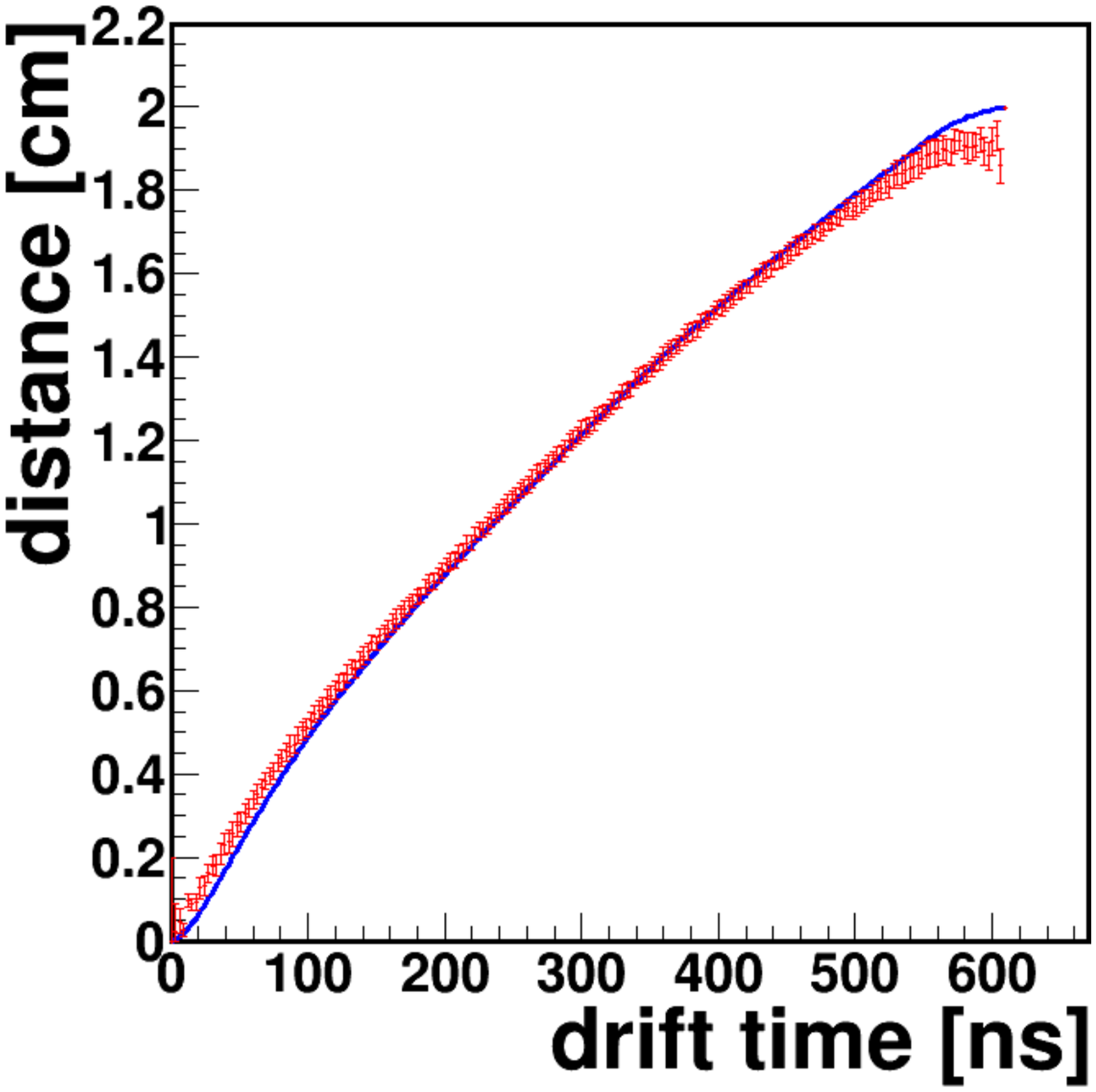}
\caption{Drift time - space relation.}
\end{subfigure}\hspace{25pt}
\begin{subfigure}[t]{0.45\textwidth}
\includegraphics[width=\textwidth]{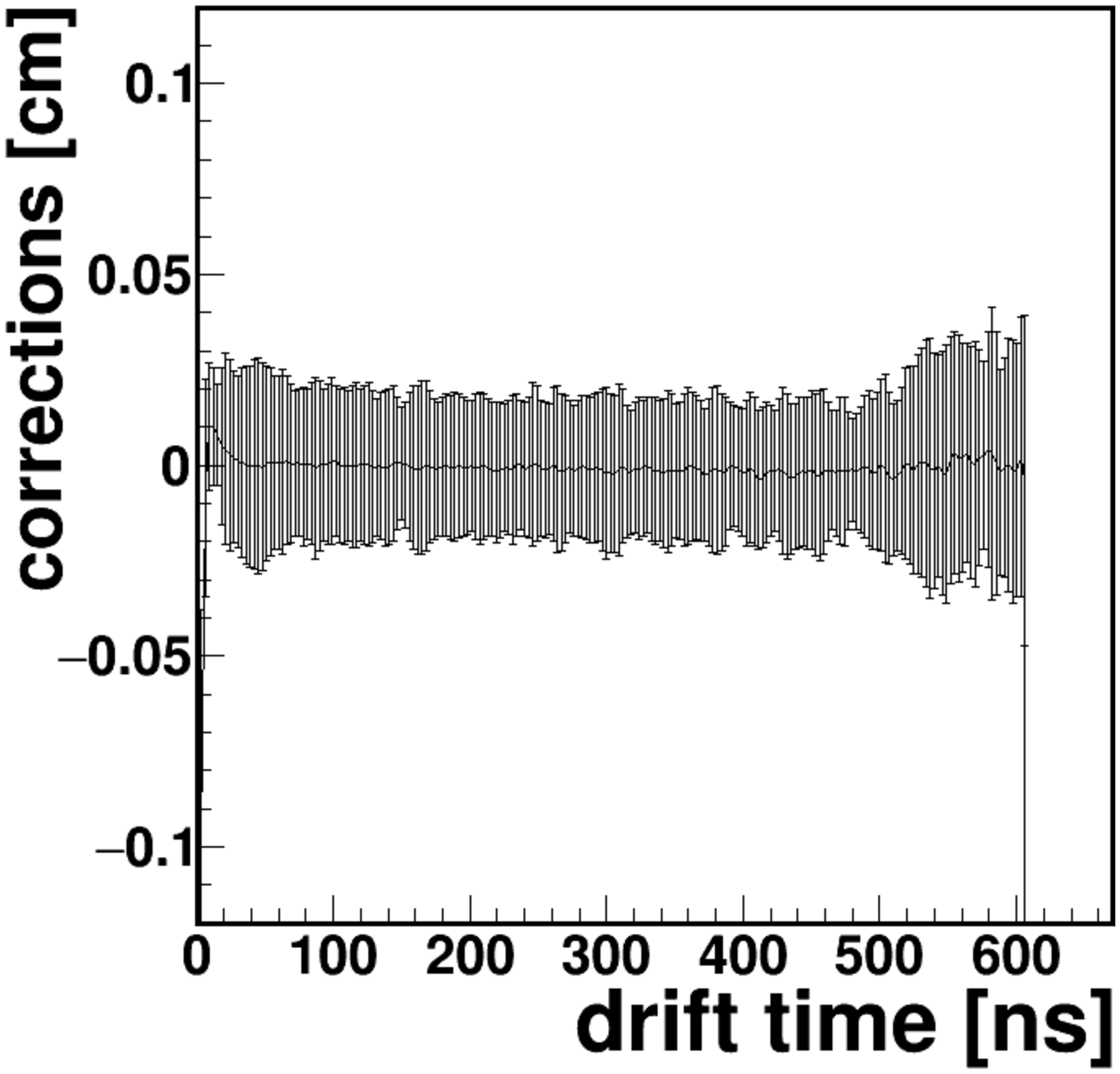}
\caption{Corrections to the drift time - space relation.}
\end{subfigure}
\caption{4th wire plane.}
\end{center}
\end{figure}
\begin{figure}[!h]
\begin{center}
\begin{subfigure}[t]{0.45\textwidth}
\includegraphics[width=\textwidth]{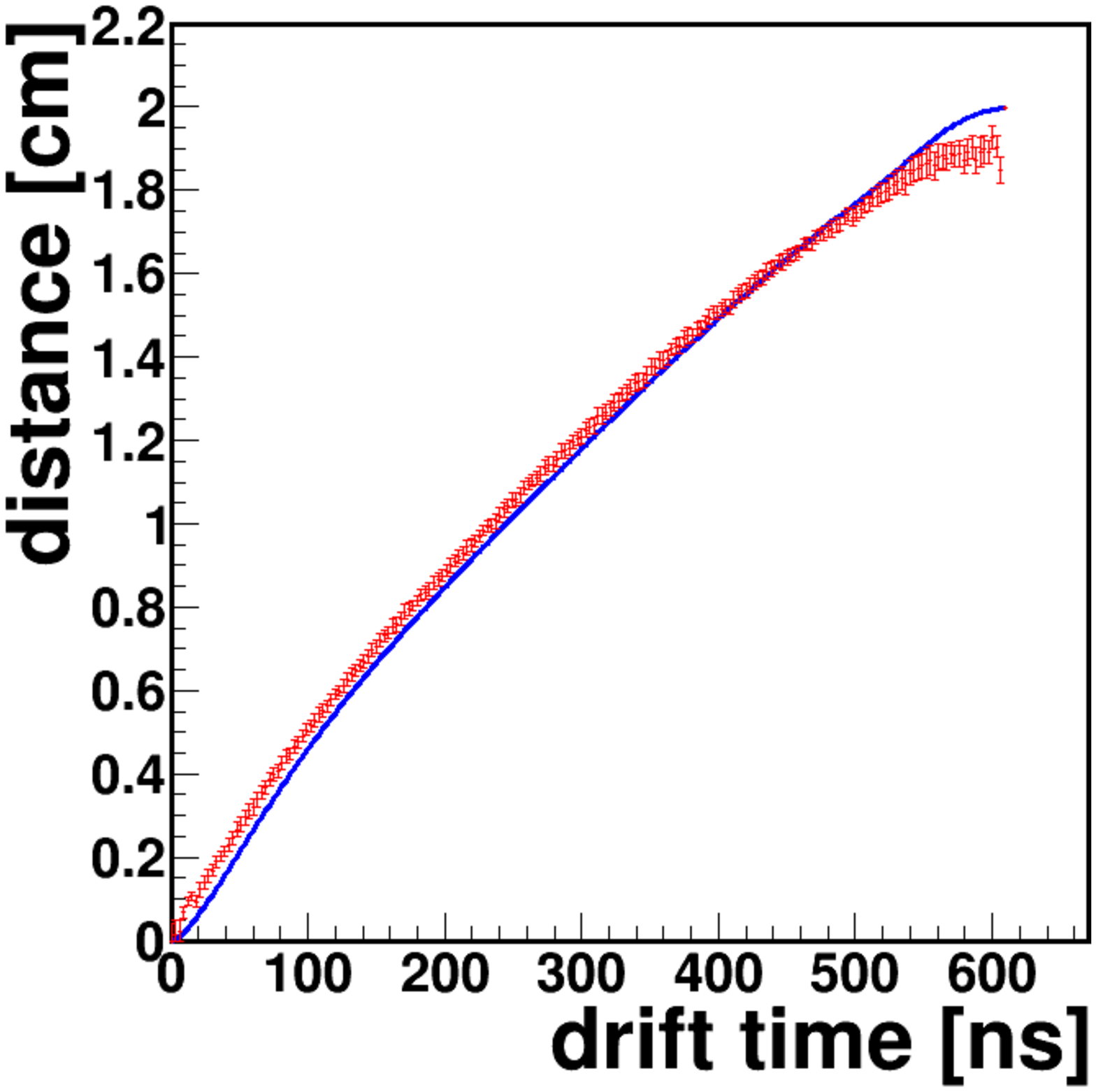}
\caption{Drift time - space relation.}
\end{subfigure}\hspace{25pt}
\begin{subfigure}[t]{0.45\textwidth}
\includegraphics[width=\textwidth]{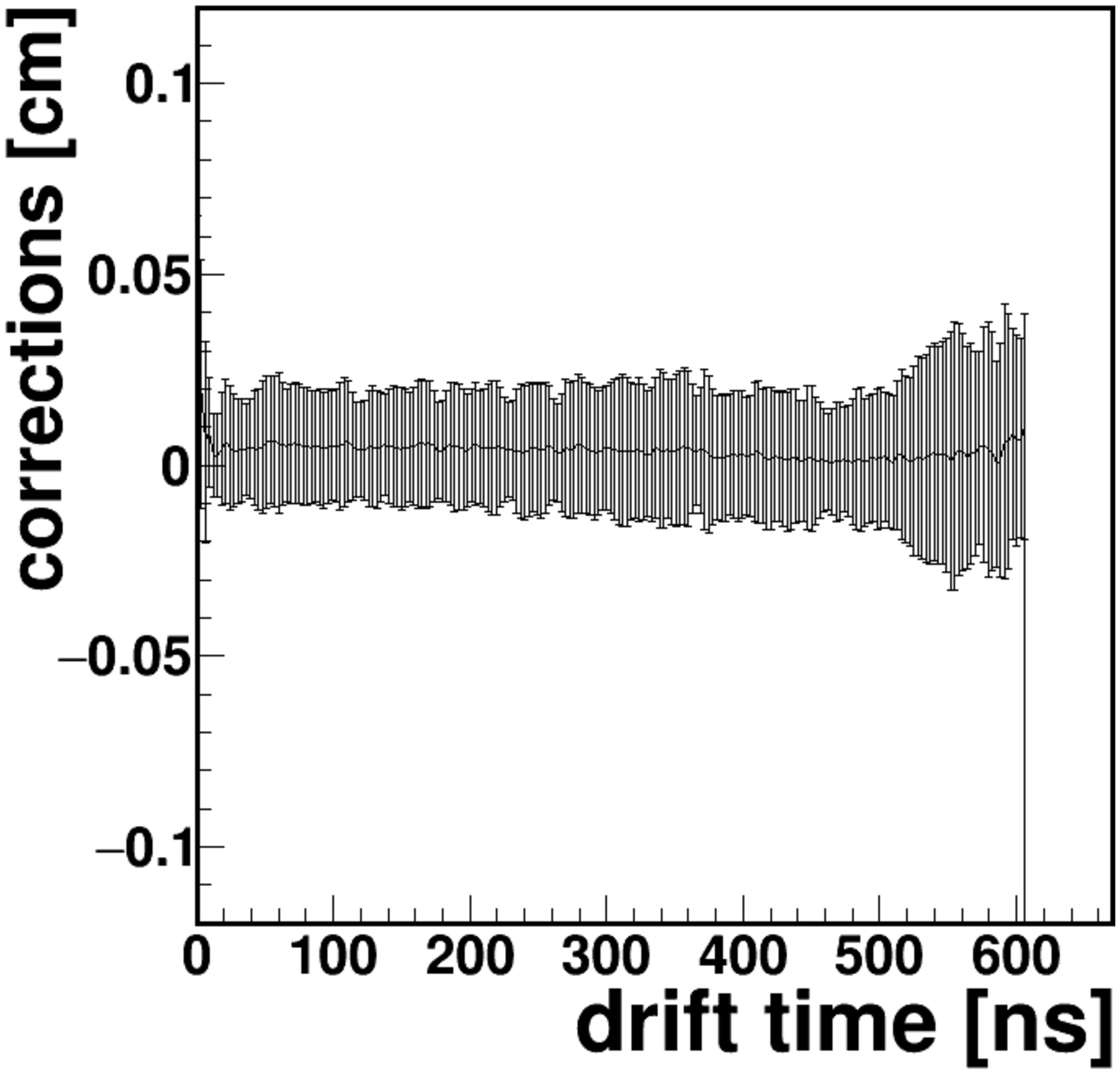}
\caption{Corrections to the drift time - space relation.}
\end{subfigure}
\caption{5th wire plane.}
\end{center}
\end{figure}
\begin{figure}[!h]
\begin{center}
\begin{subfigure}[t]{0.45\textwidth}
\includegraphics[width=\textwidth]{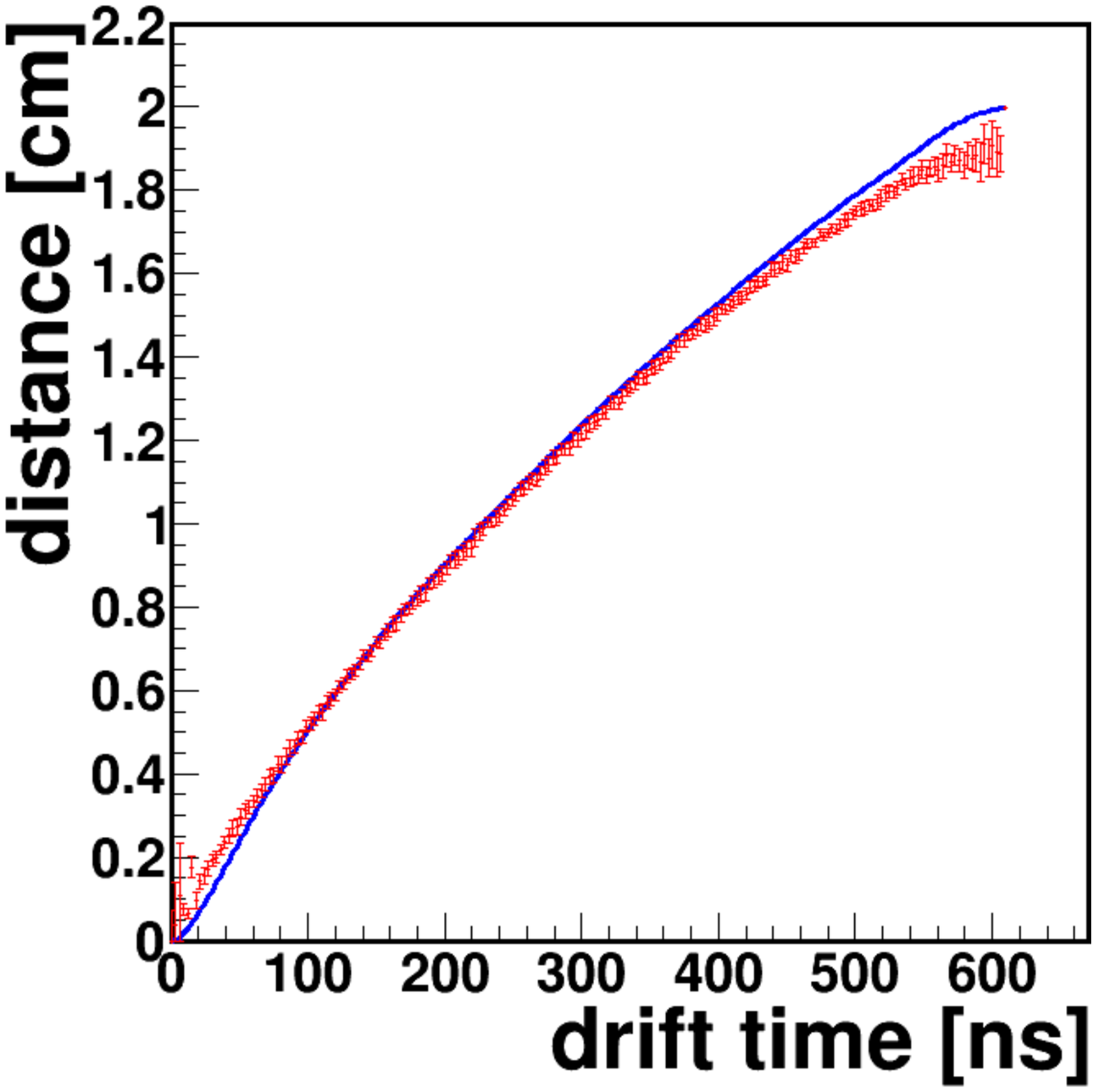}
\caption{Drift time - space relation.}
\end{subfigure}\hspace{25pt}
\begin{subfigure}[t]{0.45\textwidth}
\includegraphics[width=\textwidth]{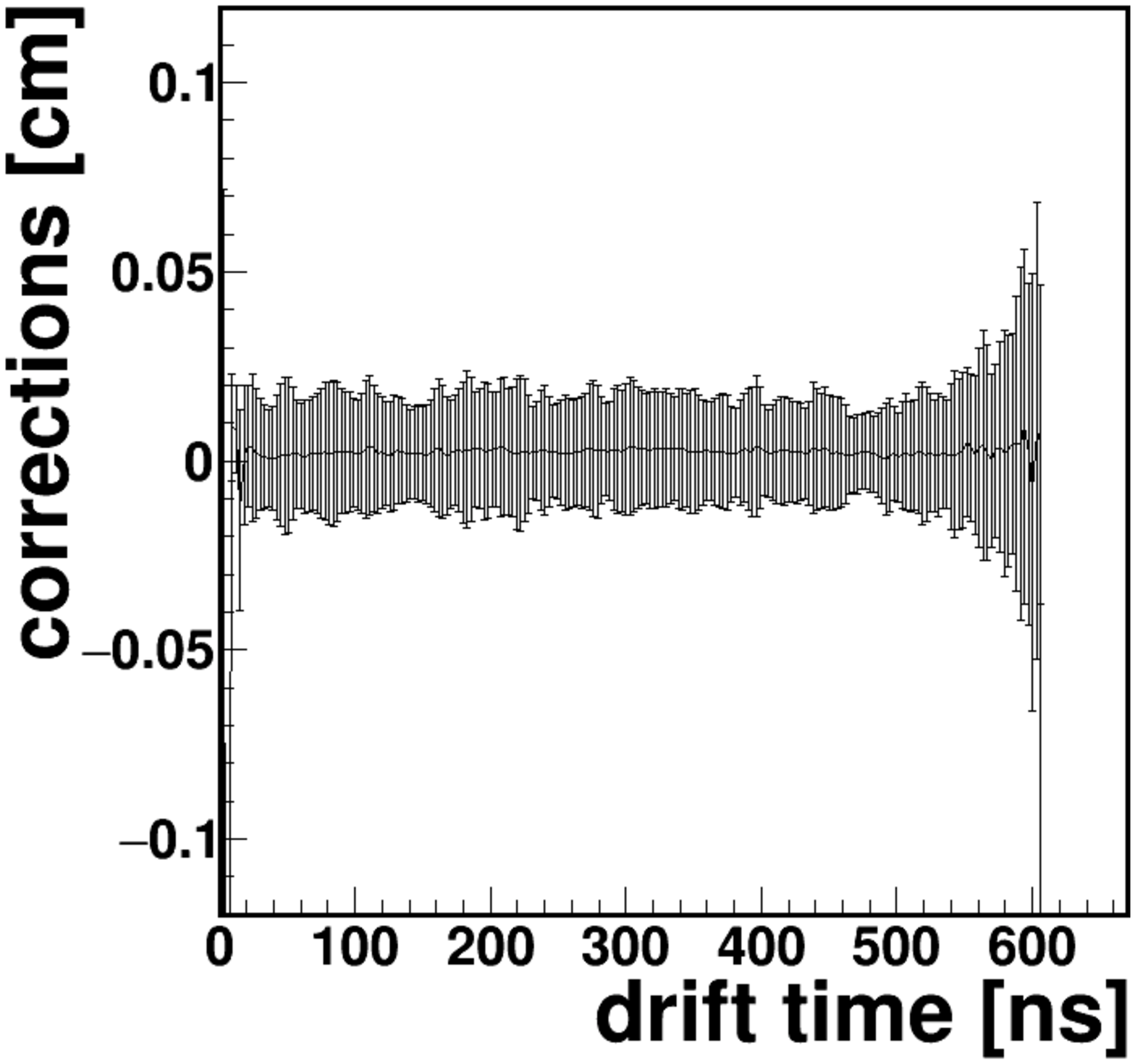}
\caption{Corrections to the drift time - space relation.}
\end{subfigure}
\caption{6th wire plane.}
\end{center}
\end{figure}
\begin{figure}[!h]
\begin{center}
\begin{subfigure}[t]{0.45\textwidth}
\includegraphics[width=\textwidth]{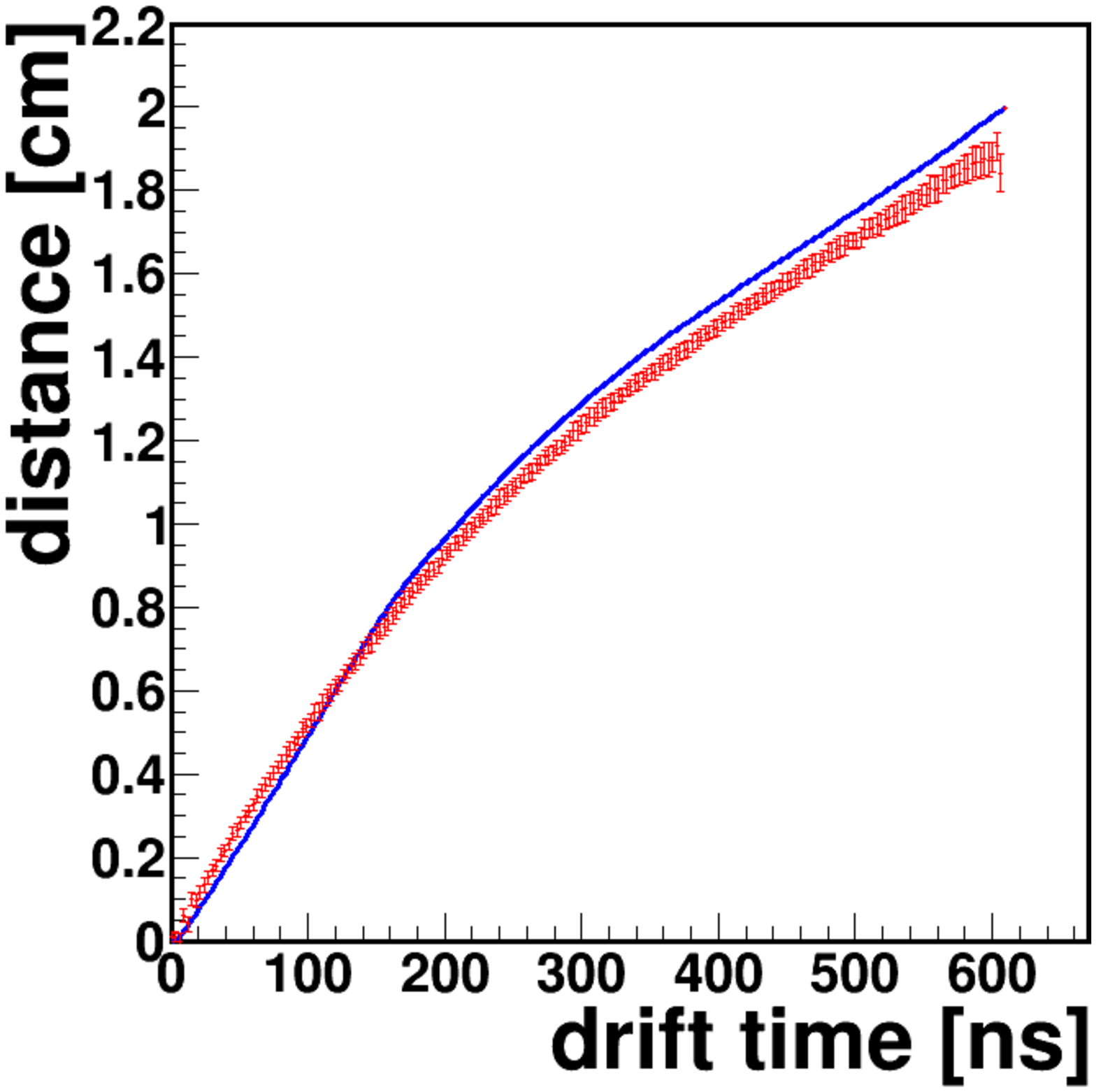}
\caption{Drift time - space relation.}
\end{subfigure}\hspace{25pt}
\begin{subfigure}[t]{0.45\textwidth}
\includegraphics[width=\textwidth]{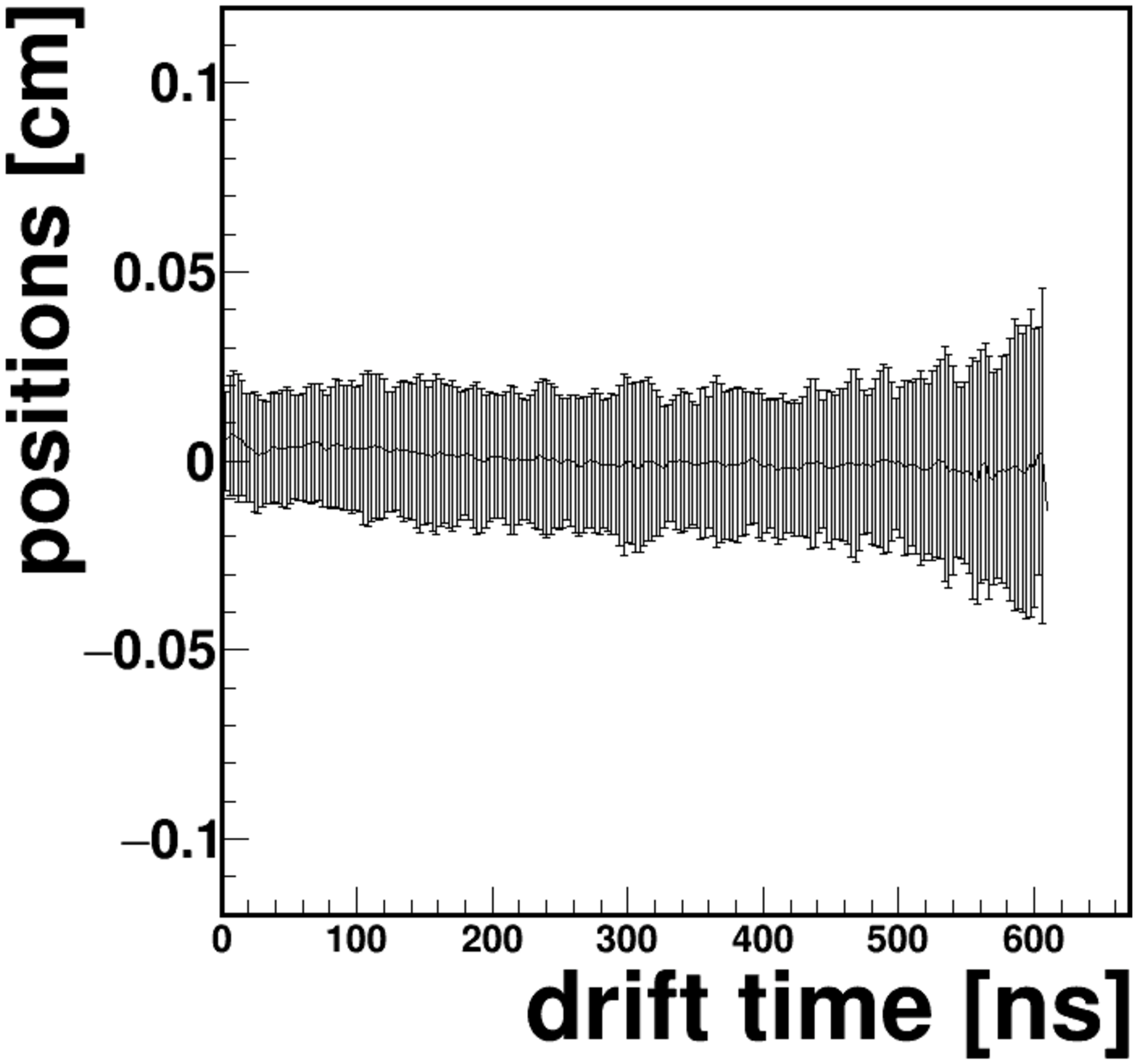}
\caption{Corrections to the drift time - space relation.}
\end{subfigure}
\caption{7th wire plane.}
\end{center}
\end{figure}
\begin{figure}[!h]
\begin{center}
\begin{subfigure}[t]{0.45\textwidth}
\includegraphics[width=\textwidth]{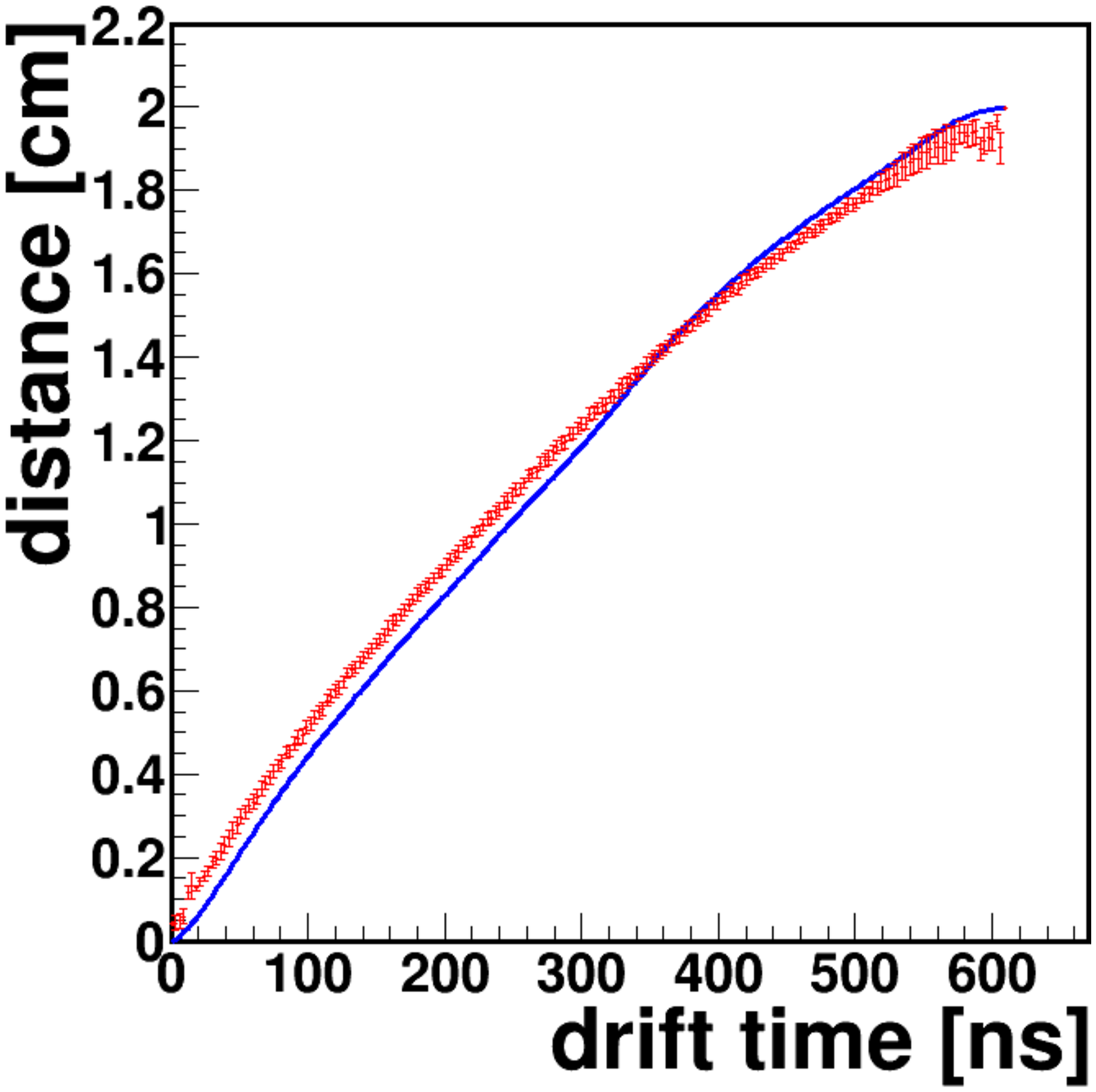}
\caption{Drift time - space relation.}
\end{subfigure}\hspace{25pt}
\begin{subfigure}[t]{0.45\textwidth}
\includegraphics[width=\textwidth]{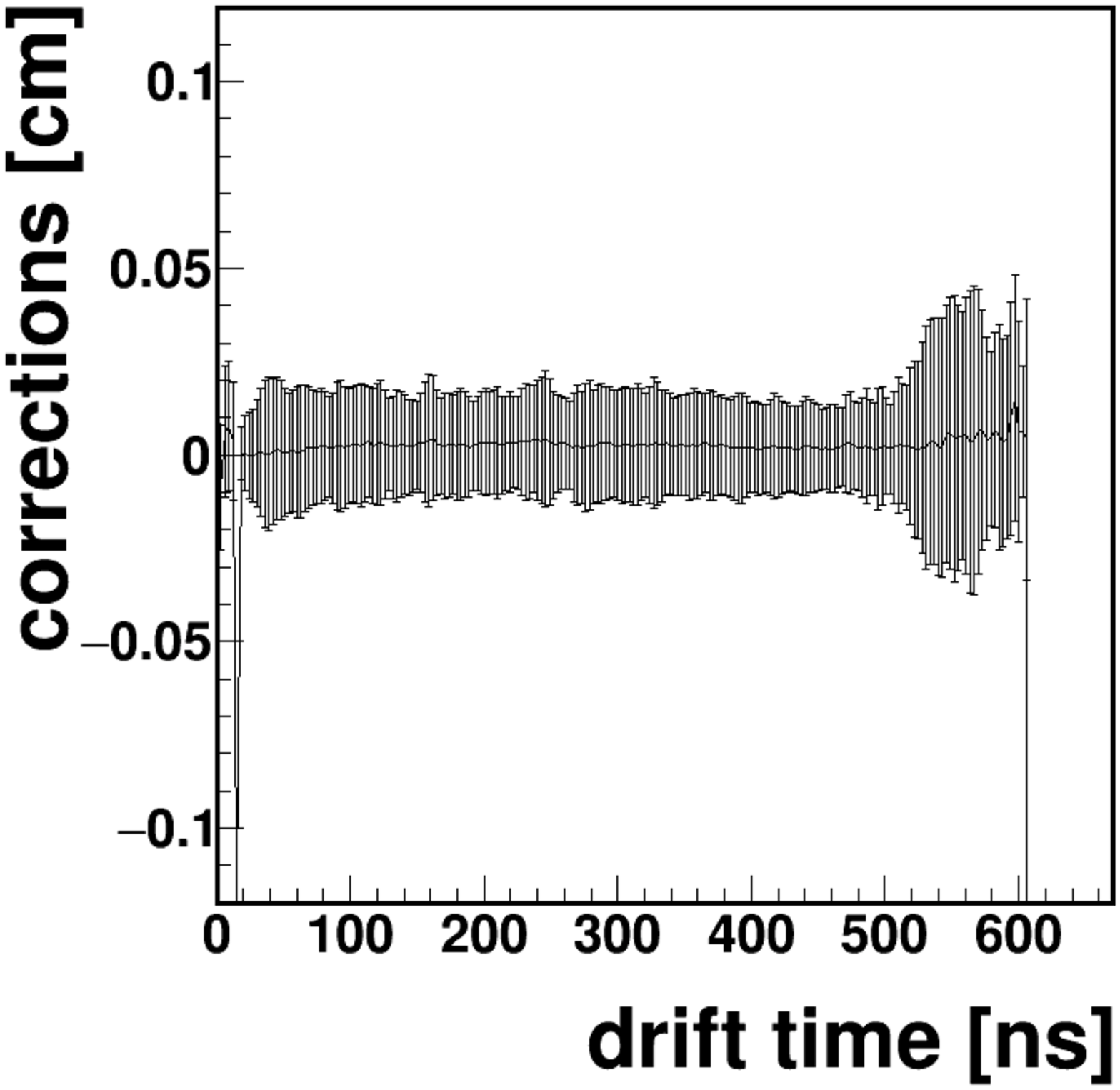}\\
\caption{Corrections to the drift time - space relation.}
\end{subfigure}
\caption{8th wire plane.}
\end{center}
\end{figure}

\end{adjustwidth*}
\end{appendices}

\FloatBarrier
\newpage
\vspace{50pt}
\begin{center}
\textbf{\textit{Acknowledgements}}
\end{center}
\hspace{20pt}\textit{I would like to express my gratitude to my supervisor, Dr Marcin Zieliński for his guidance, a~lot of valuable advice and support during my master studies and all his help in preparation of this thesis. }
\vspace{15pt}

\textit{I wish to thank Dr Dieter Grzonka for his readiness to help, patience in answering my questions and hospitality during my stays in the Forschungszentrum J{\"u}lich.}
\vspace{15pt}

\textit{I also am grateful to Prof. Paweł Moskal for the opportunity to join his research group and extend my knowledge in various aspects of experimental physics.}
\vspace{15pt}

\textit{I am very grateful to Prof. James Ritman for allowing me to work in the Forschungszentrum J{\"u}lich Institute for Nuclear Physics and supporting my stays there.}
\vspace{15pt}

\textit{Furthermore, I would like to thank all members of the P349 analysis group for their ideas and suggestions during our meetings.}
\vspace{15pt}

\textit{Last but not least, many thanks to my colleagues from Kraków for the great atmosphere during last few years.}
\vspace{15pt}

\textit{I would like to acknowledge the financial support by the Polish Ministry of Science and Higher
Education and by DAAD from resources of Bundesministerium für Bildung und Forschung (BMBF) through grant no. PPP-PL 57155292, by the Polish Ministry of Science and Higher Education through the scholarship for the best students and through the grants no. 7150/E-338/M/2015 and 7150/E-338/M/2017, by the Rector of the Jagiellonian University through the scholarship for the best students, by the Marian Smoluchowski Kraków Research Consortium “Matter–Energy–Future (KNOW) through the scholarship for the students of the Faculty of Physics, Astronomy and Applied Computer Science. I gratefully acknowledge the support given by the Forschungszentrum J{\"u}lich FFE Funding Program of the J{\"u}lich Center for Hadron Physics.}
\newpage
\bibliographystyle{unsrt}
\addcontentsline{toc}{chapter}{Bibliography}
\bibliography{mgr}

\end{document}

%% file: Motivation.tex
\chapter{Motivation} \label{chapt:motivation}
First experiments with proton beams of energies in the range a~few GeV were successfully performed in 1950's when the concept of synchrotron was realized. The experiments with proton beams in the fixed target mode lead to many notable results like discoveries of antiprotons~\citep{pbar-observation}, antineutrons~\citep{antineutrons}, $J/\psi$~\citep{Jpsi} and observation of CP violation~\citep{cp-violation}. Nowadays, the maximum energy available in the proton colliders is in the range of TeV~\citep{he-cern} which allows for a~search of new particles and tests of the Standard Model predictions~\mbox{\citep{higgs1,higgs2}}. Furthermore, the techniques of proton beam preparation and its interactions are understood well enough to allow for e.g. medical applications in the proton therapy~\citep{hadrons}.

Acceleration of a~polarized beam is more difficult due to the presence of depolarizing resonances. A~polarized proton beam was first accelerated at the Zero Gradient Synchrotron~\citep{zgs} operated between 1964 and 1979 where energies of up to 12 GeV were reached. Since then efforts have been made to provide efficient polarized proton sources and to develop methods of preserving the polarization during the acceleration which requires a~precise knowledge about the spin dynamics in the electromagnetic field. Recently studies with high energy polarized protons have been undertaken at e.g. \mbox{RHIC~\citep{rhic}}.

On the other hand, the existence of antiprotons was experimentally demonstrated at the Bevatron particle accelerator in 1955 where protons were collided with a~stationary target and masses of negatively charged secondary particles were determined~\citep{pbar-observation}. Then the first storage and cooling of antiprotons was performed as late as in 1978~\citep{ice} by the Initial Cooling Experiment at the European Organization for Nuclear Research CERN which opened a~way to antiproton physics at low energies.

Still, a~variety of physical effects can only be investigated via measurements with polarized antiprotons. One of unsolved problems is the proton structure and the origin of its spin. It was shown that only a~small fraction of the proton spin comes from the spin of quarks \cite{EMC_proton_spin}. The remaining contribution is believed to come from the spin of gluons and orbital angular momentum of quarks and gluons. 
In view of the QCD parton model \cite{qcd_parton_model}, a~polarized proton can be described by three functions: quark distribution, helicity distribution and transversity distribution. Quark and helicity distributions are well known and their measurements are possible via deep inelastic scattering (DIS) \cite{dis}. This is not possible in case of transversity due to its chiral-odd nature. One approach to the direct measurement of transversity is the double polarized antiproton-proton scattering. This is one of the subjects of the research of the PAX-Collaboration (Polarized Antiproton Experiment) \cite{PAX_letter_of_intent}.

The ability of controlling beam and target spin degrees of freedom would also allow for investigation of reactions properties and mechanisms which are now inaccessible. For example, until now there are no experimental data for the spin-spin dependence of the total antiproton-proton scattering cross-section.
Furthermore, in case of antiproton-proton reactions with the beam and target both polarized, one can selectively populate quantum states: singlet and triplet states in case of anti-parallel and parallel antiproton-proton spin configurations, respectively.

A known source of polarized antiprotons is the parity violating weak decay $\bar{\Lambda} \rightarrow \bar{p} \pi^{+}$ in which the resultant $\bar{p}$ helicity is $(64.2 \pm 1.3)\%$. This fact was used in the only experiment with polarized antiprotons so far performed in FERMILAB \cite{Only_experiment_polpbar}. In this experiment analyzing power in the inclusive $\pi^{+}$ and $\pi^{-}$ production was measured. The incident proton beam of momentum equal to $800$ GeV/\textit{c} produced $\bar{\Lambda}$ hyperons which decayed into $ \bar{p}$ and $\pi^+$. The measurement of the momentum of $\bar{\Lambda}$ and its decay products allowed to reconstruct the kinematics and to determine the transversal and longitudinal polarization components on the event by event base. The $\bar{p}$ momenta were equal to about 200 GeV/\textit{c}. Their polarization was equal to $45\%$ but the particles did not form a~beam with properties useful for further studies.

For the time being, there is no convenient method for the production of a~well-defined polarized antiproton beam with high intensity. The most popular proposal is a~filtering method which benefits from the spin dependence of nuclear reactions cross sections.

This spin filtering was first proposed for protons in 1968 \cite{spin_filtering_1968}. In this method an unpolarized beam circulating in the storage ring repetitively passes through a~polarized gaseous target. Part of the beam is lost due to the nuclear scattering but since cross sections for parallel and anti-parallel spin orientations of interacting particles are different, one spin direction is depleted more than the other. The experimental verification of this idea was performed in 1993 at the TSR in Heidelberg \cite{spin_filtering}. An unpolarized \mbox{$23$ MeV} proton beam was circulating in the ring and passing through a~polarized hydrogen gas target. A~set of measurements performed between $30$ and $90$ minutes of filtering time confirmed the growth of the polarization degree. After 90 minutes of circulation the polarization was equal to about $2\%$, although the beam intensity was equal to about $5\%$ of the initial intensity.

For antiprotons the principle of spin filtering method should remain unchanged, however, it was shown that especially for antiproton beam phase space cooling of the beam would be necessary \cite{kkilian_psc}. However, due to the lack of the experimental data on the spin dependent part of the total antiproton-proton scattering cross section any further predictions about achievable beam properties are limited. Additional difficulties in the preparation of a~filter facility may arise from the fact that longitudinal polarization effects are expected to be larger than transversal polarization effects \cite{PAX_proposal_ADRing}. 

Besides that, other methods like atomic beam sources (with trapped anti-hydrogen atoms), Stern-Gerlach effect, dynamic nuclear polarization in flight, stochastic techniques, channeling through a~bent crystal and induced synchrotron radiation has been proposed. An overview of these methods can be found in~\citep{bodega_bay,e_steffens_workshop,e_steffens_workshop2}. Some of them were already discarded due to expected low beam intensities or low degree of polarization. In~other cases lack of experimental data makes it impossible even to estimate the expected efficiency of the proposed method.

It would be a~simple alternative to the mentioned approaches if antiprotons had a~non zero polarization degree when produced \cite{kkilian_prod_process}. An indication of such a~possibility comes from experiments in which particles e.g. $\Sigma$-hyperons \cite{sigma_pol} and $\Lambda$-hyperons \cite{lambda_pol} produced in the collisions of high energy unpolarized protons with an unpolarized solid target show a~significant degree of polarization. Of course, the hyperon production cannot be directly compared to antiproton production because in the hyperon case the polarization is induced due to the strange quarks behavior which are not present in the antiproton case. However, until now there were no dedicated experimental studies performed in this direction for antiprotons. The goal of the P349 experiment is to test whether the production process can be itself a~source of antiproton polarization \cite{dgrzonka_article}. Experimental proof of such an effect would allow for planning new experiments in existing (CERN/AD) and developed (FAIR) facilities.

The main aim of this work is to perform the calibration and charged particle tracks reconstruction of one of the drift chambers which was used in the P349 experimental setup. A~precise track reconstruction is a~necessary step towards the asymmetry determination and therefore determination of the polarization degree of produced antiprotons.

%% file: MeasurementOfPolarization.tex
\chapter{Measurement of polarization} \label{chapt:MeasurementOfPolarization}
The aim of the P349 experiment is to determine the asymmetry of scattered antiprotons and on this basis to determine the degree of antiproton polarization. Experimentally it is performed by two subsequent scattering processes. Firstly, the antiprotons are produced colliding a~proton beam of momentum equal to 24~GeV/\textit{c} on a~solid target in the reaction $pN \rightarrow pNp\bar{p}$. The momentum spectrum of antiprotons is peaked at around 3.5~GeV/\textit{c} which is consistent with a~pure phase space distribution for proton-antiproton production in a~quasi-free proton-nucleon scattering. Antiproton beam transverse polarization is investigated by means of a~secondary scattering of the produced antiprotons on an unpolarized liquid hydrogen analyzer target.

\section{Analyzing power in the P349 experiment}
For the polarization determination a~measurement in a~kinematic region with known and sufficiently big analyzing power $A_y$ is necessary. For high-energy $pp$ scattering a~suitable process is elastic scattering in the Coulomb-Nuclear Interference region~\citep{CNI-region-prediction} where the analyzing power is rather low but well known from theory and confirmed by experiment~\citep{akchurin-article}. In this section we will present the reasoning which leads to estimation of the analyzing power in case of elastic $\bar{p}p$ scattering at the energy range available in the P349 experiment.

The interaction of two hadrons can be described in the helicity frame as a~mixture of strong and electromagnetic interactions.  The differential cross-section of an elastic scattering process \mbox{A + B $\rightarrow$ C + D} is given by~\citep{okada-thesis}:
\begin{equation}
\dfrac{d\sigma}{d\Omega} = \Sigma_{\lambda_A, \lambda_B, \lambda_C, \lambda_D} \vert\left\langle\lambda_C\lambda_D\vert\phi\vert\lambda_A\lambda_B\right\rangle\vert^2,
\end{equation}\label{eq:cross-section-general}
where $\phi$ indicates the matrix in spin space describing the scattering process in the spin space and $\lambda_i$ ($i = A, B, C, D$) denote spin states of respective particles. $\phi$ is a~function of the total center-of-mass energy squared.

Five independent helicity amplitudes can be introduced:
\begin{equation}\label{eq:helicity-amps}
\begin{aligned}
\phi_1(s,t) = \left\langle +\hspace{3pt}+ \vert\phi\vert +\hspace{3pt}+ \right\rangle, \\			
\phi_2(s,t) = \left\langle +\hspace{3pt}+ \vert\phi\vert -\hspace{3pt}- \right\rangle, \\ 
\phi_3(s,t) = \left\langle +\hspace{3pt}- \vert\phi\vert +\hspace{3pt}- \right\rangle, \\ 
\phi_4(s,t) = \left\langle +\hspace{3pt}- \vert\phi\vert -\hspace{3pt}+ \right\rangle, \\ 
\phi_5(s,t) = \left\langle +\hspace{3pt}+ \vert\phi\vert +\hspace{3pt}- \right\rangle, \\ 
\end{aligned}
\end{equation}
where $+$ and $-$ denote the spin states $+\frac{1}{2}$ and $-\frac{1}{2}$ of particles, $s$ and $t$ are Mandelstam variables: center-of-mass energy squared and four momentum transfer squared, respectively. In this notation $\phi_1(s,t)$ and $\phi_3(s,t)$ are non-spin-flip amplitudes, $\phi_5(s,t)$ is a~single spin-flip amplitude, $\phi_2(s,t)$ and $\phi_4(s,t)$ are double spin-flip amplitudes. 

The spin averaged differential cross section for an unpolarized beam $\frac{d\sigma}{dt}$  can be  expressed in terms of the helicity amplitudes (\ref{eq:helicity-amps}):

\begin{equation}
\dfrac{d\sigma}{dt} \sim \vert\phi_1(s,t)\vert^2+\vert\phi_2(s,t)\vert^2+\vert\phi_3(s,t)\vert^2+\vert\phi_4(s,t)\vert^2+4 \vert \phi_5(s,t)\vert^2.
\end{equation}
Taking into consideration $A_y$ one can write:
\begin{equation}
A_y\dfrac{d\sigma}{dt} \sim -\text{Im}\left \lbrace (\phi_1(s,t)+\phi_2(s,t)+\phi_3(s,t)+\phi_4(s,t))\phi^{*}_5(s,t) \right\rbrace.
\end{equation}

The high-energy $pp$ elastic scattering ($ \sqrt{s} \gg m_p$) is dominated by the Coulomb interaction for $\vert t \vert \ll 0.003$ (GeV/\textit{c})$^2$, and for very large $\vert t \vert$  hadronic interaction is the dominant one. At the kinematic region where  $\vert t \vert \approx$ 0.003 (GeV/\textit{c})$^2$ the strength of strong and electromagnetic interactions becomes comparable. 

By neglecting higher orders of electromagnetic terms and taking into account only one-photon exchange and assuming additivity of hadronic and electromagnetic amplitudes\footnote{The helicity amplitudes $\phi_i(s,t)$ can be written as a~superposition of hadronic and electromagnetic amplitudes: $\phi_i(s,t) = \phi^{had}_i(s,t) + \phi^{em}_i(s,t)$.} the unpolarized differential cross-section in the CNI region decomposes into a~sum of hadronic ($\sigma^{had}$), electromagnetic ($\sigma^{em}$) and interference ($\sigma^{int}$) contributions as follows:
\begin{equation}
\dfrac{d\sigma}{dt} = \dfrac{d\sigma^{had}}{dt} + \dfrac{d\sigma^{em}}{dt} + \dfrac{d\sigma^{int}}{dt}.
\end{equation}
Analogously, the expression for $A_y$ can be written in the form:
\begin{equation}\label{eq:Ay-interference-cross-section-corr}
A_y\dfrac{d\sigma}{dt} = \left(A_y\dfrac{d\sigma}{dt}\right)^{had} + \left(A_y\dfrac{d\sigma}{dt}\right)^{em} +\left(A_y\dfrac{d\sigma}{dt}\right)^{int}.
\end{equation}
\begin{figure}
\begin{center}
\includegraphics[width=0.5\textwidth]{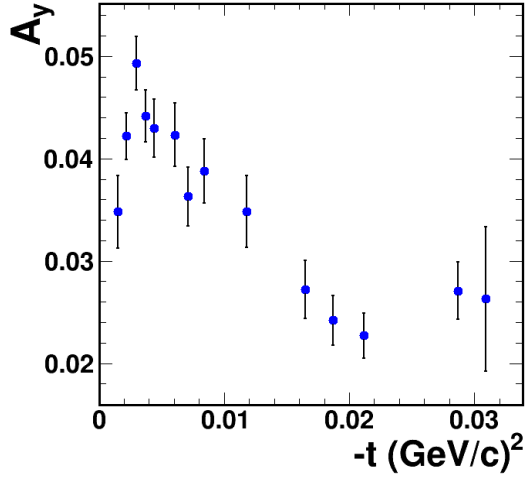}
\end{center}
\caption{Values of the analyzing power $A_y$ in the squared four-momentum transfer region  0.001 $< \vert t \vert < 0.032 $ (GeV/\textit{c})$^2$. Adapted from~\citep{okada-article}.} \label{fig:100Ay}
\end{figure}

In the equation \ref{eq:Ay-interference-cross-section-corr} the interference term comes from the interaction of nuclear non-spin-flip amplitudes: $\phi^{had}_1, \phi^{had}_3$ and electromagnetic single spin-flip amplitude caused by the interaction between charge and magnetic moment: $\phi^{em}_5$~\citep{akchurin-article,okada-article}. In view of a~polarization measurement, the interference term in the given kinematic region should be as big as possible.

If a~single photon exchange is assumed then $A^{em}_y = 0$. For high energies and small $\vert t \vert$ one can assume that $A^{had}_y \sim \sqrt{\dfrac{t}{s}} \approx 0$. Therefore, in the CNI region the dominant contribution to the analyzing power comes from the interference term $A^{int}_y$:
\begin{equation}
A^{int}_y = \dfrac{\sqrt{3}}{4} \dfrac{t_p}{m} \dfrac{(\mu-1)}{2},
\end{equation}
where $\mu$ denotes magnetic moment, $m$ stands for proton mass and $t_p$ indicates the four-momentum transfer. In this case, the maximum $A_y$ is reached for the four-momentum transfer $t_p$ equal to:
\begin{equation}
t_p = -8\pi \dfrac{\sqrt{3}\alpha}{\sigma_{\text{tot}}},
\end{equation}
where $\alpha$ is the fine structure constant and $\sigma_{\text{tot}}$ describes the total cross-section.~\citep{akchurin-article}

For $t_p \approx 0.003$ (GeV/\textit{c})$^2$ the total cross-section is 40 mb and the maximum analyzing power is about $A^{int}_y \approx 4.5 \%$~\citep{akchurin-article,dgrzonka_article}. This result was proved experimentally in the elastic scattering  of a~100~GeV/\textit{c} proton beam on a~polarized atomic hydrogen gas target where a~maximum analyzing power of 4-5\% was reached for $\vert t \vert \approx 0.003 $ (GeV/\textit{c})$^2$ (see Fig.~\ref{fig:100Ay})~\citep{okada-article}.

The theoretical predictions for the cross sections and spin dependent parameters are based on the parameters of potentials describing nucleon-nucleon (N-N) interactions. The real part of  $\rm\bar{N}$-N potential is obtained by the G-parity transformation of N-N under which all G-parity-odd contributions change sign~\citep{pbarn-interaction}. In the discussed kinematic region the expected analyzing power for $\bar{p}p$ scattering is equal to -4.5\% (only the spin-flip electromagnetic amplitude $\phi_5^{em}(s,t)$ changes the sign due to G-parity transformation~\citep{pbarn-interaction,dgrzonka_article}). This result is consistent with the experimental value of analyzing power equal (\mbox{-4.6$\pm$1.86)\%} obtained in a~measurement with a~185~GeV/\textit{c} polarized antiproton beam~\citep{akchurin-article-185}.
\begin{figure}
\begin{center}
\includegraphics[width=0.5\textwidth]{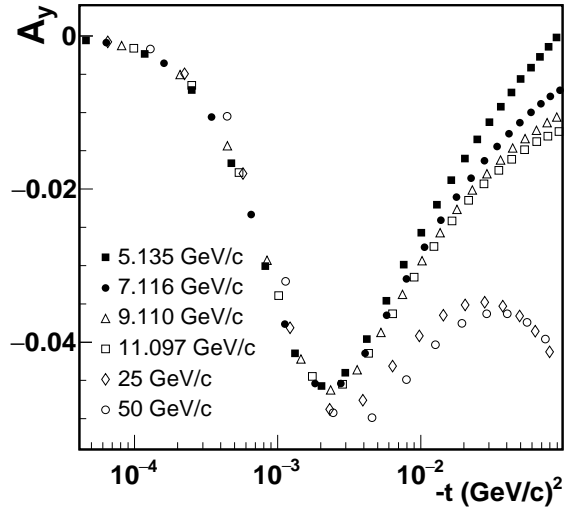}
\end{center}
\caption{Preliminary results of $A_y$ calculation. Maximum absolute value of analyzing power in a~wide range of energies are equal to about 4.5\%. Figure adapted from~\citep{Haidenbauer-prv-com}.}\label{fig:Ay-in-P349}
\end{figure}

In the P349 experiment the primary antiproton beam momentum is about 3.5~GeV/\textit{c} and therefore the assumptions made while calculating the $A_y$ may not be valid any more due to e.g. possible presence of additional amplitudes. However, according to preliminary predictions in the one boson exchange model with N-N potential adjusted to existent $\bar{p}p$ data between momenta of 50~GeV/\textit{c} down to 5.135~GeV/\textit{c}, the resulting analyzing power is comparable to the high energy case and reaches its maximum equal to about $-4.5\%$ for $\vert t \vert \approx 0.002$ (GeV/\textit{c})$^2$~\citep{Haidenbauer-prv-com}. The corresponding scattering angle is in the range of 10 to 20~mrad. Furthermore, in the Fig. \ref{fig:Ay-in-P349} it is visible that in the range \mbox{$ 0.001$ (GeV/\textit{c})$^2 < \vert t \vert < 0.005$ (GeV/\textit{c})$^2$} $A_y$ weakly depends on the primary $\bar{p}$ beam momentum. 

\FloatBarrier
\vspace{-10pt}
\section{Experimental determination of polarization}
\begin{figure}
\begin{center}
\begin{subfigure}[c]{0.45\textwidth}
\begin{center}
\includegraphics[width=\textwidth]{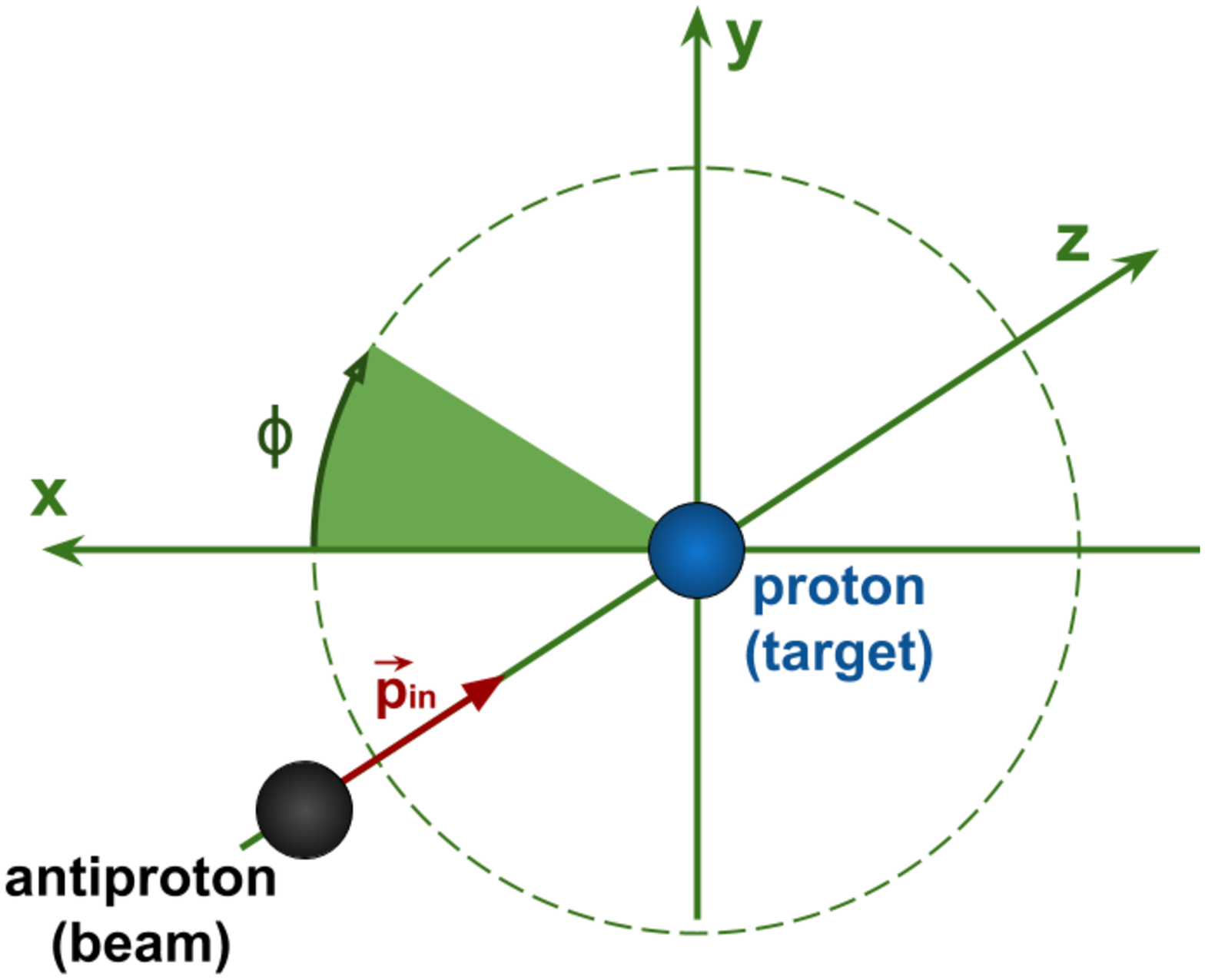}
\end{center}
\end{subfigure}
\begin{subfigure}[c]{0.49\textwidth}
\begin{center}
\includegraphics[width=\textwidth]{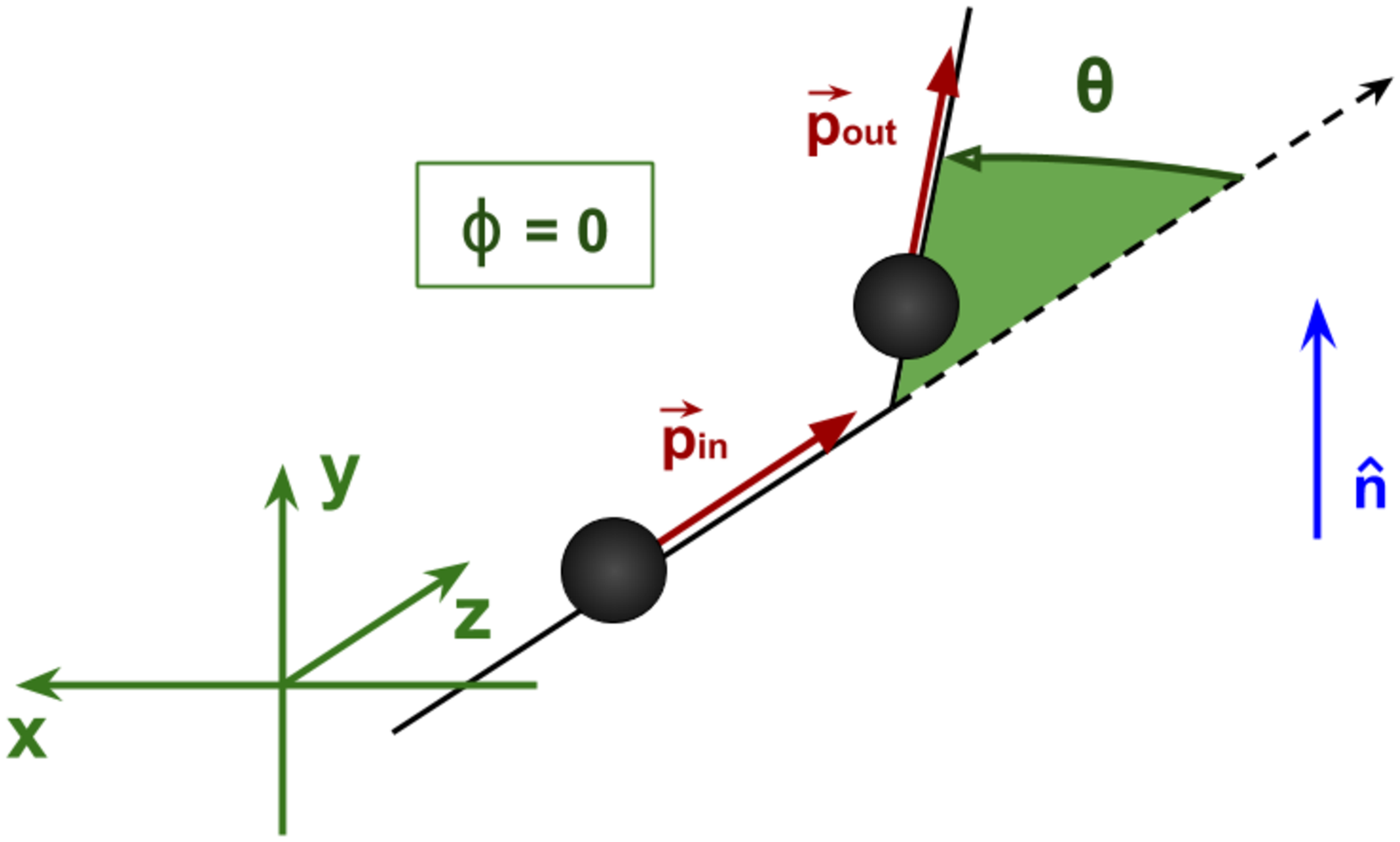}
\end{center}
\end{subfigure}
\end{center}
\caption{Coordinate frame for the elastic scattering of polarized antiproton on an unpolarized target: the azimuthal angle $\phi$ is measured in the $xy$ plane with respect to the positive side of the $x$-axis. $z$-axis is along the beam momentum ($\vec{p}_{in}$), $y$-axis (polarization axis) is parallel to the normal to the scattering plane $\hat{n}$ ($\hat{n} = \frac{\vec{p}_{in}\times\vec{p}_{out}}{\vert \vec{p}_{in}\times\vec{p}_{out} \vert}$, where $\vec{p}_{out}$ is the momentum of the scattered antiproton) and $x$-axis is chosen so that the coordinate system is right-handed. $\phi$ and $\theta$ are the azimuthal and polar angles of the scattered antiproton. As scattering occurs in the $xz$ plane $\phi = 0$ corresponds to scattering to the left and $\phi = \pi$ corresponds to the scattering to the right side.}\label{fig:coordinate-system}
\end{figure}
The cross section for the scattering process of a~transversely polarized spin $\frac{1}{2}$ particle on an unpolarized target proton is defined as~\citep{steffens}:
\begin{equation}
\sigma = \sigma_0(1 + A_y P \cos\phi), \label{eq:analysing-cross-section}
\end{equation}
where $\sigma_0$ is the cross section for an unpolarized beam scattering, $A_y$ is the single-spin asymmetry with respect to $y$-axis, $P$ is the beam polarization and $\phi$ - the azimuthal angle. The coordinate system and definitions of angles are shown in the Fig. \ref{fig:coordinate-system}. This kind of the coordinate system is referred to as a~projectile helicity frame~\citep{steffens}.

For this coordinate system one can rewrite the cross sections for a~left (right) scattering by $\sigma_L$ ($\sigma_R$) resulting from Eq. \ref{eq:analysing-cross-section}:
\begin{equation}
\begin{aligned}
\sigma_L = \sigma_0(1 + A_yP),\\
\sigma_R = \sigma_0(1 - A_yP).
\end{aligned}
\end{equation}
Therefore, introducing a~quantity called asymmetry $\epsilon$ one obtains:
\begin{equation}
A_yP = \epsilon = \dfrac{L-R}{L+R}.
\end{equation}
The asymmetry is an observable measured experimentally.

Monte-Carlo simulations were performed to estimate the accuracy of the asymmetry measurement $\delta\epsilon = \sqrt{\left(\dfrac{\delta\epsilon}{\delta L}\right)^2+\left(\dfrac{\delta\epsilon}{\delta L}\right)^2}$~\citep{dieter-report}. Assuming a~polarization of 20\% and an analyzing power equal to 4.5\%, the asymmetry resulting from the analysis of 2.5$\cdot$10$^5$ Monte Carlo events is equal to $\epsilon = 0.012\pm24\%$. The number of the scattering events corresponds to the expected statistics observed in the measurements.

%% file: ExperimentalSetup.tex
\chapter{P349 Antiproton Polarization Experiment at CERN} \label{chapt:ExperimentalSetup}

The P349 Antiproton Polarization Experiment was performed in December 2014 and June/July 2015 in the European Organization for Nuclear Research (CERN). The experimental setup was located in the Proton Synchrotron (PS)~\citep{PS-summary} East Experimental Area~\citep{EA-summary}.

\section{Antiproton production} \label{chapt:ExperimentalSetup-pbar-production}
In the East Experimental Area the 24~GeV/$c$ primary proton beam delivered from the Proton Synchrotron is split into four beamlines: T8 - T11. The beamlines deliver primary beam to the irradiation facilities (T8) and guide secondary particles of different momenta to the experimental facilities (T9 - T11)~\citep{LS1-EA-modifications}.

\begin{figure}
\begin{center}
\includegraphics[width=0.75\textwidth]{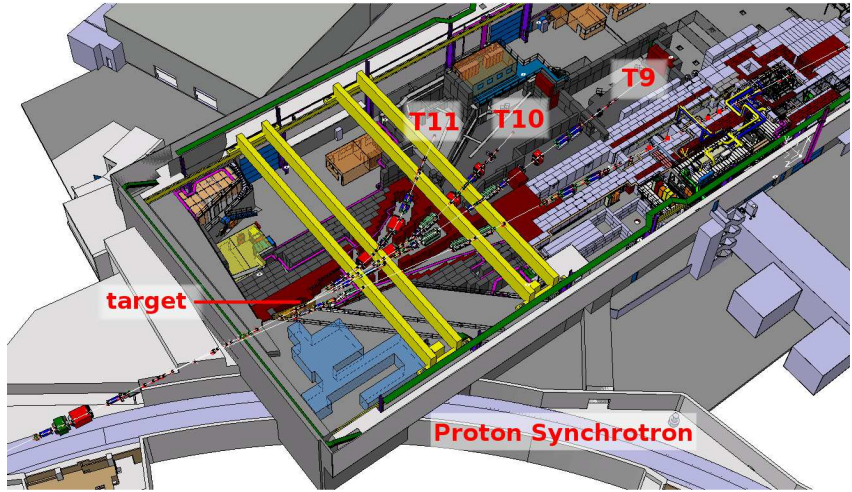}
\end{center}
\caption{Layout of the PS Experimental East Area. A~24~GeV/$c$ proton beam from PS is scattered on a~solid target. Three beamlines lead to the experimental facilities: T9, T10, T11. They differ from each other by available energies and production angles of secondary particles. Adapted from~\citep{east-area-3d-layout}.}\label{fig:east-area-description}
\end{figure}

The incident proton beam is delivered in the form of spills with a~flat maximum of about 400 ms length and proton number flux in the order of $2.5 \cdot 10^{11}$ particles per spill~\citep{prez-particle-flux}. Secondary beams are obtained by irradiation of the solid target common for beamlines T9 - T11 (see Fig. \ref{fig:east-area-description}). In the time of the P349 experiment the iridium target was used.

\begin{figure}
\begin{center}
\includegraphics[width=0.65\textwidth]{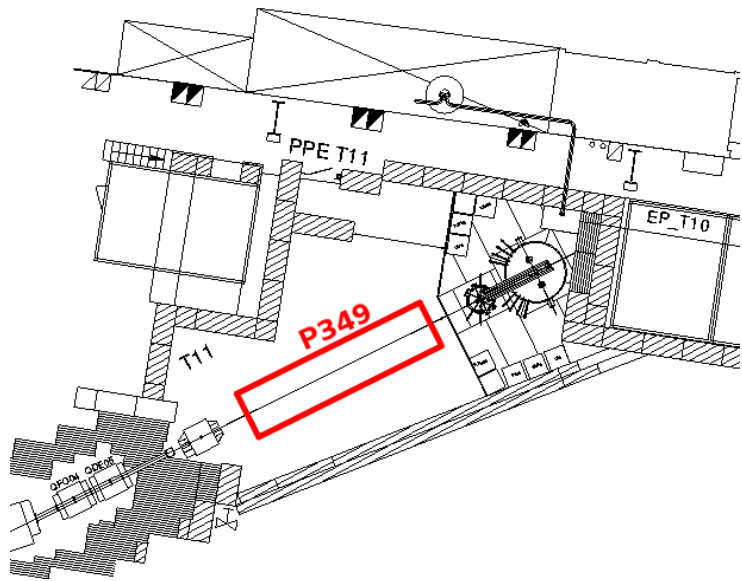}
\end{center}
\caption{Placement of the P349 experimental setup in the T11 beamline area. Adapted from~\citep{t11-p349-scheme}.}\label{fig:T11-P349}
\end{figure}

The P349 experimental setup was placed in the T11 beamline (see Fig. \ref{fig:T11-P349}) which provides secondary particles of maximum momentum equal to 3.5~GeV/$c$, at a~production angle of about 150~mrad with an acceptance of about $\pm$3~mrad horizontally and $\pm$10~mrad vertically~\citep{T11Guide}. The T11 beamline settings allow to control the particles charge and momenta.

Antiprotons are produced out of collisional energy in the $pN \rightarrow pN\bar{p}p$ reaction. Measurements of charged secondary particles of momenta equal 4~GeV/$c$ at laboratory angle of 127~mrad induced by 24~GeV/$c$ proton beam showed that antiprotons constitute about $8.5\permil$ of all negatively charged particles of about $0.5\cdot10^6$ per spill, therefore about 4000 antiprotons per spill are expected. The dominant background are pions: the $\rm \pi^+/\bar{p}$ ratio is equal to about $9\permil$. For the P349 experiment similar values are anticipated~\citep{dgrzonka_article,production-at-24GeVc}.

\section{Experimental Setup} \label{sect:ExperimentalSetup-detectors}
The P349 experimental setup was operated in air. Its central part was an analyzer target for antiprotons secondary scattering. This was a~15 cm long cell filled with liquid hydrogen. In order to determine the polarization a~precise knowledge about the left-right asymmetry in the reaction of antiproton-proton elastic scattering $\bar{p}p \rightarrow \bar{p}p$ is needed, therefore the detection setup was optimized in terms of particles identification and precise track reconstruction.

\begin{figure}
\begin{center}
\begin{subfigure}[c]{\textwidth}
\begin{center}
\includegraphics[width=0.7\textwidth]{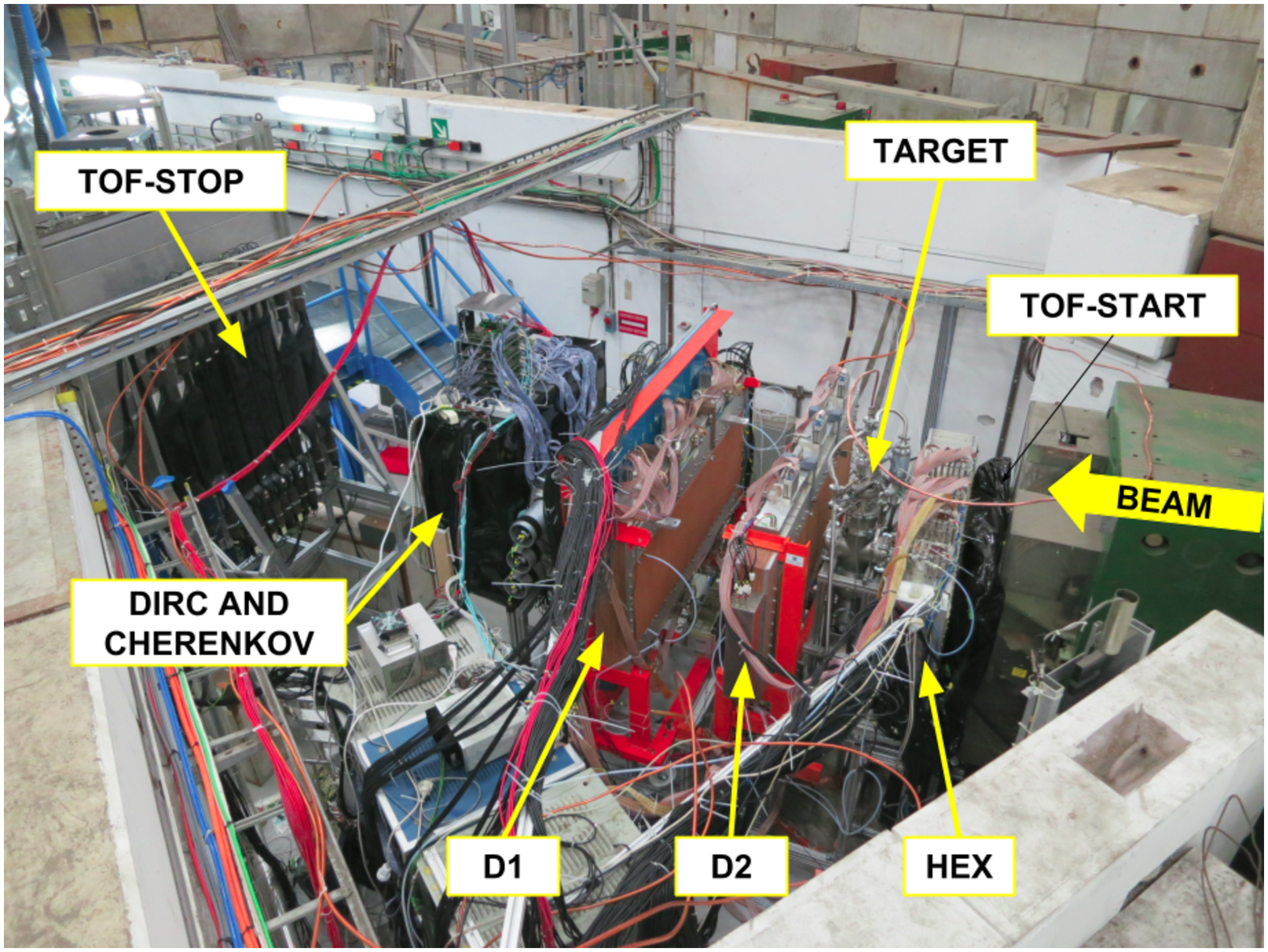}
\caption{The photograph of the experimental setup. Courtesy of D. Grzonka.}
\end{center}
\end{subfigure}
\begin{subfigure}[c]{\textwidth}
\includegraphics[width=\textwidth]{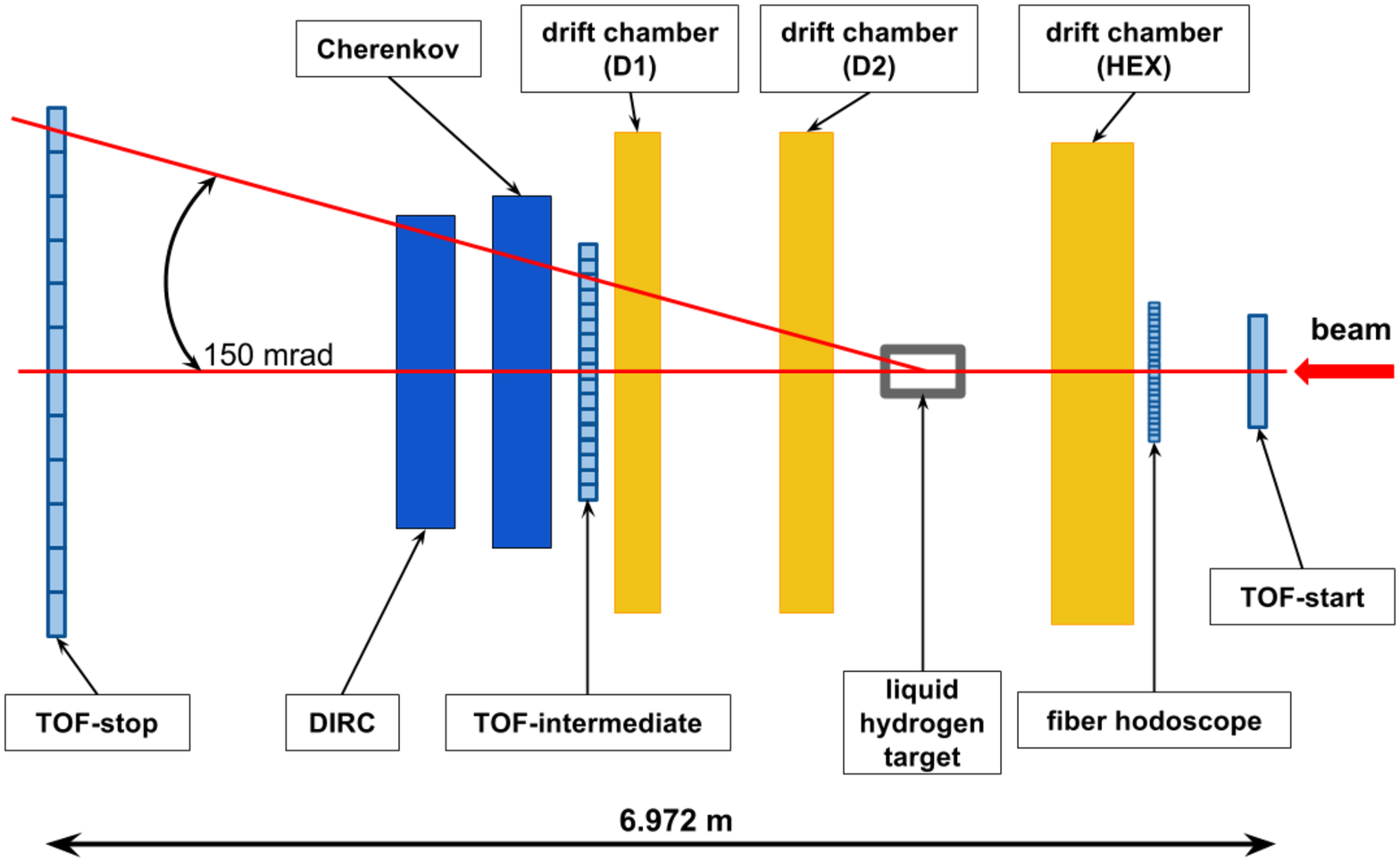}
\caption{The scheme of the detector arrangement in the horizontal plane. The beam comes from the right side.
The total angular range covered by the system is about 150~mrad (horizontally as well as vertically) and the
relevant scattering angle for the asymmetry measurement is about \mbox{20~mrad}.}
\end{subfigure}
\caption{}\label{fig:experimental_setup}
\end{center}
\end{figure}

For the setup dimensions and detectors arrangement see Fig. \ref{fig:experimental_setup}.

For the data acquisition TRB boards~\citep{trb1,trb2} from the GSI Helmholtz Center for Heavy Ion Research were used. In the drift chambers amplifier cards with discriminators which digitized signals were included and connected directly to the TRB boards. The signals from scintillators and Cherenkov detectors were amplified and digitized by PADIWA boards~\citep{padiwa} connected to the TRB boards. For each detector signal leading and trailing edge of the signal was registered which allows the determination of signal amplitude by the time over threshold method. 

\subsection*{Scintillators}
In the P349 experimental setup there were three scintillating detectors used for TOF measurement and triggering purposes.

TOF-start detector was a~single scintillating paddle, TOF-stop and TOF-intermediate consisted of twelve 10 cm - wide and sixteen 1.3 cm - wide scintillators, respectively. Each scintillator was readout at both ends with vacuum photomultipliers to provide information about a~point of interaction of particle along a~scintillator.

Furthermore, TOF-start and TOF-stop detectors were included in the trigger logic. Condition for an event to be saved was at least one signal from these detectors in a~\mbox{100 ns} window. 

\subsection*{Fiber hodoscope}
\begin{figure}
\begin{center}
\begin{subfigure}[t]{0.5\textwidth}
\includegraphics[width=\textwidth]{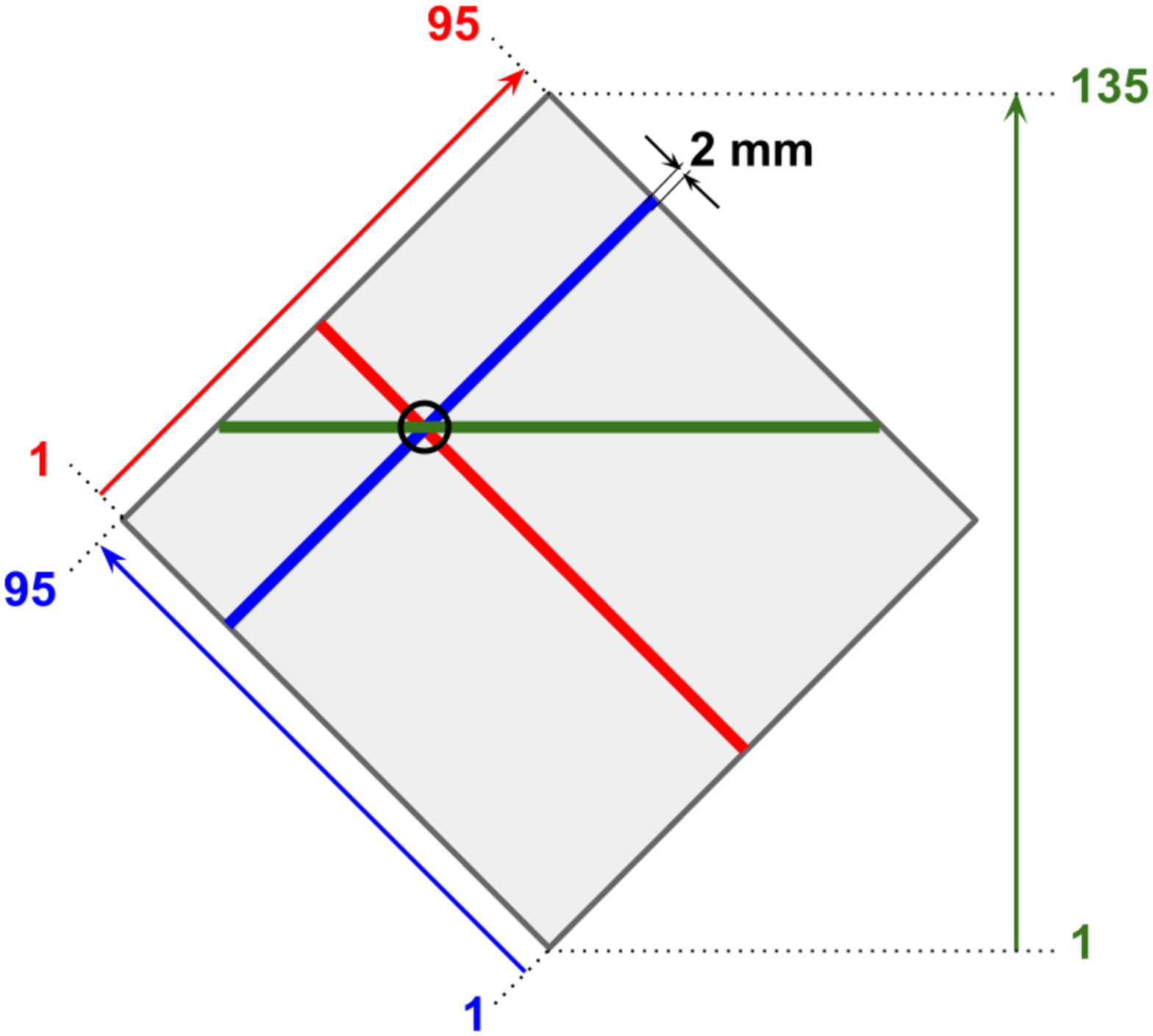}
\caption{}\label{fig:fiber_scheme}
\end{subfigure}
\begin{subfigure}[t]{0.44\textwidth}
\includegraphics[width=\textwidth]{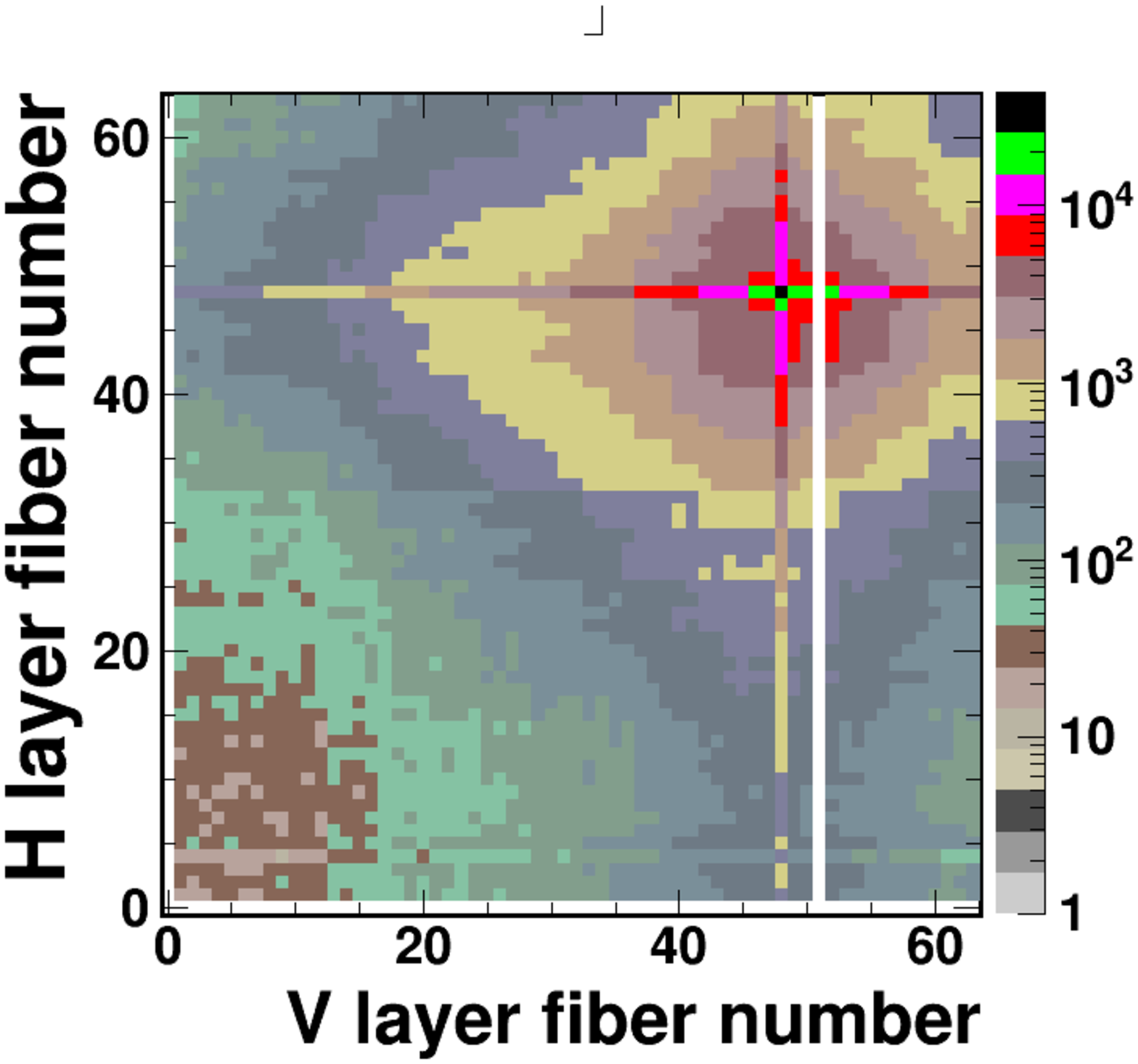}
\caption{}\label{fig:fiber_beam_prof}
\end{subfigure}
\end{center}
\caption{a) a~fiber hodoscope scheme with fibers numbering included. The point through which the particle passes the detector is determined as the point common for three fibres from three different layers (marked with black circle). In the experimental data high multiplicities for the single layers are observed (in a~single event there are typically three fibers or more with signals registered in each layer). b) Beam profile obtained by taking into account all events from inclined layers (ignoring information from the horizontal layer results in the perpendicular structures visible in the histogram). Data were collected only from the fibers with numbers up to 70.}\label{fig:fiber_det}
\end{figure}

The fiber hodoscope consisted of three layers of 2 mm wide scintillating fibers: one horizontal and two inclined at angle of $45\degree$ and $-45\degree$. Its purpose was beam profile monitoring. A~scheme of the detector and the method of the event reconstruction are shown in the Fig.~\ref{fig:fiber_scheme}. The obtained beam profile based on the signals registered in the inclined layers only is presented in the Fig.~\ref{fig:fiber_beam_prof}. 

\subsection*{Cherenkov detectors}
In order to suppress the expected high pion background a~Cherenkov detector with a~refractive index of an aerogel equal to 1.030 was used.

Measurement of the asymmetry in the CNI region requires identification of antiprotons scattered under small angles. Momenta of weakly scattered particles are close to the momenta of initial particles. The refractive index of the aerogel was chosen in a~way that the Cherenkov light production was a~process distinctive for weakly scattered pions: for the momentum equal to 3.5~GeV/$c$ the threshold refractive index for pions is  1.0008 and for antiprotons 1.035.

\begin{figure}
\begin{center}
\includegraphics[width = 0.45\textwidth]{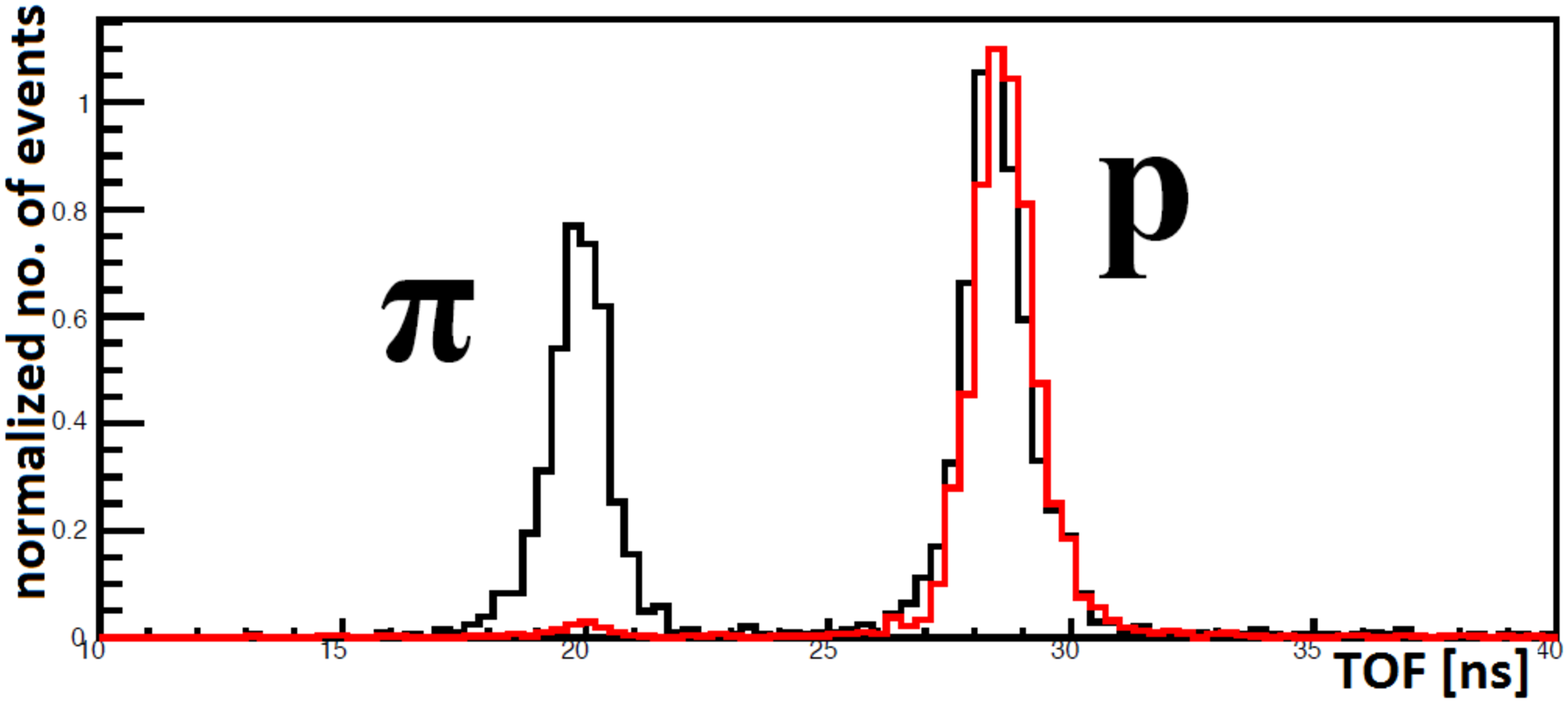}
\includegraphics[width = 0.45\textwidth]{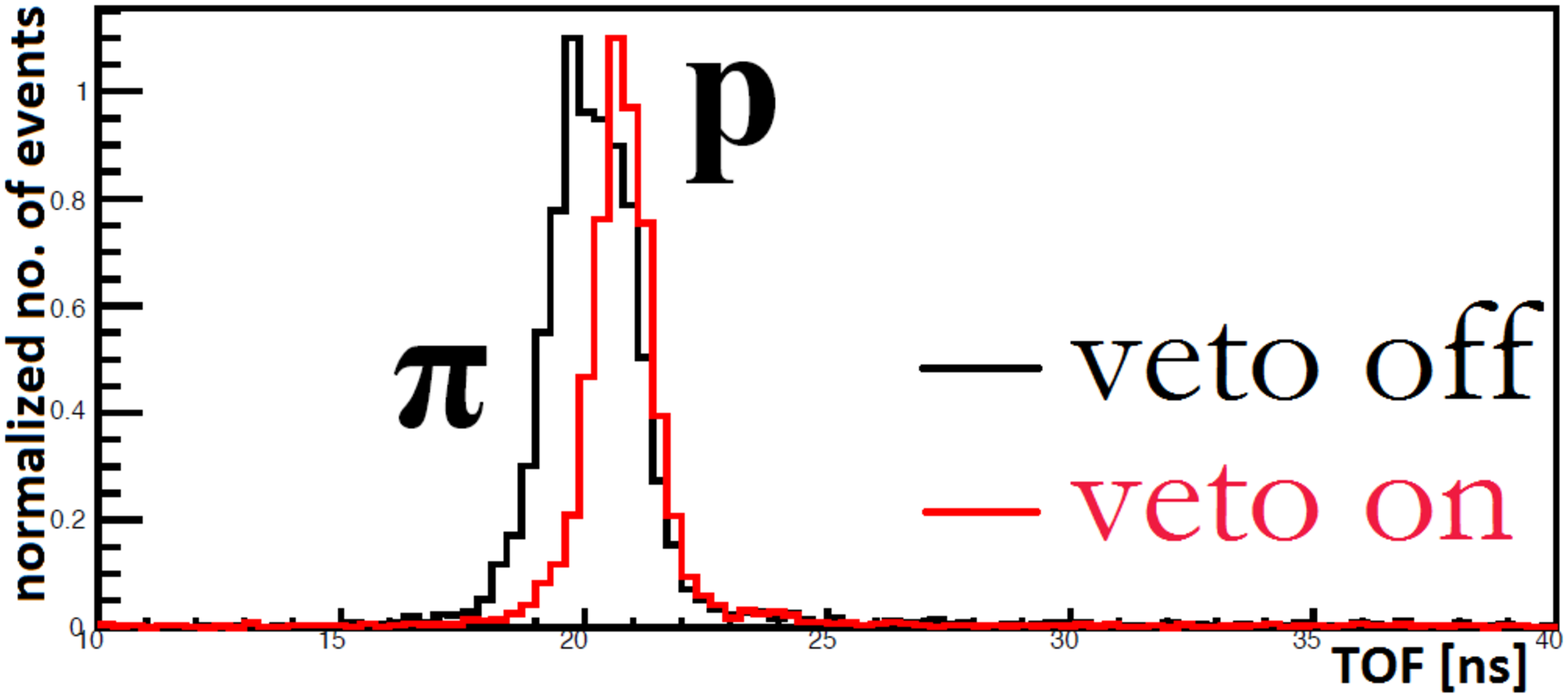}
\end{center}
\caption{Arbitrarily normalized time of flight (TOF) spectra for data from the test measurements with the beamline set to positively charged particles of momenta equal to \mbox{1~GeV/$c$} (left) and \mbox{3~GeV/$c$} (right). Results with the veto from aerogel Cherenkov detector off and on are marked with black and red lines, respectively. For momenta equal to 1~GeV/$c$ it is possible to separate pions with TOF method. For the momenta equal to 3~GeV/$c$ (right) additional information is necessary for separation. Adapted from ~\citep{dieter-private-communication}.}\label{fig:pion-online-reduction}
\end{figure}

The signal from the aerogel Cherenkov detector was included into the trigger logic as a~veto for online background reduction. Its effectiveness was shown in the test measurements with the beamline set to positively charged particles (see Fig. \ref{fig:pion-online-reduction}).

For offline particle identification the DIRC detector with Plexiglas as a~radiator was used. Here also antiprotons produced light and the particle identification is done by reconstruction of the Cherenkov angle. The expected separation between protons and pions at 3.5~GeV/$c$ is 7.8$\sigma$~\citep{dgrzonka_article}.

\FloatBarrier
\subsection*{Drift chambers}
Drift chambers are gaseous detectors used for charged particles track reconstruction~\mbox{\citep{particle-detectors,lectures-jsmyrski}}. This description focuses on planar drift chambers as detectors of this type were used in the P349 experiment.

The principle of the drift chamber operation is the measurement of time between the passage of a~charged particle through the detector and the signal registration at the sense wire (signal wire, anode). A~distance $d$ between the particle track and the nearest sense wire is calculated based on the measured drift time $t_d$ and knowledge of the drift velocity of electrons $v_e(t)$ (drift velocity can vary along the drift path): $ x = \int v_e(t) dt$~\citep{particle-detectors}. A~function which provides a~distance $d$ for a~given drift time is called a~drift time - space relation.

\begin{figure}[!ht]
\begin{subfigure}[c]{0.4\textwidth}
\begin{center}
\includegraphics[width=\textwidth]{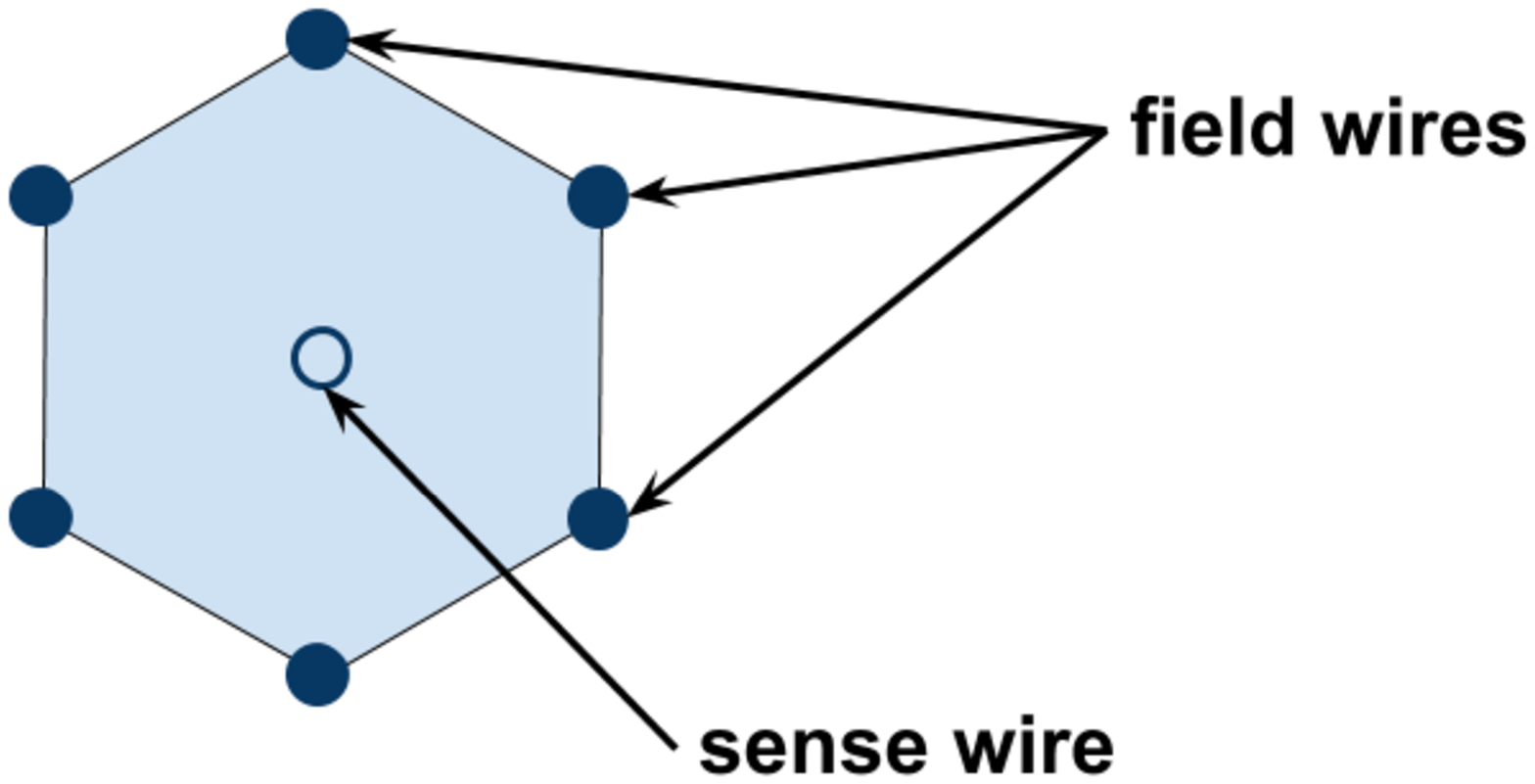}
\end{center}
\vspace{45pt}
\caption{}\label{fig:hex-cell}
\end{subfigure}
\begin{subfigure}[c]{0.55\textwidth}
\begin{center}
\includegraphics[width=\textwidth]{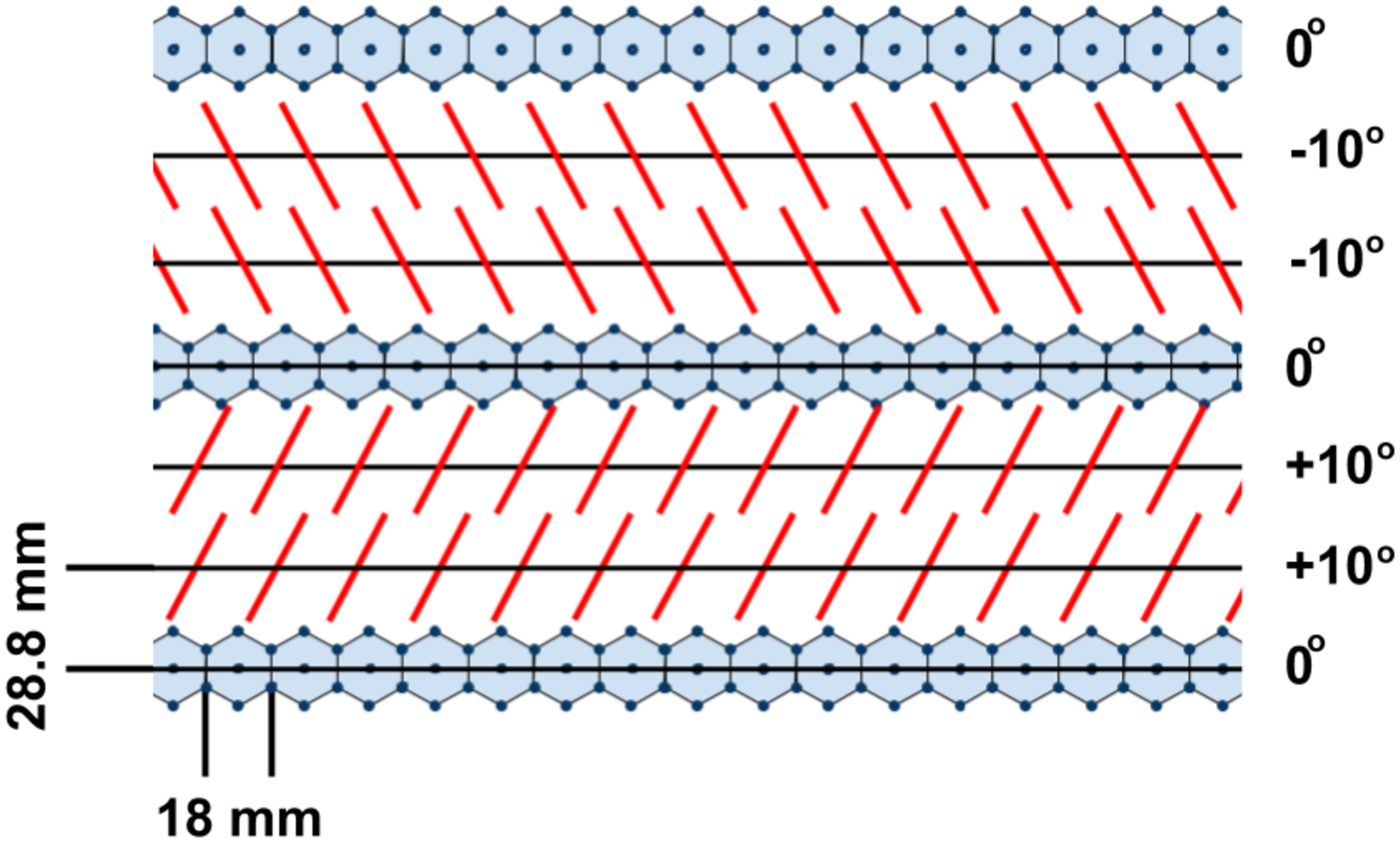}
\end{center}
\caption{}\label{fig:hex-geo}
\end{subfigure}
\caption{a) Hexagonal cell. b) Schematic arrangement of the wire planes in the HEX drift chamber.}\label{fig:hex}
\end{figure}

\begin{figure}
\begin{center}
\begin{subfigure}[t]{0.48\textwidth}
\begin{center}
\includegraphics[width=\textwidth]{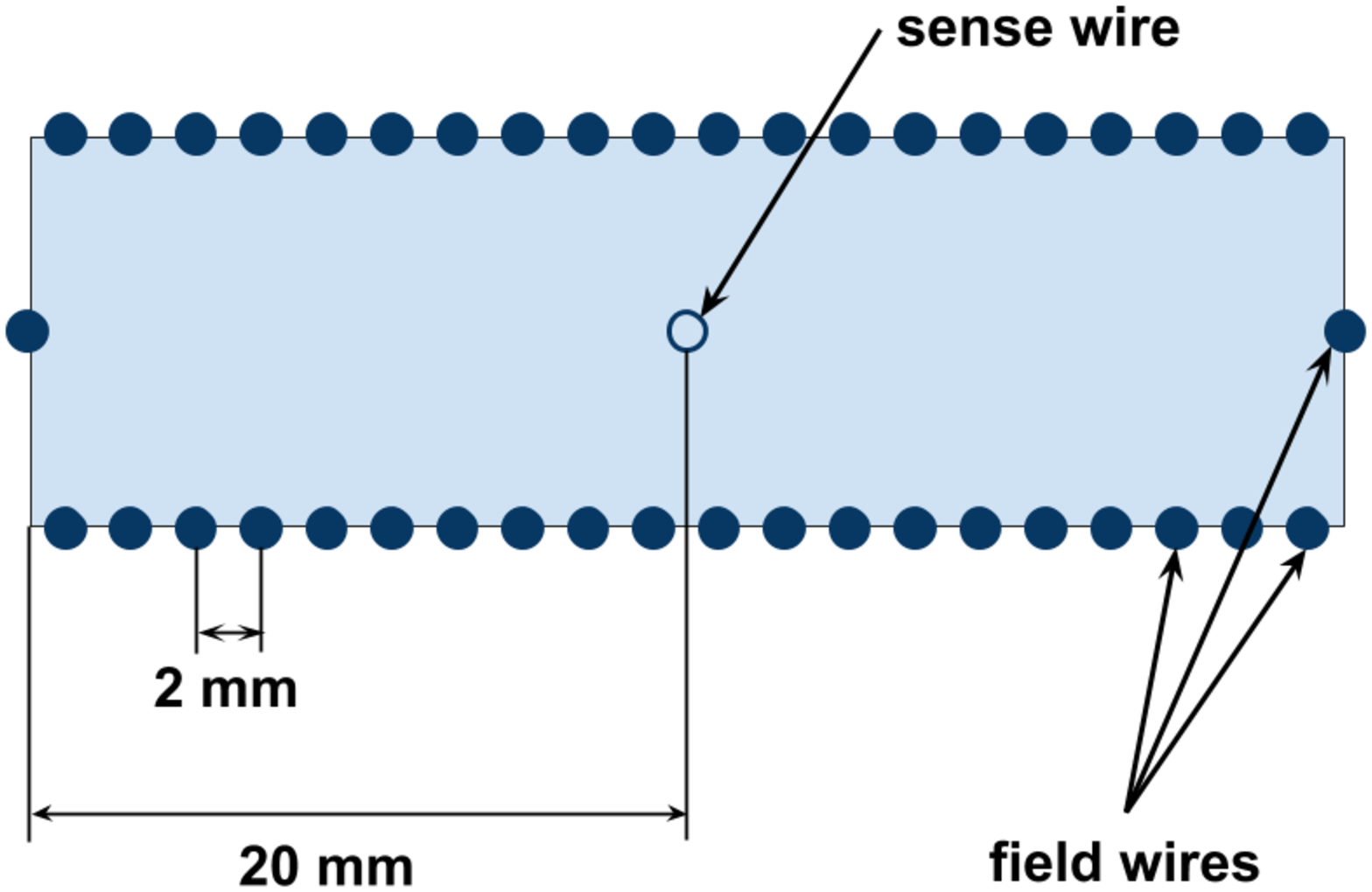}\\
\end{center}
\caption{}\label{fig:d1d2-cell}
\end{subfigure}
\begin{subfigure}[t]{0.6\textwidth}
\begin{center}
\includegraphics[width=\textwidth]{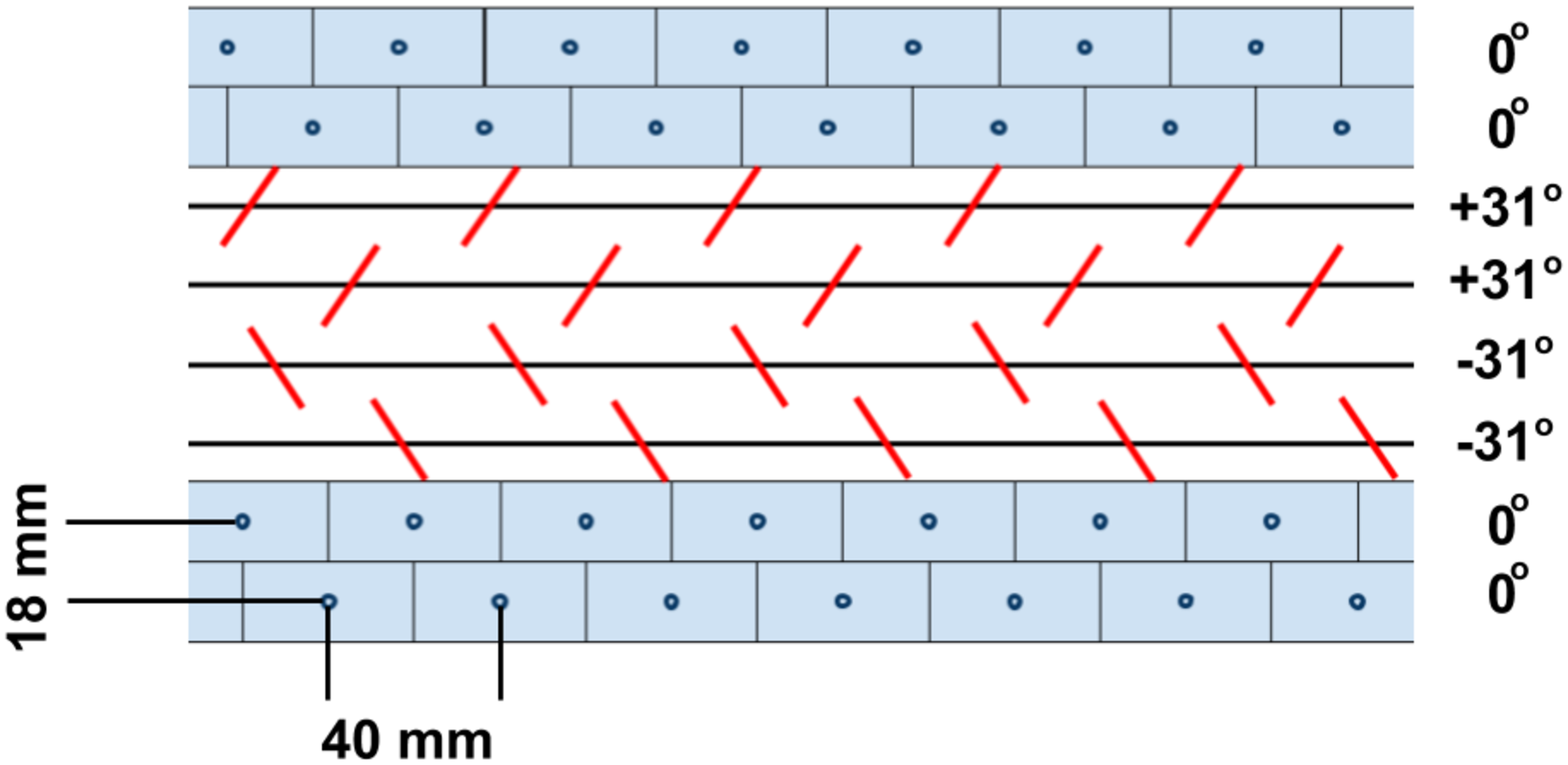}
\end{center}
\caption{}\label{fig:d1d2-geo}
\end{subfigure}
\caption{a) Rectangular cell. b) Schematic arrangement of the wire planes in the D1 drift chamber. Wire planes in the D2 are arranged in the same way as the first six planes of D1 (counting from bottom of the picture).}\label{fig:d1d2}
\end{center}
\end{figure}

Each sense wire is surrounded by field wires (cathodes) forming a~cell (see Fig. \ref{fig:hex-cell}, \ref{fig:d1d2-cell}). The cells are organized in layers in a~way that in a~given layer all sense wires are parallel to each other (see Fig. \ref{fig:hex-geo}, \ref{fig:hex-geo}).

\begin{figure}[!h]
\begin{center}
\includegraphics[width=0.7\textwidth]{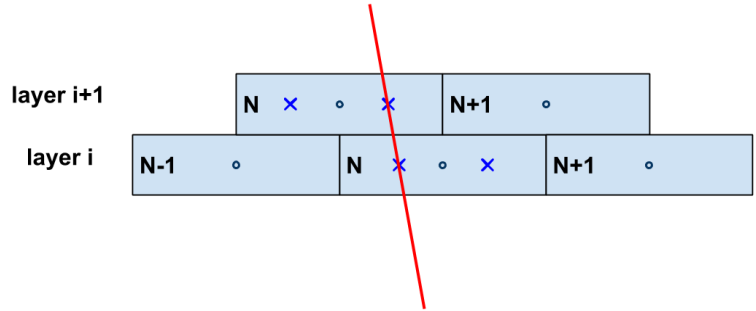}
\end{center}
\caption{The idea of resolving the left-right ambiguity in two subsequent planes with the same wire orientation. The horizontal positions of the sense wires in these planes are shifted with respect to each other by half of the drift cell width. All possible points of the particle passage (obtained from the time measurement) are marked with blue crosses. The combined information from both wire planes makes the possible choice of points which belong to the particle trajectory (red line) unambiguous. With a~cell numbering introduced as shown in the figure, one expects that particles of trajectories almost perpendicular to the drift chamber plane produce signals in the pair cells with the same numbers or cells for which the number in the $i+1$-th wire plane is greater by one than in the $i$-th plane.}\label{fig:left-right-ambiguity}
\end{figure}

For a~3-dimensional trajectory determination at least three planes with three different directions of sense wires are required. Furthermore, a~time measurement in a~single cell does not provide information whether the particle passed on the left or right side of the anode (left-right ambiguity). To resolve this ambiguity pairs of layers with the same wire orientation are used where one layer is shifted with respect to the other by half of the drift cell width (see Fig. \ref{fig:left-right-ambiguity}).

When a~charged particle passes through a~drift chamber a~primary ionization occurs and pairs of electron-ion are created. If the energy of electrons from this process is big enough, they further ionize the gaseous medium and ionization clusters are created. The electrons produced in this process drift towards the anode wire in the field provided by the high voltage between field wires and sense wires.

In the vicinity of the anode wire (at distances comparable to its diameter, typically about 10 - 30 $\mu m$) a~free electron is accelerated so that it gains the energy sufficient for ionization and an avalanche formation occurs. Electron-ion pairs are created almost at the same place in the process $ e^- + a~\rightarrow e^- + A^+ + e^- $. The charge multiplication continues until the external field is reduced below a~critical value due to the presence of the positive ions. Produced electrons drift towards the anode wire and the ion cloud drifts towards the cathode.

Drift chambers are filled with gas mixtures which usually consist of noble (e.g. Ar, Xe) and organic gases (e.g CO$_2$, CH$_4$)~\citep{particle-detectors}. The proportions of the gases present in the mixture (also contamination like water) together with the electric field in a~cell determine the drift time - space relation.

The main factors affecting the resolution of drift chambers is the diffusion or the electrons drifting to the cathode and spatial distribution of the ionization clusters along the primary ionization path~\citep{lectures-jsmyrski}.

In the P349 experiment tracking was based on a~set of three drift chambers. All drift chambers were filled with an Argon-CO$_2$ mixture.

The drift chamber with a~hexagonal drift cell structure (HEX, see Fig. \ref{fig:hex}) was placed before the target for the primary particles track determination. It consisted of seven wire planes: three with straight and four with inclined wires (two pairs of planes inclined at angles of $\pm10\degree$)~\citep{hex-drift-chamber-article}.

Scattered particles tracks were measured with a~set of two drift chambers with rectangular cells of similar construction (D1 and D2, see Fig. \ref{fig:d1d2}). Together they consisted of 14 wire planes: six with straight wires and eight with inclined wires (four pairs of planes inclined at angles of $\pm31\degree$).

%% file: TrackIdentification.tex
\chapter{Track identification and event selection criteria} \label{chapt:TrackIdentification}
The procedures described in this thesis are based on the events registered in the drift chamber D1. The motivation for this choice was the fact that drift chamber D1 has more straight wire planes than D2 and HEX which made possible preliminary tests with information from these wire planes only. However, due to analogous construction of all drift chambers, prepared procedures can be easily adapted to larger number of layers or for different drift chamber.

In order to prepare methods of drift chamber calibration and track reconstruction, the identification of events with a~single particle passing through the drift chamber is needed, although, neither knowledge about the type of particle nor information whether it was scattered or not are necessary. As prior to this analysis collected data were not investigated in view of particles identification or event categorization, the simplest event selection criteria guaranteeing the presence of a~single track were applied:
\begin{itemize}
\item one signal in the TOF-start detector,
\item one signal in the TOF-stop detector registered later than the signal from TOF-start,
\item exactly one cell with signal in each wire plane of the D1 drift chamber.
\end{itemize}

Furthermore, the relative position of cells with signals in successive pairs of wire planes with the same wire orientation was taken into account. For the analysis only events with signals in the neighboring cells were chosen (see Fig. \ref{fig:left-right-ambiguity} and \mbox{Chapt. \ref{chapt:ExperimentalSetup}}).

From events chosen for calibration it was necessary to require exactly one cell with signal in each wire plane (see Chapt. \ref{chapt:CalibrationProcedure}). Methods of track reconstruction and calibration were tested on the same data sample, nevertheless, for the track reconstruction the event selection conditions can be loosen with respect to the calibration event sample (see Chapt.~\ref{chapt:track-reco}).

%% file: CalibrationProcedure.tex
\chapter{Calibration procedures} \label{chapt:CalibrationProcedure}
A complete drift chamber calibration procedure including all drift chambers can be divided into three separate steps:
\begin{itemize}
\item determination of the drift time offsets of sensitive wires,
\item adjustment of drift time - space relations in a~given period of measurement (e.g.~a~day of measurement),
\item fixing the relative detector positions.
\end{itemize}
Points 1 and 2 were completed and are described in the sections \ref{sec:dt-offset} and \ref{sec:dt-calibration} of this thesis. The third point requires information from other detectors and its description is not included in this thesis.

\section{Drift time offset determination}\label{sec:dt-offset}
\begin{figure}[!h]
\begin{center}
\begin{subfigure}[c]{\textwidth}
\begin{center}
\includegraphics[width=0.45\textwidth]{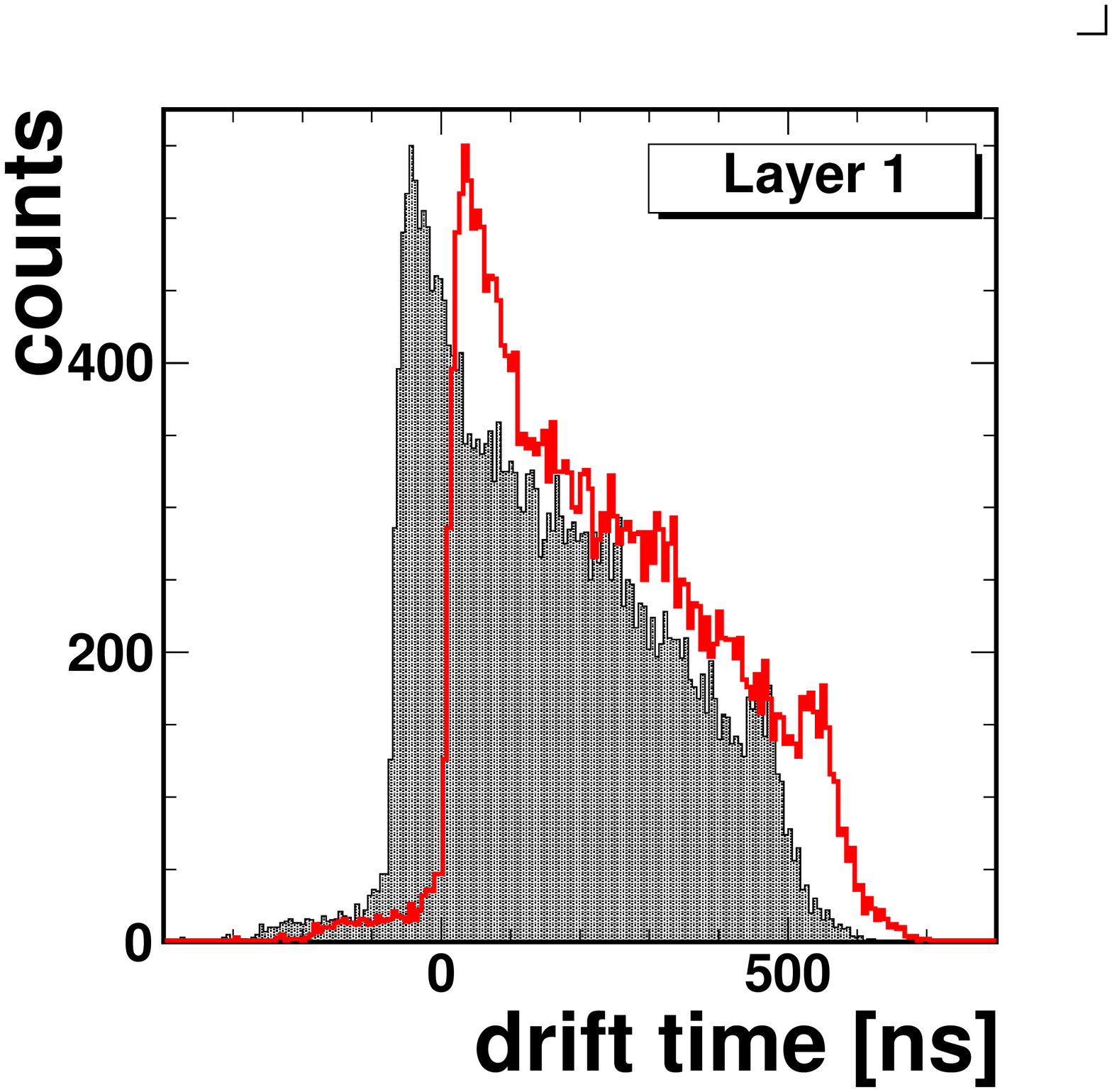}
\includegraphics[width=0.45\textwidth]{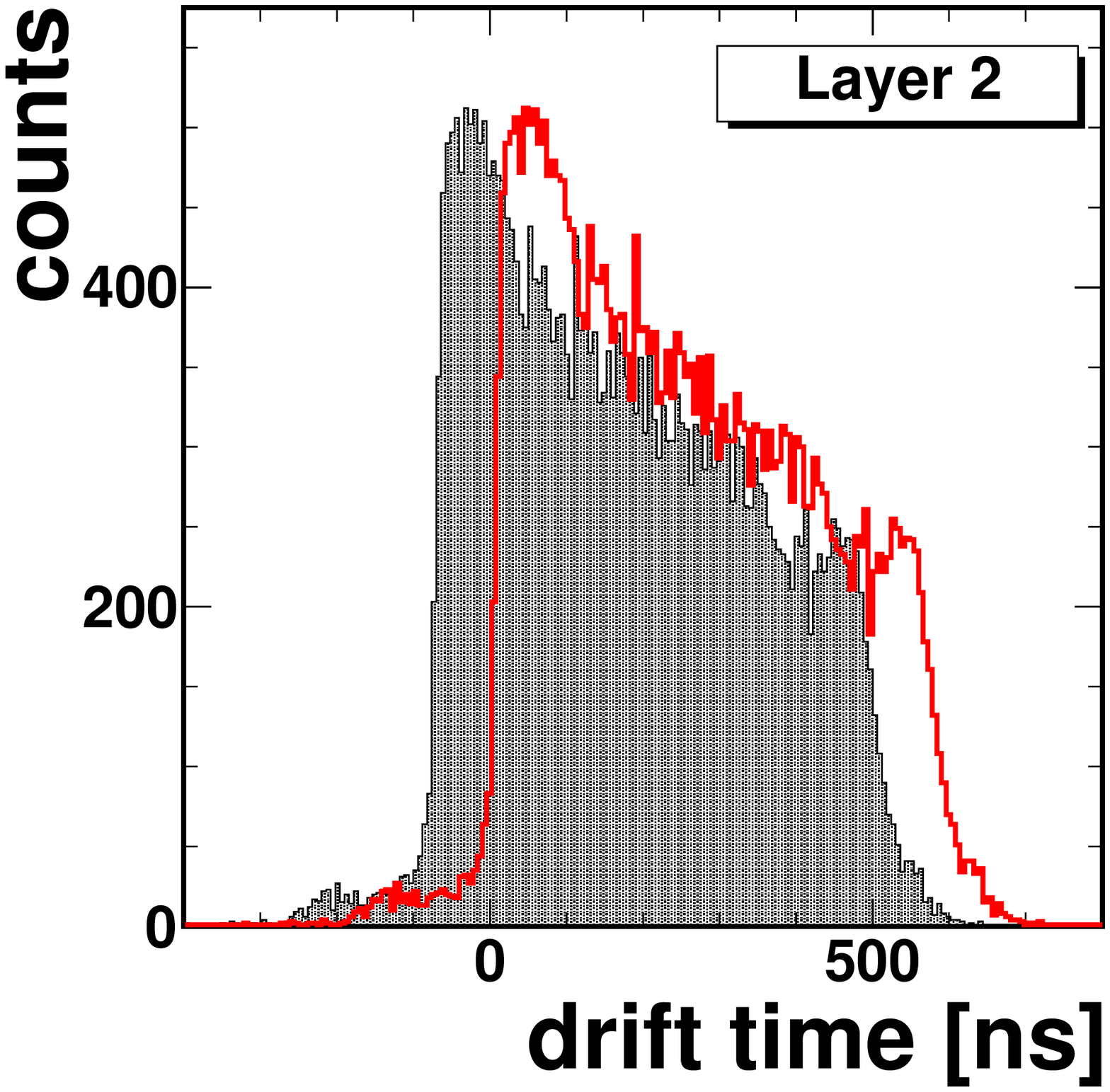}
\caption{}\label{fig:drift-time-offset-determination}
\end{center}
\end{subfigure}
\begin{subfigure}[c]{0.45\textwidth}
\begin{center}
\includegraphics[width=\textwidth]{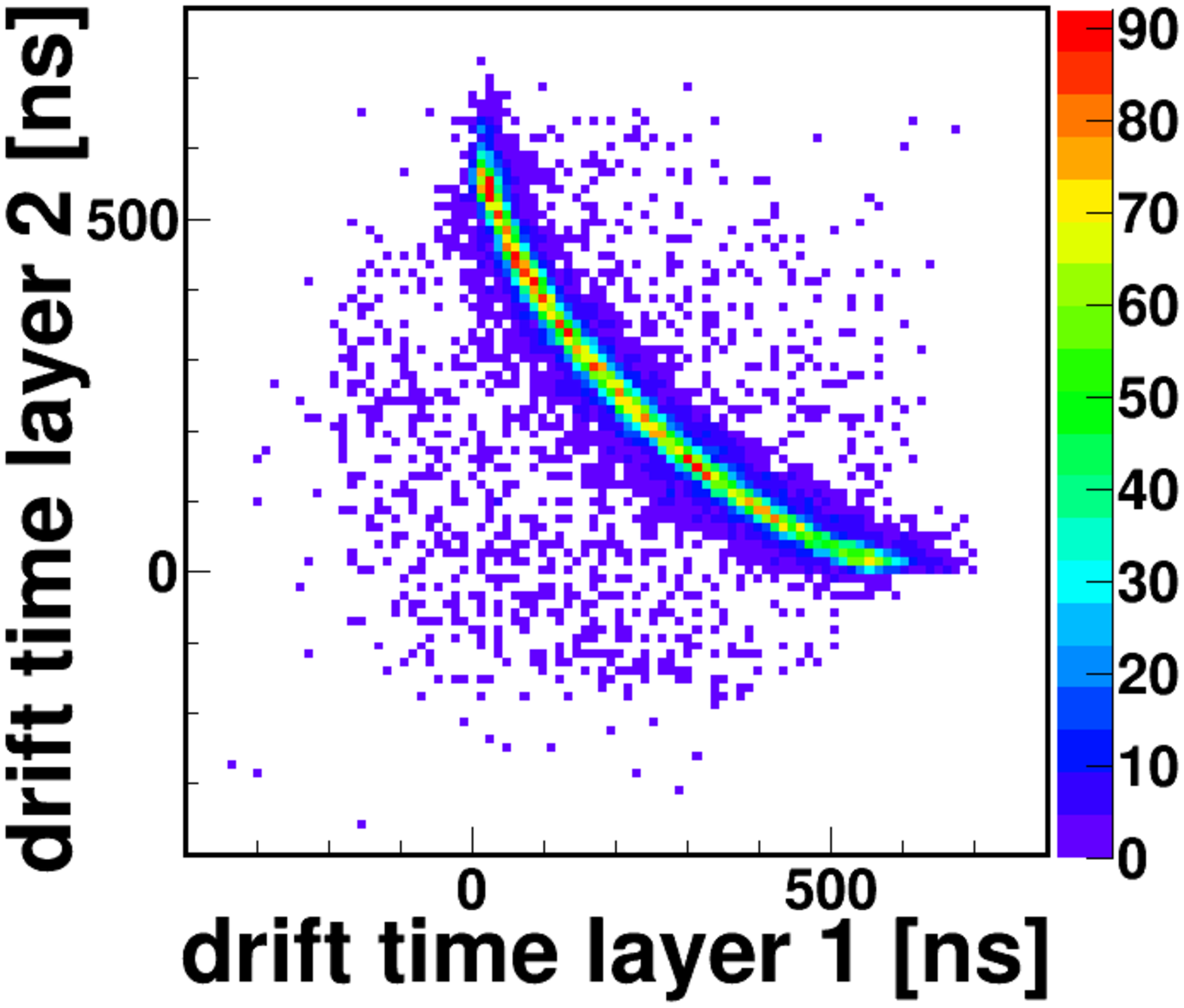}
\caption{}\label{fig:drift-time-offset-anticorr}
\end{center}
\end{subfigure}
\end{center}
\caption{Drift time spectra from the neighboring cells of subsequent wire planes with the same wire orientation (first and second layer) of the D1 drift chamber. The spectra before and after shifting by the offset are indicated with gray shading and red line, respectively. Zero is assumed to be the beginning of the shifted spectrum. Events with negative values of drift times after shift are random coincidences excluded from further analysis. b) Anti-correlation between drift times registered in the cells in layer 1 and layer 2. Events apart from the histogram maximum are random coincidences (background events).}\label{fig:drift-time-offset}
\end{figure}

In the first step the starting point of the drift time spectrum for each wire was determined and the spectrum was shifted by the appropriate offset (see Fig.~\ref{fig:drift-time-offset-determination}).

The drift times need to be extracted from the measured TDC values $t_{\text{TDC}}$. The $t_{\text{TDC}}$ can be expressed as a~sum:
\begin{equation}
t_{\text{TDC}} = t_{\text{real}} + t_{\text{driff}} + t_{\text{offset}} - t_{\text{trigger}},
\end{equation}
where $t_{\text{real}}$ is the time when a~particle passed through the drift cell, $t_{\text{driff}}$ denotes drift time, $t_{\text{offset}}$ consists of delays from electronics and $t_{\text{trigger}}$ indicates the time of the trigger common for all detectors.

The $t_{\text{real}}$ is connected with the time of flight of a~particle between the TOF-START detector and the D1 drift chamber. The distance between these detectors is in the order of 3 m while the velocities are: 0.966 $c$ for protons or antiprotons and 0.9992 $c$ for pions (both of momentum equal to 3.5 GeV/$c$). Resulting times of flight for pions and protons (or antiprotons) differ by about $0.3$ ns. Therefore, differences in the time of flight of different particles are assumed to be negligible in comparison to the drift time range (about 600 ns).

The histogram in the Fig.~\ref{fig:drift-time-offset-determination} shows an exemplary spectrum of times of signal registration in a~single wire of the D1 drift chamber corrected for the trigger time from the START detector ($t_{\text{trigger}}$). Red line indicates the beginning of the drift time spectrum. The drift time offset $t_{\text{offset}}$ is chosen so that the drift time spectrum begins in zero.

\section{Drift time - space calibration}\label{sec:dt-calibration}
The calibration procedure has to be done iteratively. This is required to optimize the drift time - space relations for all layers of the D1 drift chamber for a~given period of measurement. These relations may fluctuate due to the changes of temperature and atmospheric pressure in the experimental hall or changes in the gas mixture composition. Data presented in this thesis base on 10 hours of data taking.

\subsection*{Homogeneous irradiation method}
In the first step of the calibration procedure an approximate drift time - space relation for each layer of the drift chamber was determined by means of the homogeneous irradiation method. The method relies on the assumption that the number of particles passing by a~given width $x$ of a~drift cell is proportional to this width, i.e.:
\begin{equation}
N = c \text{d}x,
\end{equation}
where $c$ is a~constant. Number of particles in the range $dx$ is unambiguously connected with the drift time range $dt$ by the drift-time space relation:
\begin{equation}
\dfrac{\text{d}N}{\text{d}t}\text{d}t = c \text{d}x.
\end{equation}
The calibration curve $x(t)$ is obtained from the relations:
\begin{equation}
\begin{aligned}
x(t) &=  \dfrac{\mathop{\mathlarger{\int}}_{0}^t\dfrac{\text{d}N}{\text{d}t}dt}{c},\\
c &= \dfrac{\mathop{\mathlarger{\int}}_{0}^{t_{max}}\dfrac{\text{d}N}{\text{d}t}\text{d}t}{d},
\end{aligned}
\end{equation}
where $d$ is equal to half width of the drift cell ($d = 2$ cm). For the purposes of this method cumulative drift time spectra from all wires in the layer were created and integration was replaced by summing over all bins of the histogram in the range from 0 to the end of the spectrum.  In case of the P349 experiment the irradiation was not equal in all cells: the main contribution to the cumulative drift time spectrum comes from a~few (4 - 5) drift cells with similar number of particles passing through in a~given time. Nevertheless, this is enough to obtain a~reliable relation for further optimization.

\begin{figure}
\begin{center}
\begin{subfigure}[c]{0.45\textwidth}
\begin{center}
\includegraphics[width=\textwidth]{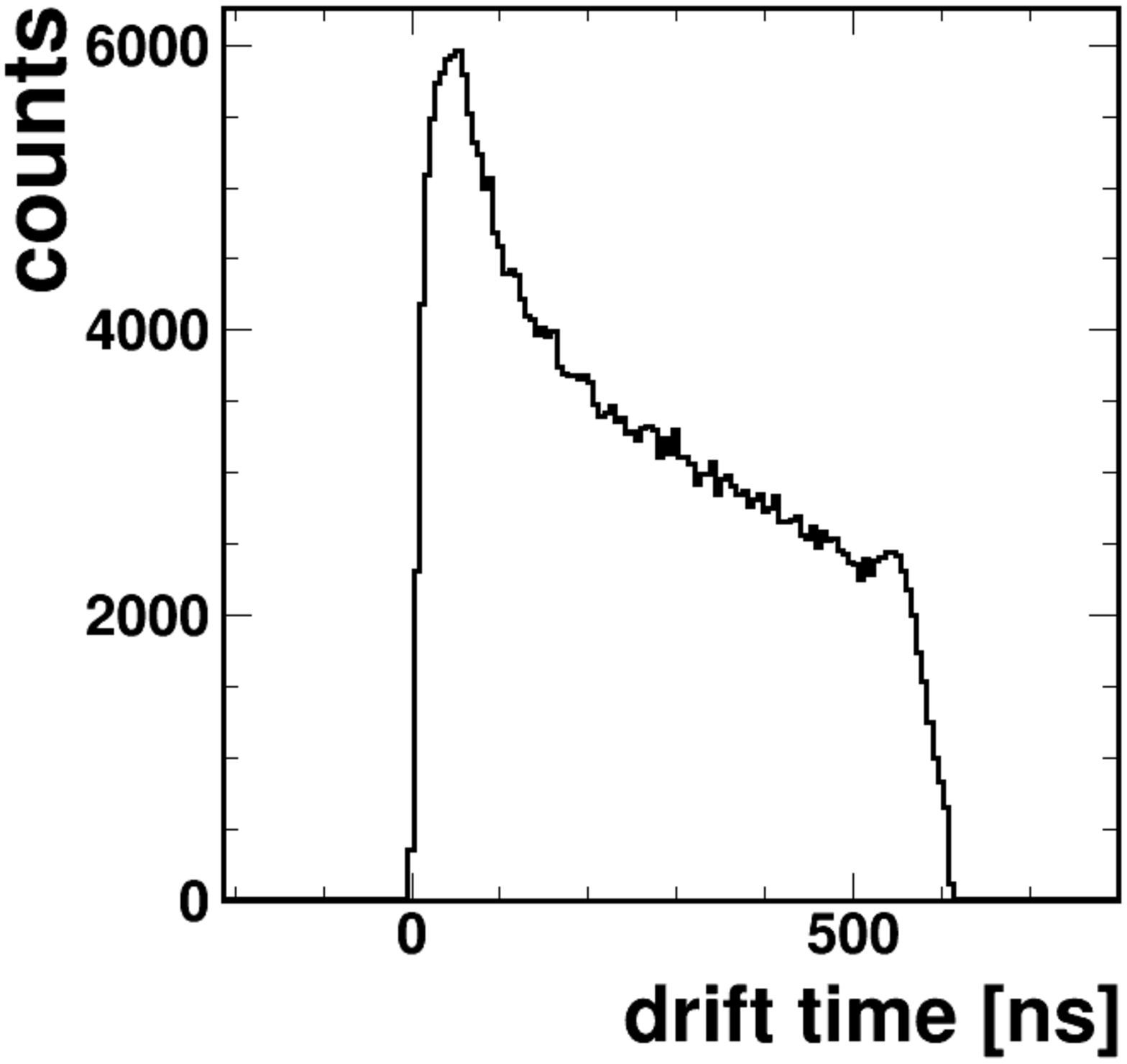}
\caption{}\label{fig:homogenous-calibration-drift-time-spectrum}
\end{center}
\end{subfigure}
\begin{subfigure}[c]{0.45\textwidth}
\begin{center}
\includegraphics[width=\textwidth]{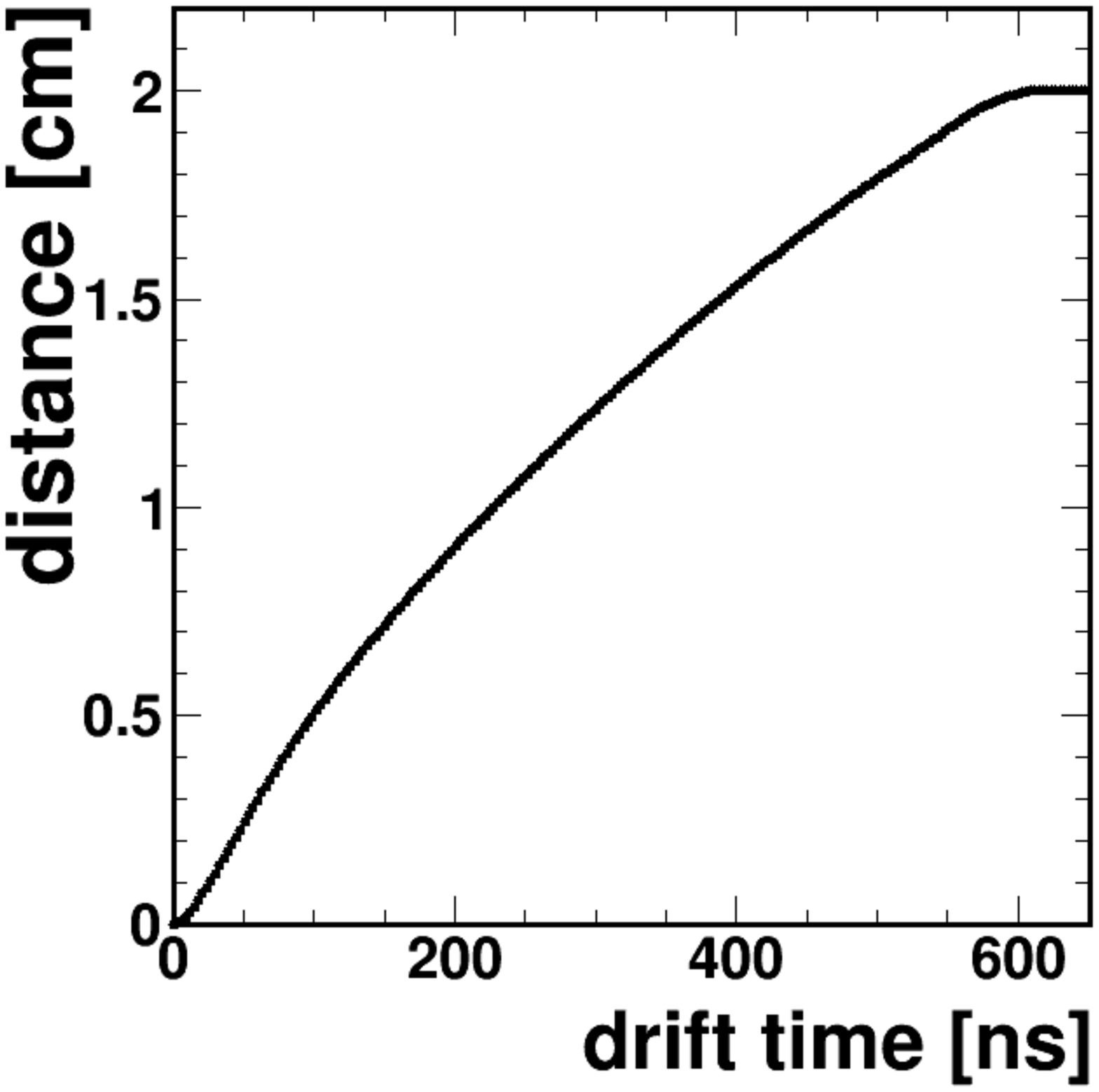}
\caption{}\label{fig:homogenous-calibration-dt-relation}
\end{center}
\end{subfigure}
\end{center}
\caption{a) a~typical cumulative distribution of the drift times spectra from all wires in one layer. b) Initial drift time - space relation obtained by means of homogeneous irradiation method for the same layer.}\label{fig:homogenous-calibration}
\end{figure}

An exemplary drift time spectrum and the resulting drift time - space relation are shown in the Fig.~\ref{fig:homogenous-calibration-drift-time-spectrum} and \ref{fig:homogenous-calibration-dt-relation}, respectively.

\subsection*{Optimization of the drift time - space relations}
The calibration is performed iteratively on the sample of events in which all layers had a~signal in exactly one drift cell.

At the beginning of each iteration the positions of the hits in all wire planes are calculated based on a~current drift time - space relations (in the first iteration the ones obtained with homogeneous irradiation method are used). Then the 3d straight tracks for all events are fitted. Tracks used in further steps of the calibration procedure for a~certain layer are obtained without including information from the considered layer (unbiased fit). 

\begin{figure}
\begin{center}
\includegraphics[width=0.45\textwidth]{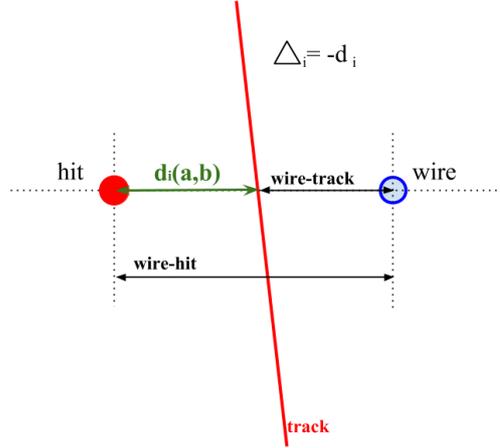}
\end{center}
\caption{An example of $\Delta_i$ position correction determination in the single event (green arrow). $\Delta_i$ is defined as a~difference between the wire-track and wire-hit distances.}\label{fig:distances-d-delta}
\end{figure}

\begin{figure}
\begin{center}
\begin{subfigure}[c]{0.47\textwidth}
\begin{center}
\includegraphics[width=\textwidth]{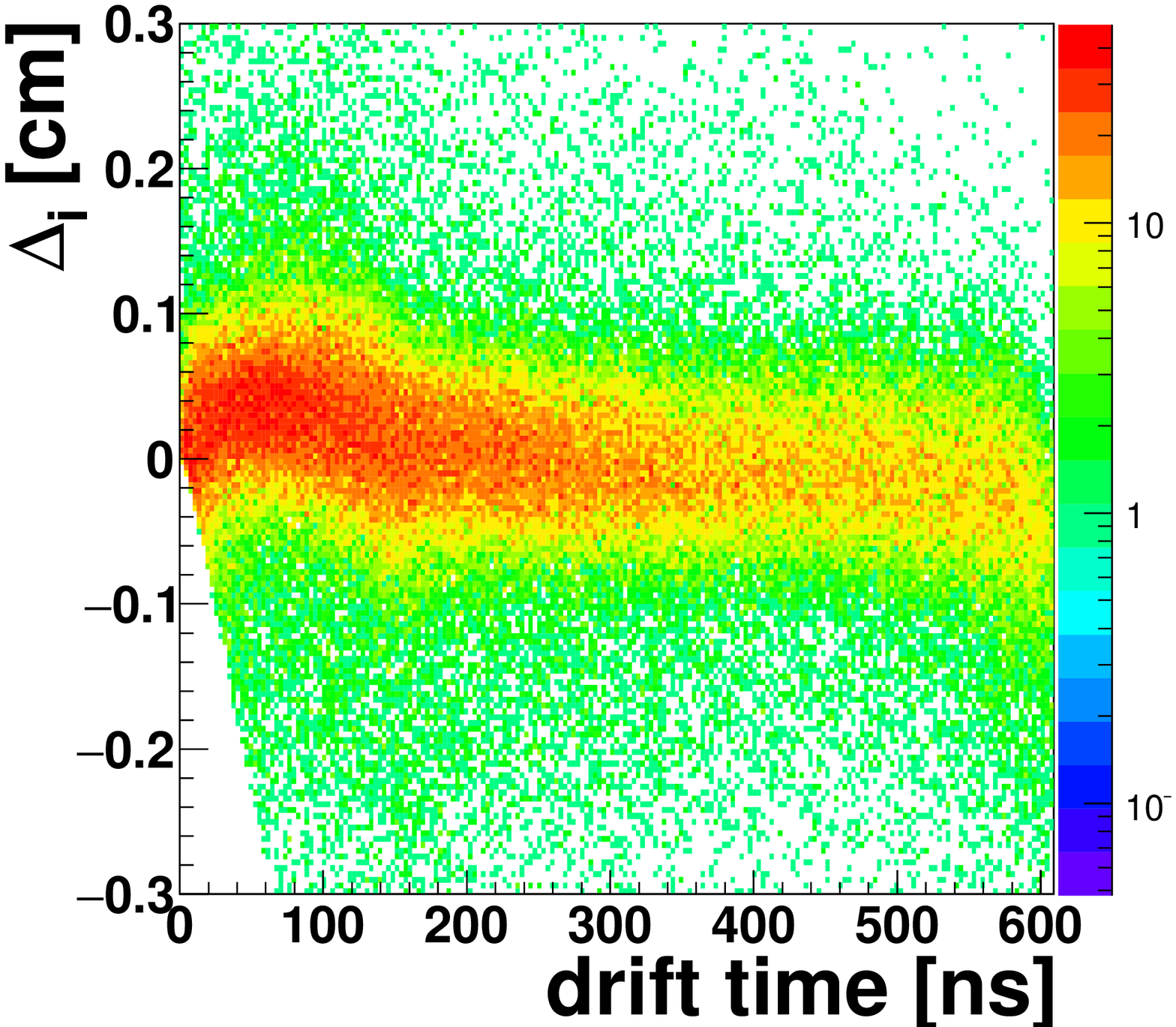}
\caption{}\label{fig:delta-projection-hist}
\end{center}
\end{subfigure}
\begin{subfigure}[c]{0.43\textwidth}
\begin{center}
\includegraphics[width=\textwidth]{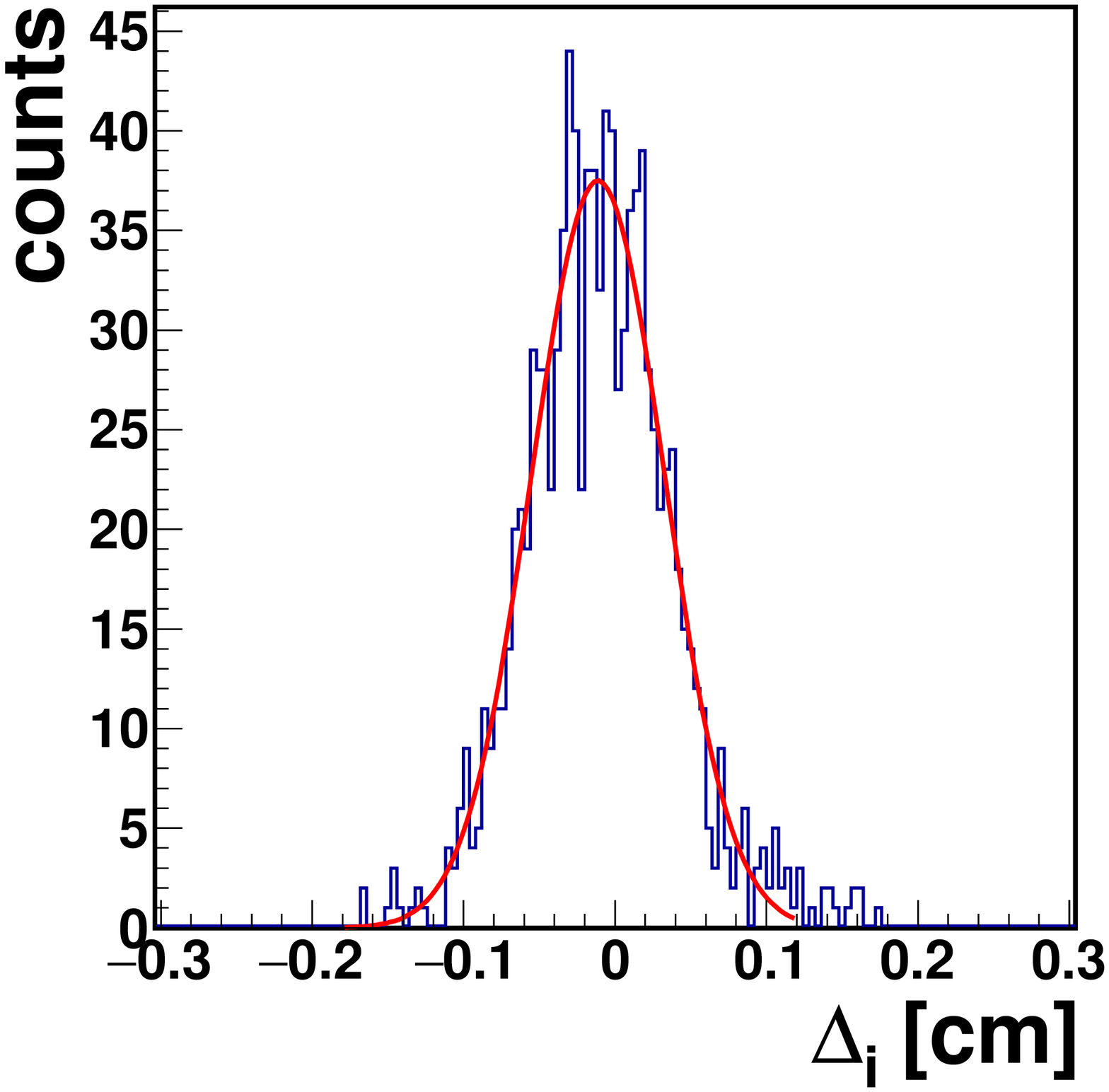}
\caption{}\label{fig:delta-projection-gauss}
\end{center}
\end{subfigure}
\end{center}
\caption{(a) Histogram of distances $\Delta_i$ vs. drift time. (b) An example of a~Gaussian function fit to the projection of a~single bin from the histogram shown in the part (a).}\label{fig:delta-projection}
\end{figure}

Further, the hit position corrections $\Delta_i$ defined as the differences between wire-track and wire-hit distances are calculated for each event (see Fig.~\ref{fig:distances-d-delta}). In order to extract the corrections of calibration curves the histograms of distances $\Delta_i$ vs. drift time are built separately for each wire plane (see Fig.~\ref{fig:delta-projection-hist}). In these histograms for each drift time bin (3 ns wide) a~projection is made onto the distance axis (y axis) and a~Gaussian function is fitted (exemplary projection and Gaussian fit is shown in the Fig.~\ref{fig:delta-projection-gauss}). For a~single time bin, the calibration curve is shifted by the mean value obtained from the fit. One standard deviation of the fitted Gaussian function is considered as the uncertainty of the position determination for the drift times belonging to the given bin. The obtained calibration is used as a~starting point for the next iteration. 

The definition of $\Delta_i$ justifies the usage of an unbiased fit. As reversed values of errors are used as weights of points while fitting, the distances between hit and reconstructed track are smaller for the layer with smaller errors in one iteration. Excluding a~layer for which a~track is reconstructed helps to avoid a~non-physical behavior of uncertainties after a~larger number of iterations. This is particularly important for the subsequent wire planes with the same wire orientation (especially when there is only one pair of planes with a~given orientation).

\begin{figure}
\begin{center}
\begin{subfigure}[c]{\textwidth}
\begin{center}
\includegraphics[width=0.45\textwidth]{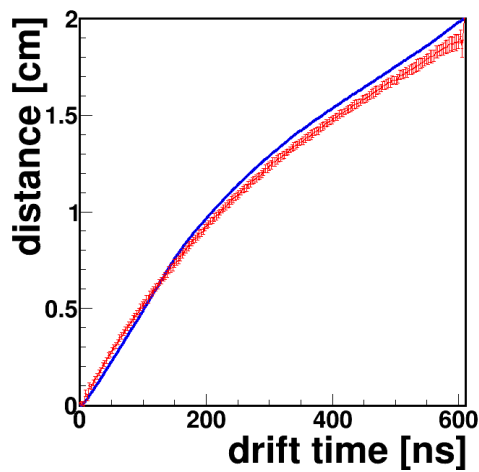}
\caption{}\label{fig:calib-res-curve}
\end{center}
\end{subfigure}
\begin{subfigure}[c]{0.45\textwidth}
\begin{center}
\includegraphics[width=\textwidth]{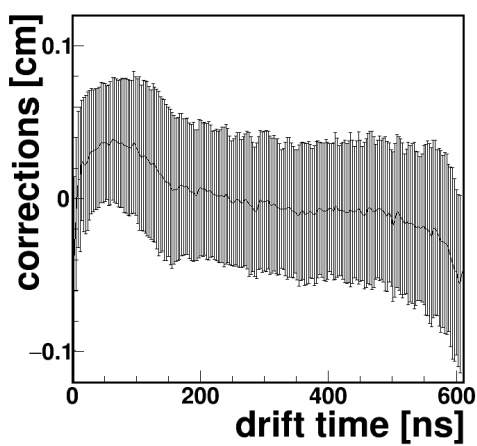}
\caption{}\label{fig:calib-res-corr-1}
\end{center}
\end{subfigure}
\begin{subfigure}[c]{0.45\textwidth}
\begin{center}
\includegraphics[width=\textwidth]{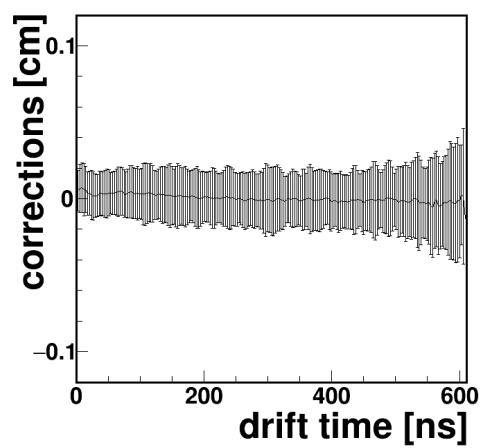}
\caption{}\label{fig:calib-res-corr-2}
\end{center}
\end{subfigure}
\end{center}
\caption{Results of calibration procedure. (a) Drift time space relations: initial (blue) and obtained after seven iterations (red). (b) Corrections of the calibration curve obtained in the 1st iteration. (c) Uncertainties of the calibration curve obtained in the last iteration (biased fit).}\label{fig:calib-res}
\end{figure}

An exemplary final calibration obtained after seven iterations is shown in the Fig.~\ref{fig:calib-res-curve} together with an initial calibration curve. With an increasing number of iterations, the corrections approach zero.

In order to determine the position resolution of final drift time - space relations firstly, a~biased fit to all layers is performed and $\Delta_i$ vs. drift time spectra are built. Uncertainties determined from the histograms are then included in another biased fit from which final uncertainties are determined in analogous way. In the Fig.~\ref{fig:calib-res-corr-2}  the values of corrections and their uncertainties are shown after the second biased fit. In case of iterations with biased fit, the position corrections are not applied.

A considerable discrepancy in position resolution for the biased and unbiased fit is visible. However, it can be explained not only by smaller number of point taken into account while fitting. Excluding a~certain layer from fit results in the bigger distance between the reconstructed track and hit position in this layer than in case of the biased fit. This causes bigger spread of the $\Delta_i$ values and therefore - bigger uncertainties of positions. The greater the uncertainty is, the less significant small variations of the distance between reconstructed track and hit and the track reconstruction is less precise (see Chapt. \ref{chapt:track-reco}). Including D2 drift chamber into the track reconstruction is supposed to reduce the described discrepancy as it will provide information from additional six layers placed at a~distance of about 50 cm from the edge of the D1 drift chamber.

The final uncertainties of positions are in the order of 150 - 200 $\rm \mu m$ in the range of drift times from about 100 ns to about 200 ns. Greater uncertainties for the drift times lower than 100 ns and greater than 500 ns result from the smaller number of hits registered: in the closest proximity of the wire this may be caused by short length of the ionization path and therefore production of a~charge insufficient for a~signal registration. For the signals from particles passing close to the drift cell edge, a~signal could be registered in more than one sense wire (which is not considered in this analysis) or not registered at all due to electron - ion recombination. If in one layer the particle passes through a~drift cell close to the sense wire, it is expected that in the subsequent layer it will pass through the cell relatively far from its sense wire and and vice versa. If the signal in one of these layers is not registered the event is not considered as useful for the calibration.

Obtained values of position resolution are comparable with the position resolution of 100 - 200 $\rm \mu$m  achieved in the COSY-11 experiment where D1 and D2 drift chambers were also used for tracking~\citep{cosy-11-res}. 
The D2 drift chamber is planned to be included in the calibration procedures as soon as its relative position optimization is finished.

%% file: TrackReconstruction.tex
\chapter{Track reconstruction} \label{chapt:track-reco}
Track reconstruction relies on the fact that the particle trajectory in the drift chambers is a~straight line. A~line in 3-dimensional space is described by the coordinates of a~point and a~vector.

In order to reconstruct a~track, first, coordinates of the point and vector are calculated analytically (see Sec. \ref{sec:track-calc}). Secondly, results of calculations are used as the initial conditions of a~numerical minimization procedure (see Sec. \ref{sec:track-fit}).

\begin{figure}
\begin{center}
\includegraphics[width=0.75\textwidth]{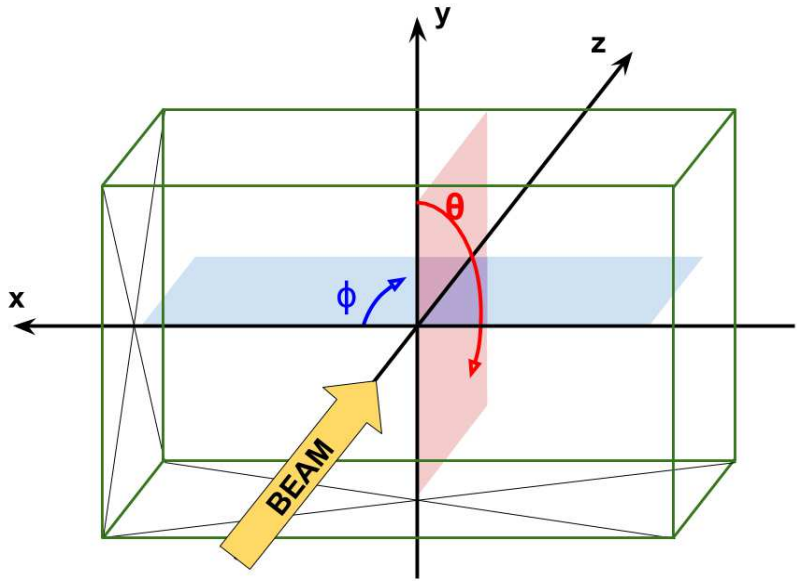}
\end{center}
\caption{Definition of the coordinate system (placed in the geometrical center of the D1 drift chamber) with indicated orientation for the $\phi$ and $\theta$ angles measurement.}\label{fig:angles}
\end{figure}

Furthermore, the coordinate system is chosen in such a~way that the beam direction defines $z$-axis, $y$-axis points up, $x$-axis direction is defined by requiring the right-handedness of the coordinate system. Plane $xy$ is parallel to the wire planes and the beginning of the coordinate system is placed in the geometric center of the drift chamber. Polar angle $\theta$ and azimuthal angle $\phi$ are defined as shown in the Fig.~\ref{fig:angles}.

The values of $z$ coordinates of hits are determined based on the wire plane number. Positions along $x$ and $y$-axis are calculated from the cell number and information about the angle of wires in a~given layer. The left-right ambiguity is resolved as it was explained in Sec. \ref{sect:ExperimentalSetup-detectors}.

\section{Analytical calculation of track parameters}\label{sec:track-calc}
For a~charged particle trajectory reconstruction in a~drift chamber an information from wire planes of three different wire orientations is needed. The analytic approach starts from the reconstruction of the event in two dimensions, separately for each orientation.

For this the plane perpendicular to the sense wires is considered. In this plane the reconstruction of the particle trajectory simplifies to a~two-dimensional line fit, in case of vertical wires in the $xz$ plane. For inclined layers there are only two wire planes for each direction, therefore the coefficients of the track can be directly calculated. For the vertical wires the coefficients are obtained from a~numerical minimization of the squared distances:
\begin{equation}
\left(\chi^2\right)_{2d} = \mathlarger{\mathlarger{\sum}}_{i=1}^{4} d_i^2,
\end{equation}
where the distances $d_i$ are measured as shown in the Fig.~\ref{fig:distances-d-delta}.

The reconstructed tracks are in fact three hit planes in 3d space: these planes are parallel to the sense wires in the $xy$ plane and their $xz$ direction is defined from the described line fit. Therefore, each plane can be described by two linearly independent vectors $\vec{u}_{i(xy)}$ and $\vec{u}_{i(xz)}$ in three dimensional space ($i$ distinguishes between hit planes, $i$ = 1 - 3). The signs of these vectors are not important and can be chosen arbitrary). Therefore a~hit plane is described parametrically as a~set of points whose vectors representing the position ($\vec{p}_{i(hit)}$) satisfy:
\begin{equation}
\vec{p}_{i(hit)} = \vec{p_0} + s \vec{u}_{i(xy)} + t \vec{u}_{i(xz)},
\end{equation}
where $s$ and $t$ are real numbers and  $\vec{p}$ is a~position of the point which belongs to the plane. The normal to the $i$-th hit plane $\hat{n}$ is calculated as:
\begin{equation}
\hat{n_i} = \dfrac{\vec{u}_{i(xy)}\times\vec{u}_{i(xz)}}{\vert \vec{u}_{i(xy)}\times\vec{u}_{i(xz)} \vert}.
\end{equation}

In the ideal case, the intersection of all planes should be a~single straight line. In~presence of experimental uncertainties and due to imperfect drift time - space relation, there are three intersection lines obtained (one per pair of different hit planes). An~intersection line of hit plane $i$ and $j$ is a~set of points parametrized by:
\begin{equation}
\vec{r}_{ij(int)} = \vec{r}_{ij(0)} + t\vec{a}_{ij},
\end{equation}
where $\vec{r}_{ij}$ is a~point which belongs to the line, $\vec{a}$ is its direction vector $\hat{a}_{ij}=\hat{n_i}\times\hat{n_j}$) and $t$ is a~real number.

Finally, the track equation is calculated:
\begin{equation}\label{eq:track-rec-calc}
\begin{aligned}
\vec{r}_{track} &= \vec{r}_0 + t\vec{a}_{track},\\
r_{0}^{k} &= \dfrac{1}{3}\mathlarger{\mathlarger{\sum}}_{i,j,i\neq j} r^k_{ij(0)},\\
a_{track}^k &= \dfrac{1}{3}\mathlarger{\mathlarger{\sum}}_{i,j,i\neq j} a^k_{ij}, \hspace{12 pt} k = {x,y,z}.
\end{aligned}
\end{equation}

For the determination of $\vec{u}_{i(xz)}$ there are at least two hits per each wire plane orientation needed, as only this allows to resolve left-right ambiguity.

The mean track as determined as in the Eq. \ref{eq:track-rec-calc} is in general not the optimal approach towards the track reconstruction as it does not include a~weighting due to the number of hits and uncertainties of hits positions from the calibration. To find the optimum track a~minimization of the squared error sum is performed.

\vspace{-20pt}
\section{Optimization of track parameters}\label{sec:track-fit}
The method described in the previous section provides an approximate equation of a~particle trajectory. In this section a~proposed algorithm of its optimization is described.

The new optimum equation of the track has to be found. It is given by: 
\begin{equation}
\vec{r'}_{track} = \vec{r'}_0 + t\vec{a'}_{track}.
\end{equation}

There are four minimization parameters chosen: $\vec{r'^x}_0$, $\vec{r'^y}_0$, $\vec{a'^x}_{track}$, $\vec{a'^y}_{track}$ as the reference point is put in $z=0$ (geometric center of the drift chamber) and vector $\vec{a'}_{track}$ can always be normalized so that $\vec{a'}_{track}= 1$.

\begin{figure}
\begin{center}
\includegraphics[width=0.6\textwidth]{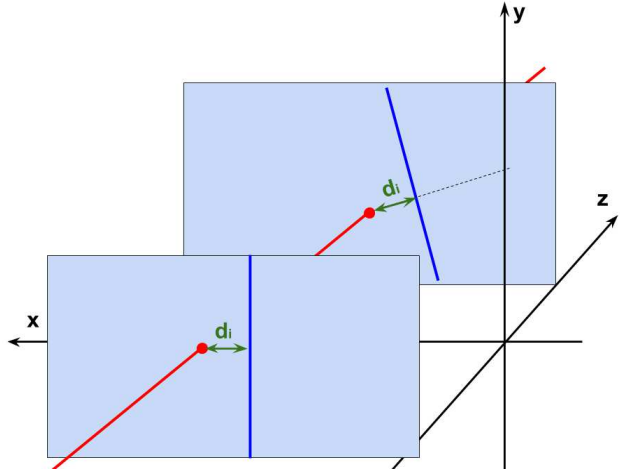}
\caption{The idea of definition of $d_i$ in a~layer with straight and inclined wires. A~track (red line) is reconstructed by minimization of the sum $d^2_i$. $d_i$ are defined in the wire planes and calculated perpendicularly to the wire (blue line).}\label{fig:dist-scheme}
\end{center}
\vspace{-11pt}
\end{figure}

The parameters are obtained by means of numerical minimization of the expression:
\begin{equation}
\left(\chi^2\right)_{3d} = \mathlarger{\mathlarger{\sum}}_{i=1}^{8} \dfrac{d_i^2(r'^x_0, r'^y_0, a'^x_{track}, a'^y_{track})}{\sigma_i^2},
\end{equation}
where $d_i$ is the distance between hit (calculated from drift time - space relation) and the point where track intersects $i$-th wire plane (as shown in the Fig.~\ref{fig:dist-scheme}). $d_i$ depends on all minimization parameters. Initial values of the parameters are taken from $\vec{r}_0$ and $\vec{a}_{track}$.

\begin{figure}
\begin{center}
\begin{subfigure}[c]{0.45\textwidth}
\begin{center}
\includegraphics[width=\textwidth]{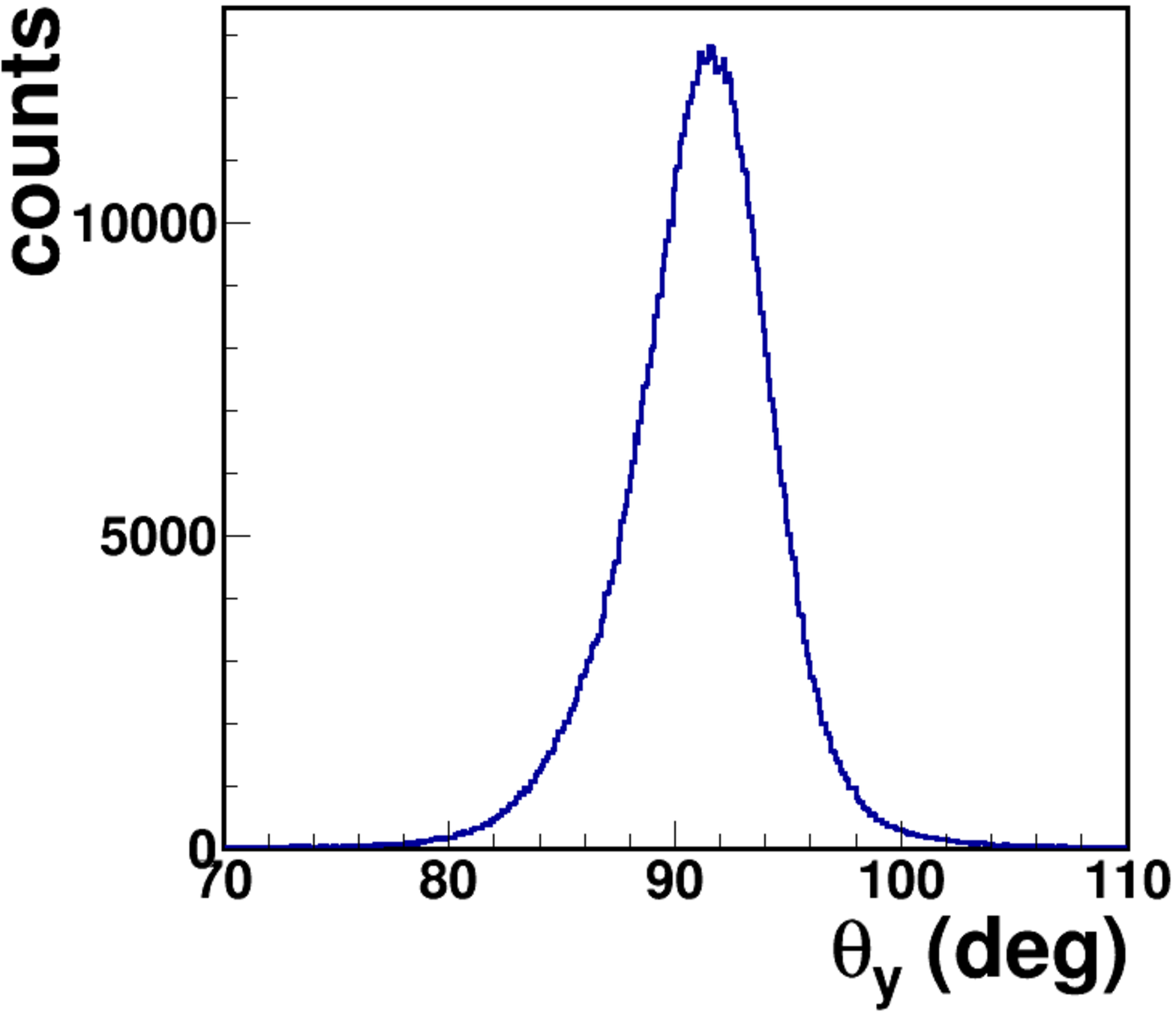}
\caption{$\theta$ distribution.}\label{fig:angle-theta}
\end{center}
\end{subfigure}
\begin{subfigure}[c]{0.45\textwidth}
\begin{center}
\includegraphics[width=\textwidth]{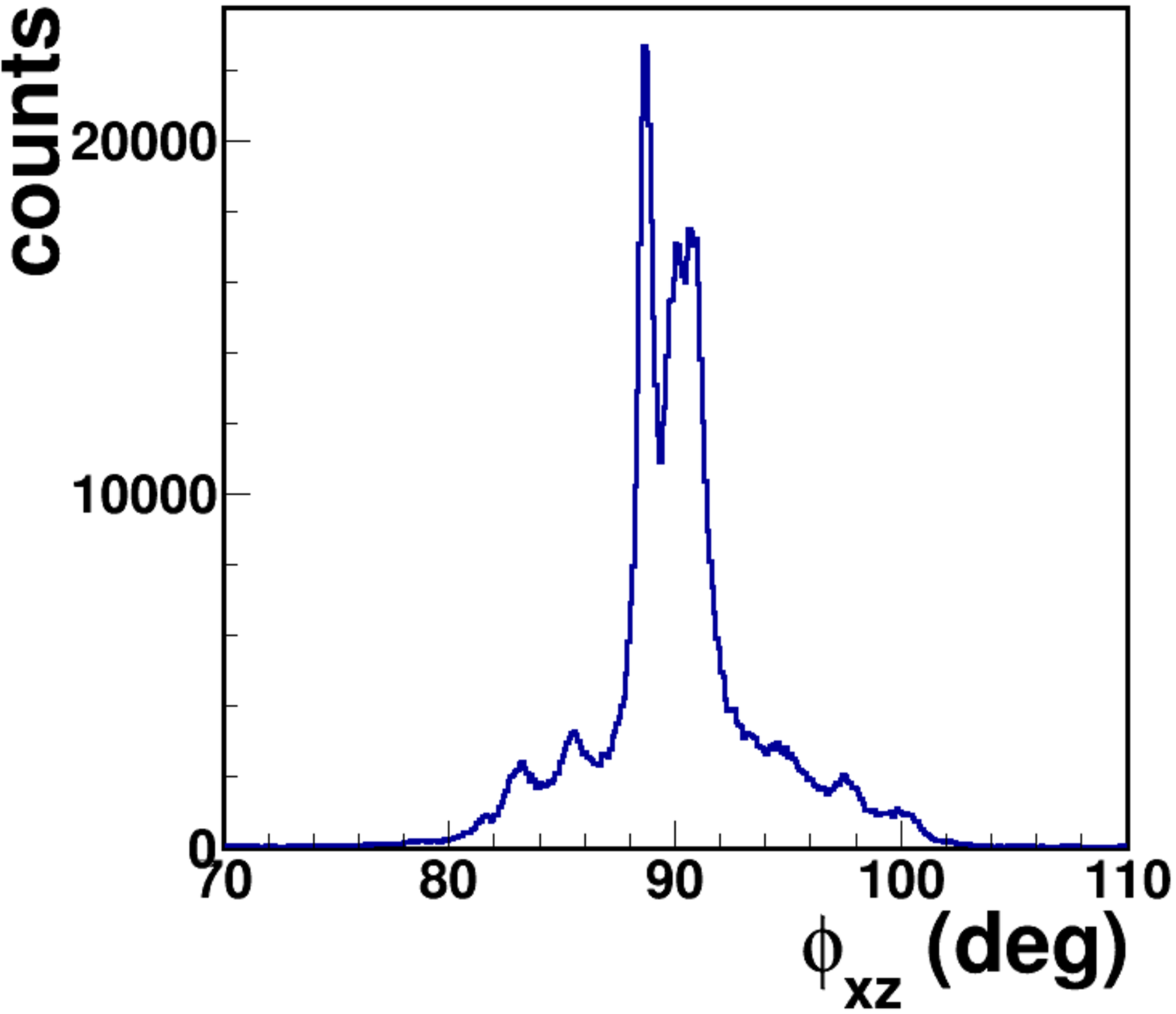}
\caption{$\phi$ distribution.}\label{fig:angle-phi}
\end{center}
\end{subfigure}
\end{center}
\vspace{-11pt}
\caption{Results of the track reconstruction. The definitions of the angles are defined in the \mbox{Fig.~\ref{fig:angles}}.}\label{fig:track-fit-res}
\end{figure}

The tracks were reconstructed for the same sample of events used for the calibration. The obtained distributions of $\phi$ and $\theta$ angles are shown in the Fig.~\ref{fig:track-fit-res}. The $\theta$ distribution is rather symmetrical with respect to $\sim$90$\rm ^o$ but it is shifted towards larger angles which shows that the beam is not perpendicular to the drift chamber. Two maxima in the $\phi$ distribution can be identified with the tracks going through the analyzing target and another group of tracks with directions different from the expected beam intensity maximum. Those tracks originate mainly from the beamline and its walls~\citep{thomas}. For the calibration purposes the origin of the tracks is unimportant. In view of identification of antiprotons scattered on the analyzing target, in the further analysis information from the fiber hodoscope and Cherenkov detector needs to be included.

This method allows to reconstruct the track without information from all layers. For instance, during the calibration procedure there are eight tracks determined for all events. In each of the eight fits a~different layer is excluded, i.e. unbiased track for $j$-th layer is obtained by minimization of the expression:
\begin{equation}
\left(\chi^2\right)_{3d}^{j} = \mathlarger{\mathlarger{\sum}}_{i=1,i\neq j}^{8} \dfrac{d_i^2(r'^x_0, r'^y_0, a'^x_{track}, a'^y_{track})}{\sigma_i^2}.
\end{equation}
The initial parameters $\vec{r}_0$ and $\vec{a}_{track}$ which are taken from the biased fit result do not change in the minimization procedure for the set of the eight unbiased fits, in a~given event.

\nopagebreak[4]Furthermore, the extension of this method for a~greater number of wire planes (from HEX and/or D2 drift chambers) is straightforward: a~sum in the $\left(\chi^2\right)_{3d}$ needs to go over contributions from all analyzed layers. As it is expected that the in the majority of events scattering in the liquid hydrogen target did not occur, this method can also be useful when performing a~calibration of the relative position of a~given drift chamber with respect to other detectors. In order to find the correct position the relevant detector has to be shifted (rotated) by a~small amount and for each position the value of the $\left(\chi^2\right)_{3d}$ has to be determined. The value of $\left(\chi^2\right)_{3d}$ plotted as a~function of the shift (angle) is expected to have a~minimum in the position closest to the real detector position during the experiment. 

%% file: Conclusions.tex
\chapter{Conclusions and plans} \label{chapt:conclusions}
The main goal of this thesis was to perform the drift time - space calibration of the D1 drift chamber and to prepare and test the procedure for a~charged particle trajectory reconstruction in 3d based on the information from this drift chamber.

Experimental data collected during about 10 h of measurement were analyzed. For the calibration only events with single track signature were chosen in order to perform a~simultaneous calibration for all eight layers of the D1 drift chamber. The 3d track reconstruction was tested on the same event sample. 

In the first step of the calibration procedure offsets of the drift time spectra were determined and cumulative drift time spectra for all layers were prepared. Then, drift time - space relations for all wire planes were obtained via the homogeneous irradiation method. As only a~few most central cells of each wire plane in the drift chamber were irradiated with high and comparable intensity these relations required further optimization.

For this purpose the iterative calibration procedure was prepared. In each of its steps the distances corresponding to given drift time bins were one by one shifted to provide better agreement of reconstructed hit positions with fitted track. As expected, with increasing number of iteration, the corrections values approached zero. After seven iterations resulting uncertainties of the position determination are in the order of \mbox{150 $\rm \mu m$} to about \mbox{220 $\rm \mu m$} for the range of drift times from about \mbox{100 ns} to \mbox{500 ns} which is consistent with results obtained in the COSY-11 experiment~\citep{cosy-11-res} where the set of D1 and D2 drift chambers was also used (but with different gas mixture). Bigger uncertainties for the remaining drift time ranges can be explained by the difficulty in registration and therefore a~worse statistics of the tracks which passed in the close proximity or far from the sense wires. 

The 3d particle trajectories were reconstructed as straight lines in 3d. In the reconstruction procedure the track parameters were first calculated analytically. This results were then used as initial conditions of the numerical minimization which aims to minimize distances between the fitted track and hit positions within the wire planes. Although the procedure was tested on the sample of events in which all wire planes had exactly one cell with signal, the procedure itself allows for track parameter determination also in case when information from some wire planes is missing.

Three dimensional track distributions were obtained. The tracks distribution in the $yz$ plane was rather symmetrical, with maximum slightly shifted from 90$\rm ^o$ which indicates that the beam direction was not perpendicular to the drift chamber plane. The resulting $\phi$ angles show that the track distribution was not symmetrical in the $xz$ plane: one part of tracks passes through the analyzer target, however a~group of tracks originates from the beamline and its walls. The origin of tracks in not important for the calibration procedure as long as the distribution of angles is not too broad. This is due to the fact that drift time - space relations might be different for tracks passing through the drift chamber at different angles.

Prepared procedures are easy to be adapted for other drift chambers. Further works on tracking will focus on analogous D2 and HEX calibration and optimization of the relative positions of all drift chambers based on reconstruction of unscattered events. Moreover, incorporation of track finding algorithms (e.g. Hough transformation) is planned as it would allow for track identification and determination of its initial parameters in case of noisy events or events with more than one particle passing through the drift chamber.

Finally, for the antiproton identification information from the Cherenkov detectors need to be included into the analysis as well and only the tracks scattered in the analyzer target need to be chosen.